\documentclass[runningheads,natbib]{svjour3}                     

\smartqed  
\usepackage{graphicx}
\usepackage{natbib}
\usepackage{mathptmx}      
\usepackage{amsmath}
\usepackage{amsfonts}
\usepackage{amssymb}
\usepackage{amsbsy}
\usepackage{color}
\usepackage{latexsym}
 \journalname{The Astronomy and Astrophysics Review}

\newcommand{\thetabn}{\theta_\mathit{Bn}}

\begin{document}

\title{\textsf{Fundamentals of collisionless shocks for astrophysical application, 2. Relativistic shocks}
}

\titlerunning{Collisionless Relativistic Shocks}        

\author{\textsf{A~M~Bykov and R~A~Treumann$^*$}\thanks{$^*$Visiting the International Space Science Institute Bern, Switzerland.}      
}

\authorrunning{A M Bykov and R A Treumann
} 

\institute{A M Bykov \at
             Ioffe Institute for Physics and Technology, 194021 St.Petersburg, Russia\\ \email{byk@astro.ioffe.ru} \\  
           \and
                 R A Treumann\at
              Department of Geophysics and Environmental Sciences, Geophysics Section, Ludwig-Maximilians-University Munich,
Theresienstr.\,37-41, 80333\,Munich, Germany \\ \email{rudolf.treumann@geophysik.uni-muenchen.de}     \and Department of Physics and Astronomy, Dartmouth College, Hanover, NH\,03755, USA \\
                                                \emph{Present address: International Space Science Institute, Hallerstrasse 6, 3012 Bern, Switzerland \\ }
}
\date{Received: date / Accepted: date}

\maketitle
\vspace{-0.3cm}
\begin{abstract}
{In this concise  review of the recent developments in relativistic shock theory in the Universe we restrict ourselves to shocks that do not exhibit quantum effects. On the other hand, emphasis is given to the formation of shocks under both non-magnetised and magnetised conditions. We only briefly discuss particle acceleration in relativistic shocks where much of the results are still preliminary. Analytical theory is rather limited in predicting the real shock structure. Kinetic instability theory is briefed including its predictions and limitations. A recent self-similar relativistic shock theory is described which predicts the average long-term shock behaviour to be magnetised and to cause reasonable power law distributions for energetic particles. The main focus in this review is on numerical experiments on highly relativistic shocks in (i) pair and (ii) electron-nucleon plasmas and their limitations. These simulations do not validate all predictions of analytic and self-similar theory and  so far they do not solve the injection problem and the self-modification by self-generated cosmic rays. The main results of the numerical experiments discussed in this review are: (i) a confirmation of shock evolution in non-magnetised relativistic plasma in 3D due to either the lepton-Weibel instability (in pair plasmas) or to the ion-Weibel instability; (ii) the sensitive dependence of shock formation on upstream magnetisation which  causes suppression of Weibel modes for large upstream magnetisation ratios $\sigma>10^{-3}$; (iii)  the
sensitive dependence of particle dynamics on the upstream magnetic inclination angle $\thetabn$, where particles of $\thetabn>34^\circ$ cannot escape upstream, leading to the distinction between `sub-luminal' and `super-luminal' shocks; (iv) particles in ultra-relativistic shocks can hardly overturn the shock and escape to upstream; they may oscillate around the shock ramp for a long time, so to speak `surfing it'  and thereby becoming accelerated by a kind of SDA; (v) these particles form a power law tail on the downstream distribution; their limitations are pointed out; (vi) recently developed methods permit  the calculation of the radiation spectra emitted by the downstream high-energy particles; (vii) the Weibel-generated downstream magnetic fields form large amplitude vortices which could be advected by the downstream flow to large distances from the shock and possibly contribute to an extended strong field region; (viii) if cosmic rays are included, Bell-like modes can generate upstream magnetic turbulence at short and, by diffusive re-coupling, also long wavelengths in nearly parallel magnetic field shocks; (ix) advection of such large-amplitude waves should cause periodic  reformation of the quasi-parallel shock and eject large amplitude magnetic field vortices downstream where they contribute to turbulence and to maintaining an extended region of large magnetic fields.
}
\keywords{Collisionless shocks \and relativistic shocks \and generation of magnetic fields \and Weibel modes \and Bell modes \and Gamma Ray Bursts \and Pulsar Wind Nebulae termination shocks \and external shocks \and internal shocks \and particle acceleration \and shock radiation \and downstream turbulence}
 \PACS{95.30.Qd \and 52.35.Tc \and 52.72.+v}
\end{abstract}

\section{\textsf{Introduction}}
\label{sec:intro}
\noindent
Key observations in astrophysics suggest that relativistic collisionless shocks play an important if not a central role in the universe. Since many of the astrophysical processes are quite violent, i.e. release large amounts of energy on short time scales, it is quite reasonable to expect that shocks do frequently form either as explosion driven blast waves, caused by ultra-relativistic outflows of matter encountering an obstacle or interacting with other flows, or by nonlinear growth and steeping processes of unstably excited waves in those flows; and that a substantial fraction of these shocks will be relativistic. These two  ways of shock formation already lead to the distinction of two types of relativistic shocks: \emph{external} shocks produced in the interaction of flow with an external medium, and \emph{internal} shocks which evolve in the absence of any external obstacle inside the flow. 

Formation of shocks is a well-proven fact in the manifestly collisionless plasma of interplanetary space which has been, is and for long time will be the only place in the Universe where large-scale collisionless shocks can be studied \emph{in situ}. Among others, one important lesson that can be learned from their observation is that none of them form on the scale of the mean free path $\lambda_\mathit{mfp}$ (or `resistive scale') which in interplanetary space with its characteristic dimension $\sim 10^{\,2}$ AU to $10^{\,3}$ AU (the size of the heliosphere) is of the order of several AU. Without any exception the widths of collisionless shocks are of the order of the ion inertial or ion gyro-radius scales, i.e. $\lesssim10^{-6}\lambda_\mathit{mfp}$. This implies that even in astrophysical systems whose global dimensions $L\gg\lambda_\mathit{mfp}$ grossly exceed their internal mean free paths, collisionless shock formation proceeds on micro-scales $\Delta_s$ that are far below the collisional mean free path with shock transition widths $\Delta_s\ll\lambda_\mathit{mfp}\ll L$.  

There are no relativistic shocks in the reach of any man-made spacecraft, however. The shocks in the solar system, the heliosphere, and the heliospheric surrounding are without any exception non-relativistic. Therefore, all information on relativistic shocks other than observational is based on theoretical arguments, ingenious though sometimes questionable speculation, numerical simulations or generalisation of our heliospheric knowledge of nonrelativistic collisionless shocks to the relativistic domain. The most contemporary and comprehensive accounts of the latter have been given in \citet{Balogh11}. Such arguments rely solely on indirect observations of either high-energy particles (Cosmic Rays), intense electromagnetic radiation emitted from certain extended objects, particularly from those that are narrow in their transverse dimension, or on Gamma Ray Bursts (GRBs). For these reasons nothing about the very structure of collisionless relativistic shocks can be said that is comparably precise to our knowledge on non-relativistic shocks.

From an astrophysical point of view lack of the precise structure of relativistic shocks might, in fact, not be as important as it is for the near-Earth non-relativistic shocks. In most cases it suffices to assume that certain global conditions are specified at a relativistic shock: its speed, Mach number, compression ratio and magnetic-field strength. This is also the historical path along which reference to relativistic shocks has in the past been gone. With these characteristics in mind that are based on global plausibility considerations, the shock serves as an instrument for the generation of the two primary astrophysical effects that can be observed by remote sensing: Cosmic Rays and radiation, respectively, energetic  (i.e. accelerated)  particles, and photons. The former are described by a set of measured numbers --  composition, abundance, anisotropy, and the properties of their energy spectrum like shape, slope, breaks, cut-offs etc. --, which contain  information about the source of Cosmic Rays, their generators and a few properties of the generators. 

Radiation in the electromagnetic spectrum provides information on the spatial location of the shocks, their geometrical shape, their relation to the astrophysical object, and on processes inside and in the environment of the shock and thus on the mechanism of shock production and possibly even the shock structure under various conditions. It is mostly the generation of  electromagnetic radiation that requires knowledge about the processes that generate and maintain shocks. This, however, is a wide field as the generation of radiation implies the presence of particles which emit radiation. At a shock, such particles must be prepared by the shock to generate the radiation, which is a most complicated and  unresolved problem. Unlike Cosmic Rays, which reach our solar system, the shock-prepared particles cannot be directly observed. 

In order to understand the production of Cosmic Rays, on the other hand, the mechanism of shock generation and shock structure are of lesser importance as long as the fraction of shock-accelerated energetic particles that are present in the shock environment remains low. When this condition is violated, then the Cosmic Rays do also contribute to shock structure and formation by ``mediating" the shock. 

Because they are very strong  radiators (much stronger than non-relativistic shocks) and therefore must be efficient particle accelerators, emissions from narrow regions in astrophysical space has led to the suspicion that very high Mach number shocks reaching into the relativistic domain cause the observed effects. 
In most of these cases the existence of relativistic shocks in astrophysics has been assumed and the conditions from non-relativistic shock calculations like, for instance, the shock compression ratio, have been extrapolated into the relativistic domain. Though this naturally is a first reasonable approach to an unexplored field, more recent numerical investigation based on particle-in-cell (PIC) simulations demonstrated that relativistic shocks exhibit particular properties which are quite different from those of ordinary collisional gasdynamic and also collisionless plasma shocks. These properties are related to the self-consistent generation of magnetic fields, production of fast particles, which tend to modify the shock, and intense radiation. The present review of some of the contemporary achievements extends a previous publication \citep{Treumann09} into the relativistic domain.

\section{\bf\textsf{Observational Evidence for Relativistic Shocks}}

Inference on relativistic shocks is obtained either from observations of high-energy radiation produced by the shock-accelerated particle component or, indirectly, from the detection of high-energy Cosmic Rays of non-solar origin. When radiation accumulates around the shock, the emitted radiation can be taken as a signature, mapping the shock location and for extended shocks also their geometry. Since radiation in the high-energy range comes from accelerated particles, all relativistic shocks are in one or another way involved into the acceleration of charged particles. It is intuitive that the efficiency of acceleration should drastically increase with shock Mach number and thus be reflected in the hardness of the accelerated particle spectrum and the highest achieved particle energies. Moreover since synchrotron emission in magnetic fields and bremsstrahlung are closely related to the hardness of the particle spectrum, observation of flat radiation spectra and radiation in the Gamma and X-ray range can be taken as indication of high relativistic Mach numbers and the presence of magnetic fields. At distant cosmological objects the redshift may move the energy down into the optical or even the synchrotron frequency range and absorption of radiation on the line of sight by interfering material may suppress part of the spectrum. 

\subsection{\bf\textsf{Pulsar Wind Termination Shocks}} 
Pulsar winds have probably the highest (or at least with certainty very high) Lorentz factors of the order of up to $\Gamma\sim10^6$. Compared with the energy of this flow, the wind itself is in good approximation cold. In addition pulsar winds are strongly magnetised with ratio of magnetic to total kinetic energy $\sigma\sim 10^{-3}-10^0$. Pulsar winds are expected to consist of electron-positron pairs $(e^+,e^-)$,  although ions may also be mixed into the winds, in case they are collimated,  particularly at their edges where ions may enter the wind from the the external medium by some mixing processes. 

\begin{figure}[t!]\sidecaption\resizebox{0.62\hsize}{!}
{\includegraphics[]{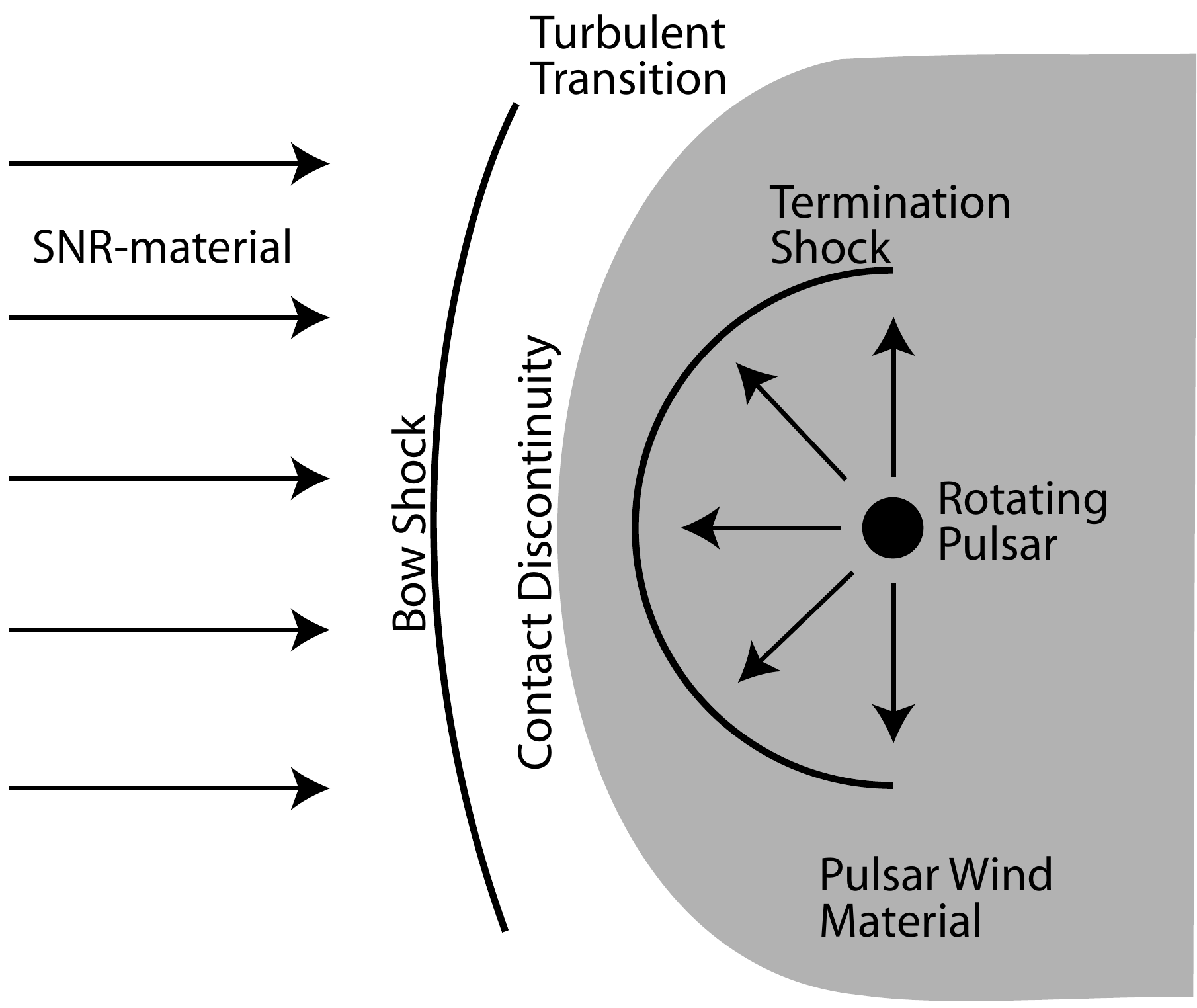} }
\caption[ ]
{\footnotesize Sketch of a pulsar-wind shock system. The figure is drawn in the pulsar system where the supernova-remnant (SNR) material streams against the ultra-relativistic outflow from the pulsar. The interaction between the wind and the SNR produces a bow shock wave behind which a turbulent transition region is located. The outer boundary of the pulsar wind is a discontinuity, either contact or tangential depending on the local conditions and on the magnetic field. Inward one expects the turbulent pulsar wind transition between the discontinuity and the termination shock of the wind  \citep[after][]{swaluwea03}. \vspace{0.1cm}}\label{chap2-fig1-pw}
\end{figure}

In most cases \citep[like the Crab pulsar, for instance; see][]{gallantea02} the rapidly rotating pulsar of angular frequency $\Omega$, rotation period $P=2\pi/\Omega$, moment of inertia $I_p$, and spin-down luminosity $L_p=I_p\Omega\dot{\Omega}$ which causes its windy outflow is embedded into the comparably slow expansion of the supernova which created the pulsar and into which the pulsar is immersed. Hence, the pulsar wind environment is the supernova wind. The interaction between the pulsar wind and  the upstream Supernova Remnant (SNR) material causes a  pulsar-wind bow-shock system, which is located outside the outer boundary of the wind. Behind the bow shock one finds some turbulent (downstream) transition region which is probably terminated by a contact discontinuity (or, if magnetised, a tangential discontinuity). Inside of the latter the pulsar wind develops a termination shock that, in the frame of the wind, moves inward toward the central pulsar. Both these shocks, the bow shock and the termination shock, are believed to be relativistic and collisionless.  

The expected physics of the pulsar wind has been described by \citet{kc84a} who, in analogy to the  non-relativistic shocks in the heliosphere, assumed that the cold pulsar wind thermalises at the termination shock (see the sketch in Figure \ref{chap2-fig1-pw}). The plasma downstream of the termination shock is hot. Its sound velocity $c_s/c\approx \sqrt{3}$ is a fraction of the speed of light, and its luminosity at the radial location $r_\mathit{TS}$ of the termination shock is $L_\mathit{\,TS}\simeq 4\pi\Gamma^2N_1r^2_\mathit{TS}m_ic^3(1+\sigma_1)$. It is believed that the magnetisation ratio $\sigma_1=B^2/\mu_0 m_iN_1c^2\Gamma\ll 1$ is small, where $N_1$ is the upstream density, $B$ magnetic field, and $m_i$ the ion (proton) mass. This yields a downstream pressure $\mathcal{P}_\mathit{\,TS}\approx \frac{2}{3}\Gamma^2N_1m_ic^2=L_\mathit{\,TS}/6\pi cr^2_\mathit{TS}$. Simple arguments then suggest \citep{swaluwea03} that this pressure is constant up to the bow shock. In principle, however, as long as the pulsar has not broken out of the SNR, the evolution of the entire system depends in the external SNR-medium which determines the conditions at the bow shock. Recently, the spectral evolution of this internal medium has been investigated by \citet{bucciantiniea11} who found an injection spectrum following a broken power law  reproducing very well the observed radiation. After break-out from the SNR into the interstellar medium the dynamics and shock properties are determined by the conditions in the interstellar gas. 

\subsection{\bf\textsf{Very-High Energy Gamma Rays}}

Another indication for the possible presence of relativistic shocks is the observation of Gamma Rays in the very-high energy range $\mathcal{E}\gtrsim$ TeV. Radiation of this kind has been observed in different manifestations in the Galaxy \citep{HESS-coll06}. The origins of diffuse TeV emission in its central region are probably dense molecular clouds which stand in the way of high-energy Cosmic Rays that most probably have been accelerated by shocks. The most probable three galactic sources of high-energy Cosmic Rays are SNRs, Pulsar Wind Nebulae (PWNe) and binary systems (BSs) \citep[for a recent of detection $> 100$ GeV gamma rays from binary pulsars cf.][]{aharonianea09} which may produce the required ultra-high energy Cosmic Rays. Extragalactic sources can be active galactic nuclei (AGNs and Blazars) \citep[][]{acciariea09}, radio and starburst galaxies \citep[][]{aceroea09}  and, of course, clusters of galaxies \citep[examples are given in][]{aharonianea09-coma,aharonian09-abell}. All arguments rely on the diffusive shock acceleration mechanism (DSA) in its various more or less sophisticated versions \citep[cf., e.g.][and others]{bt85,be87,md01,drury01}. 

\subsubsection*{\textsf{Galactic sources}}
\begin{figure}[t!]
\centerline
{\includegraphics[width=0.5\textwidth]{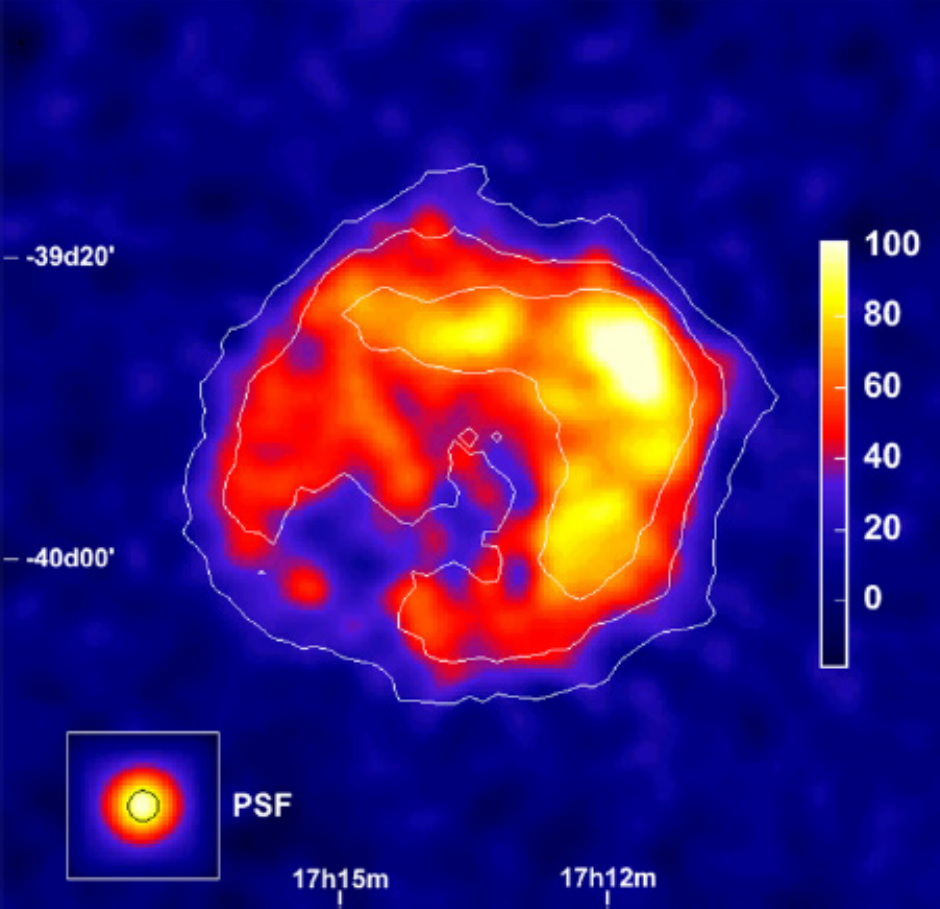} 
\includegraphics[width=0.5\textwidth]{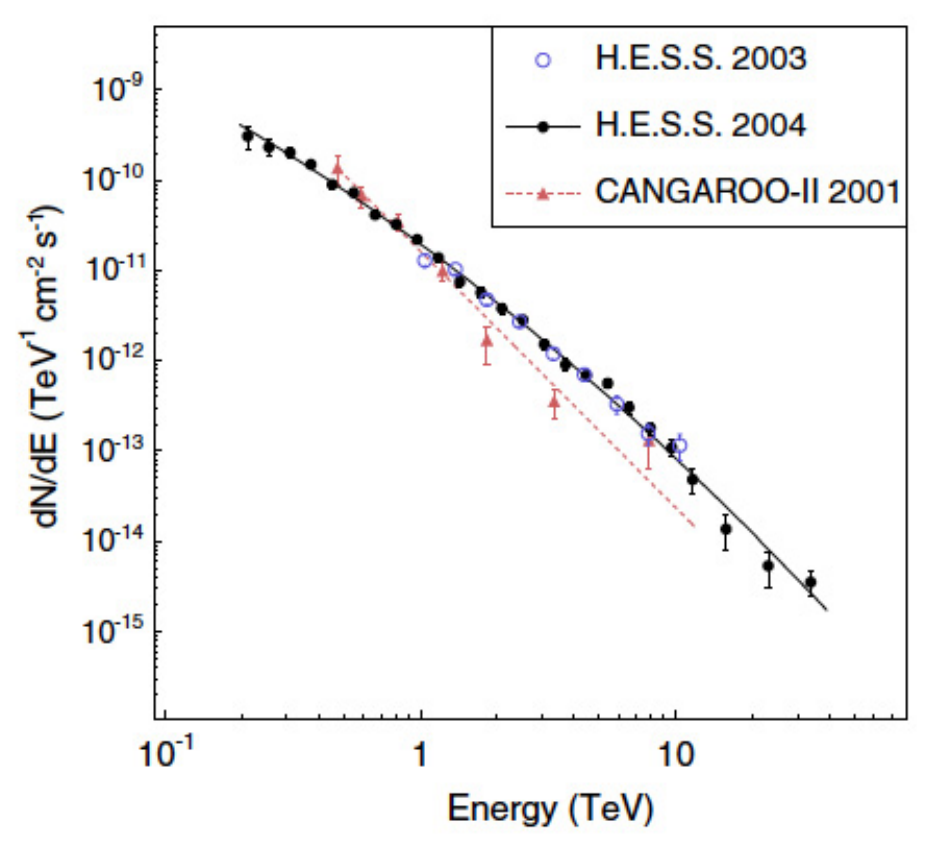} }
\caption[]
{\footnotesize {\it Left}: gamma-ray image above 300 GeV obtained with the HESS telescope array in 2004. this image shows the young SNR source RX J17.13.7-3946 which was observed in China as a `guest star' in 393 AD. Colour coding is linear based on photon counts. In the lower left lower corner of this picture, which was adapted from  \citet{aharonianea06} the point spread function is indicated. The high energy gamma emission shows an asymmetric shell structure that is typical for the interaction of the SNR blast shock wave with an external molecular cloud. {\it Right}: Interestingly the spectrum  is very hard, and is not a simple power law; moreover, it shows no angular dependence. A tentative model explanation has been given by assuming a modification of the diffuse shock acceleration theory \citep{malkovea05}. }\label{relsh-fig-RXJ1713}
\end{figure}

Approximately $10\%$ of the total mechanical energy release in the Galaxy stems from SNRs. This suffices  to generate the high-energy Cosmic Rays at the observed energy density of $10^{-13}$ J m$^{-3}$.  For an example observed recently with Chandra see \citet{bambaea03}. Young SNRs, in particular, are strong gamma-ray emitters both by the hadronic Cosmic Ray component as by $\gg10$ TeV electrons, which emit synchrotron or inverse Compton radiation and are thus strong candidates for relativistic shock acceleration.  An example of the gamma radiation from a young (i.e. a historical) SNR is shown in Figure \ref{relsh-fig-RXJ1713}.

Terminating pulsar winds are the other connection between galactic very-high energy gamma rays and shocks. Pulsar winds are ultra-relativistic. Ending up in the diffuse Pulsar Wind Nebula they generate a termination shock whereby electrons will probably be accelerated up to 100 TeV energy or even beyond. These electrons should produce TeV gamma rays when undergoing inverse Compton scattering on the diffuse radiation fields and push the radiation energy up into the very-high energy gamma range. Clearly, inverse Compton scattering by the abundant Cosmic Microwave Background photons will also create lower energy radiation below 100 GeV. Radiation of this kind has been observed from the Crab. Its spectrum is unique in the sense that it extends over 21 decades in frequency or energy, from radio frequencies to very-high energy gamma rays implying that electrons are accelerated up to $\gtrsim 10^3$ TeV. In a stochastic diffuse acceleration process they assume random pitch angles so that the bulk of the electron energy is released as radiation in the strong pulsar nebula magnetic field of $\sim100\mu$G as synchrotron emission. At the same time inverse Compton scattering generates the very-high energy gamma emission. Besides the Crab nebula, which is exceptional  in its radiation, {\small CANGAROO} and {\small HESS} have identified four other Pulsar Wind Nebulae which are also efficient gamma emitters: Vela X \citep{aharonian06-vela}, MSH 15-51 \citep{aharonian05-msh}, the PSR J1826-1334 nebula \citep{aharonian06hess-j}, and Kookaburra \citep{aharonian06kook}. The radiation source in all these cases is clearly separated from the corresponding pulsars indicating that the emission does not come from the pulsar itself but rather from the nebula caused by the deceleration of the pulsar wind at the remote location of the reverse termination shock as was suggested in the simple pulsar wind {\small MHD} model of \citet{kc84a,kc84b}.

In pulsar nebulae the particles are accelerated in the external shock region, but in compact binaries one expects complete thermalisation of the plasma. Yet, the observation of very-high energy gamma rays from such objects as well suggests that this assumption might not be true. As long as one sticks to the shock acceleration picture the resolution of this puzzle can be seen in a number of \emph{internal shocks} located inside the jet and close to the compact object which serve as efficient accelerators for particles. This can be the case when the object does, indeed, eject jets. Such objects, called micro-quasars, have been found in the galaxy, too \citep{mirabel94}.  The binary object LS I+61 303 \citep{albertea06} belongs to this category of objects, it has radio emitting jets located at a distance of $\approx$2 kpc. It emits gamma rays up to energies of 4 TeV,  possibly indicating that external termination (or internal) shocks are created. The gamma-ray luminosity of this source above $\approx$200 GeV is sufficiently high at $\approx10^{27}$ J s$^{-1}$ to support this assumption.

Finally, of the galactic sources the central galactic engine near Sgr A*, a probably supermassive black hole and its environment, is of interest. TeV gamma radiation with no time variability has been observed from this region where, however, a pulsar is also in closer vicinity, where ejection of material  could cause shock generation in jets and termination shocks \citep{aharneron05a,aharneron05b}. 

\subsubsection*{\textsf{Extragalactic sources}}
As mentioned above, potential extragalactic candidates for relativistic shock signatures in very-high energy gamma rays are AGNs, radio and starburst galaxies, and galaxy clusters. 

\begin{figure}[t!]
\centerline
{\includegraphics[width=0.52\textwidth]{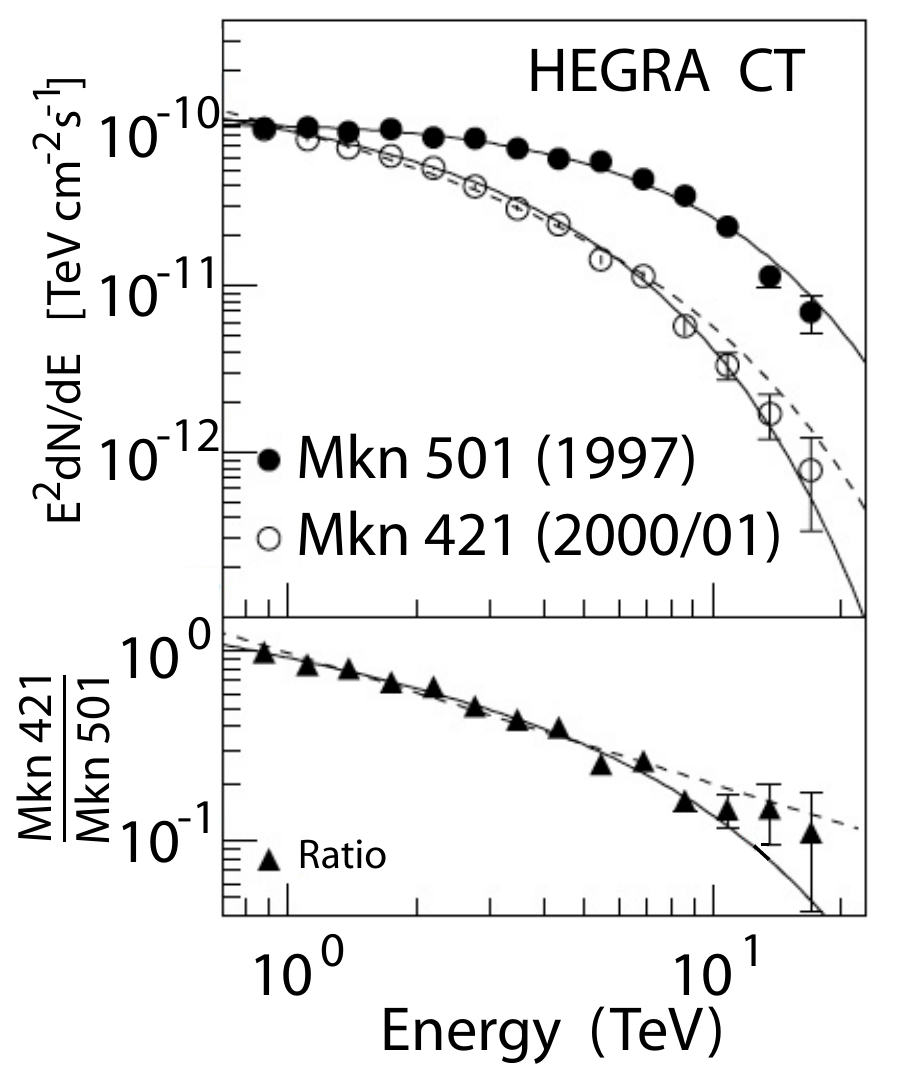} 
\includegraphics[width=0.45\textwidth,height=0.58\textwidth]{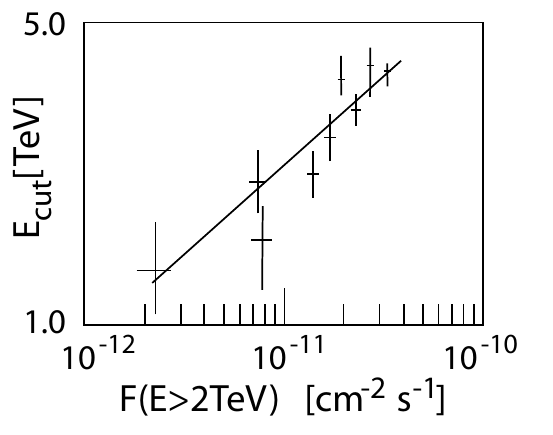} }
\caption[]
{\footnotesize {\it Left}: gamma-ray spectra for Mkn 501 and Mkn 421 during gamma flare states. Such spectra can be composed of exponentially truncated power laws \citep[adapted from][]{aharonianea02}. The lower part of the figure shows the ratio of the emission spectra of the two objects.  {\it Right}: The spectral cut-off energy as function of gamma photon irradiation for Mkn 421 showing that the irradiance $F(\mathcal{E}>2)$ TeV increases  with truncation energy $\mathcal{E}_{cut}\equiv\mathcal{E}_0$ according to a weak power law \citep[data taken from][]{aharonianea05}.  }\label{relshock-fig-Mkn501}
\end{figure}

Of the AGN candidates, AGNs with relativistic jets close to the line of sight (with $\Gamma$ few times 10 and $\sigma\sim 10^{-3}$ or less), so-called Blazars and among them the BL Lac objects, are the strongest emitters of observed radiation.\footnote{We may note at this point that it was the observational evidence from relativistic explosions in AGNs that led \citet{bm76} to develop their ultra-relativistic (collisional) self-similar (non-magnetised) fluid model of blast-wave driven ultra-relativistic shocks on which most of the subsequent literature was based.} This is believed to be caused by relativistic Doppler-boosting due to the line-of-sight orientation of the jets. Emission in these objects covers the entire range from radio to gamma rays. TeV gamma rays have been detected from the BL Lac objects. Their radiation is highly variable in brightness and polarisation which is a signature of the presence of ultrarelativistic electrons, which must have been accelerated to these energies. Variability in gamma flares in some of these objects has been found to be as fast as several minutes \citep[for instance in Mkn 501, as reported by][]{albertea07}. In the flaring state the gamma-ray spectra of these objects are generally very hard. They can well be described by an exponentially truncated power law d$\,N_\gamma/$d\,$\mathcal{E}\propto \mathcal{E}^{-\alpha}\exp(-\mathcal{E}/\mathcal{E}_{\,0})$ with power $\alpha\sim 2$ and truncation energy $\mathcal{E}_{\,0}$. This is shown in Figure \ref{relshock-fig-Mkn501}. Such spectra are typical for shock-accelerated particles causing the radiation. More distant BL Lacs seem to have steeper photon spectra probably caused by intergalactic absorption. However, generally, the steepness  seems to be intrinsic, allowing for hard spectra but not harder than $\alpha=\frac{3}{2}$. From energy requirements and statistical mechanical theory such a power-law index is an absolute lower for any phase space distribution.

One principal point to be noticed in all these observations is that interpretation of the gamma emission by shock-accelerated particles encounters serious difficulties when we assume that the spectra are caused by hadrons, i.e.  from protons or other baryons. The emission must be caused mostly by leptons (electrons and positrons), and this poses the problem of how these particles are efficiently accelerated to the observed extremely high energies by shocks. This is a general problem which has not yet been solved satisfactorily. It requires either direct leptonic acceleration by shocks, which is a problem in itself, and possibly might imply a coupling with other acceleration mechanisms until shock acceleration may set on for leptons, or it requires an efficient redistribution of energy from high energy shock-accelerated hadrons to leptons.

Other candidates are radiogalaxies. TeV gamma rays have been detected from the nucleus  of the nearby galaxy Centaurus A  and from M87 (which is 16 Mpc away in the Virgo cluster). In M87 the gamma rays may be produced either in the assumed central supermassive black hole or a hot knot (HST-1) roughly 100 pc displaced from the nucleus. The latter possibility is not improbable as the variability of the gamma emission correlates with that of the hot knot. Since hot knots are believed to be related to the termination of jets flowing out of the central engine, relativistic shocks would naturally be involved in the generation of this radiation, again implying violent particle acceleration.

In starburst galaxies like M82 and NGC 253 there is plenty reason for strong winds, shock formation, particle acceleration, and generation of very-high energy gamma emission simply because of the high expected supernova rate. Indeed, gamma rays up to TeV energy have been detected from these objects. Further candidate objects are ultra-luminous infrared galaxies like Arp220 (located at $\approx70$ Mpc). Finally, galaxy clusters containing large numbers of either of these objects are clearly candidates of relativistic shocks. Relativistic shocks in those clusters are, however, probably not to be found in the intracluster medium, they are expected rather to  belong to the galaxies which are members of the cluster. Observation of such shocks is however hindered by the comparably high thermal X-ray flux from clusters and the high absorptivity of the intracluster plasma which might prevent the observation of gamma rays.

\subsection{\bf\textsf{Gamma Ray Bursts}}  

Gamma Ray Bursts (GRBs) are a particular class of gamma-ray sources. We consider them here separately from the above-mentioned  gamma-ray evidences for relativistic shocks of Lorentz factors $\Gamma>10^2$ with grossly unknown magnetisation ratios $\sigma$. As their name tells, they are brief gamma-ray splashes that last only for short times (minutes to days) and are singular events, probably of  extra-galactic origin. Since their discovery in the sixties and their identification with remote events in the eighties, GRBs have been in the focus of astrophysical interest partly because of their enormous brightness and luminosity, their time variability, nearly homogeneous distribution over the sky, and partly because of the suspicion that they would have cosmological relevance. Meanwhile a large amount of review articles on GRBs has accumulated in the scientific literature \citep[as for a selection cf., e.g.,][]{gw99,piran99,piran04,piran05b,meszaros02,meszaros06,waxman06,nakar10}. Gamma Ray Bursts are believed to be emitted  when stellar-mass black holes form in the Galaxy (or in other galaxies as well) by accreting stellar mass objects. Birth of such black holes is  accompanied by relativistic outflows for which numerous observational evidence has accumulated from interstellar radio scintillations \citep{frailea97}, apparently superluminous motions in afterglows \citep{taylorea04}, and afterglow light curves \citep{harrisonea99}.

\begin{figure}[t!]\sidecaption\resizebox{0.65\hsize}{!}
{\includegraphics[ ]{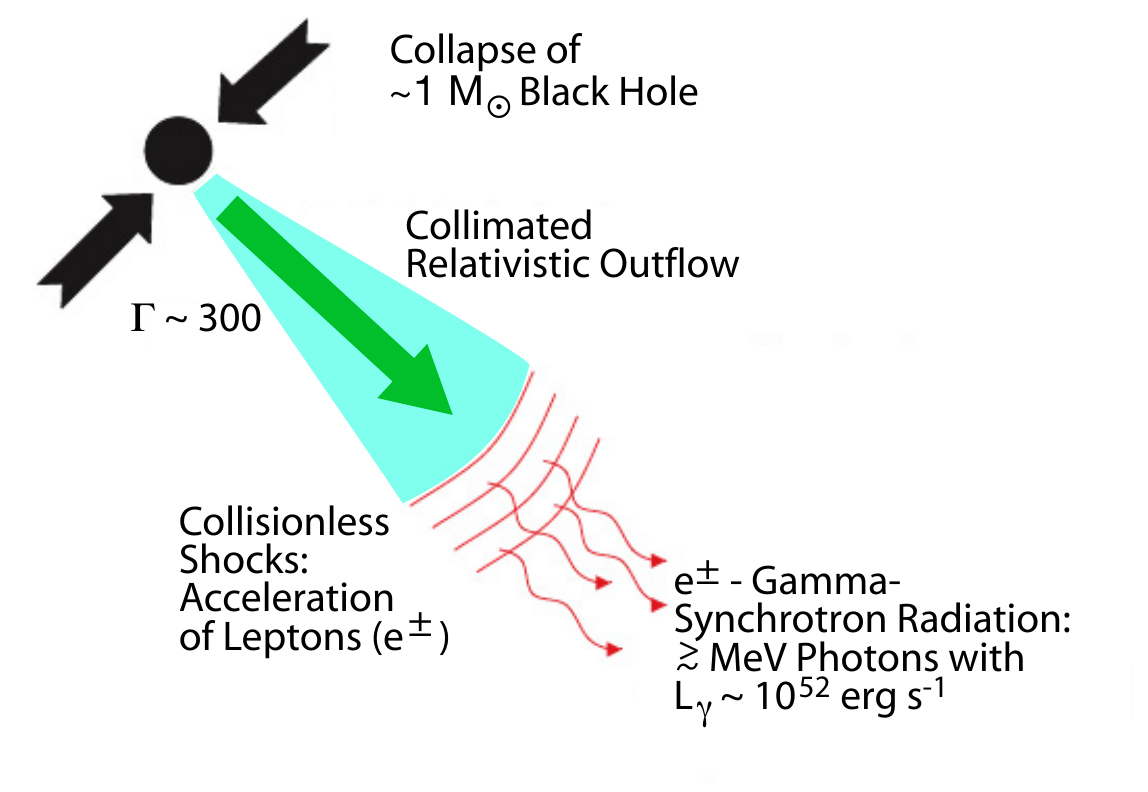} }
\caption[]
{\footnotesize A so-called `Fireball' scenario in which a GRB is caused by the relativistic outflow from the merger of two massive objects \citep[after][]{waxman06}. The relativistic Lorentz factor is assumed to be of order $\Gamma\sim 300$. Gamma rays may emerge with high luminosities of the order of $L_\gamma\sim 10^{45}$ J s$^{-1}$ from internal shocks in the highly relativistic outflow as synchrotron emission from electrons at $\gtrsim$MeV photon energy. When the flow interacts with the environment, an external collisionless shock is generated which causes the synchrotron GRB afterglow. \vspace{-0.1cm}}\label{chap1-fig1-wavediag}
\end{figure}

Though it still remains unclear what the real source of GRBs  ultimately is, the most probable models of generation of Gamma Ray Bursts are based on the assumption of ultra-relativistic collisionless shocks \citep{waxman06} that should be produced at least at some later stage in highly collimated ultra-relativistic outflows. From that point of view, Gamma Ray Bursts offer themselves as the ideal test case for ultra-relativistic shocks, their effects and properties. This view has been strongly pushed in the literature. Here we briefly review the supporting arguments. We will later, in the discussion of the theory of collisionless shocks, return to GRBs as a field of application of ultra-relativistic collisionless shock theory which is mainly based on numerical {particle-in-cell (PIC)} simulation techniques {\citep[for an early comprehensive review cf.][]{dawson1983} which today can be applied to quite large numbers of particles per cell, extended boxes, up to three spatial dimensions and even inclusion of realistic ion-to-electron mass ratios $\mu\equiv m_i/m_e=1840$.} 

\subsubsection*{\textsf{Progenitor, central engine, outflows, jets, radiation}}
Observationally GRBs can be divided into long (L) and short (S) GRBs. Only the former have been shown to be related to supernovae, exploding within $\sim$1 day of the supernova (SN). Such a connection was ruled out for the latter. It is thus suggested that LGRBs occur in host galaxies with high star formation rate. Establishing such a connection immediately implies that LGRBs are at least  accompanied by external shocks which, like in SNs, are generated when the outflow blast interacts with the interstellar medium in the host galaxy. In addition, internal shocks contained in the outflow may also exist, but a large part of particle acceleration and radiation in the observed spectrum is probably generated in the external shock. Estimates of the relativistic Lorentz factor of the outflow univocally agree that $\Gamma> 30$ if not $\Gamma\gg100$ as suggested by the detection of GeV gamma-ray photons. These values certainly correspond to angularly highly collimated (i.e. beamed) outflow as otherwise the released energy would exceed $10^{48}$ J. Beaming has been confirmed by afterglow observations of GRB 030329. 

What concerns the location of the prompt emission of radiation, it seems to be certain that it comes from a radial interval of $(10^{9}\, \mathrm{cm\ to}\ 10^{10}\, \mathrm{m}) < r < (10^{14}\, \mathrm{cm\ to}\ 10^{15}\, \mathrm{m})$  away from the central object, converting $\gtrsim10\%$ of the kinetic energy of the outflow into radiation, most of it into sub-MeV gamma radiation. This range suggests that it is emitted from the outflow itself, probably from internal processes as, for example, particles accelerated by internal shocks and emitting synchrotron radiation. This seems to be the case, because observation has ruled out inverse Compton radiation, as this would require too strong magnetic fields inside the jet. 

Since the connection to SNs is missing for SRBs, their origin is tentatively attributed to the merger of either two neutron stars or a neutron star and a black hole. Also magnetars or proto-magnetar systems are in the focus. Under such circumstances high energy relativistic jets will be emitted from the central engine which would explain the high energy release of up to $\approx10^{46}$ J \citep[for the recent observation of the most distant cosmological GRB cf.][]{chandra10}. All these assumptions are highly hypothetical so far; they could possibly be confirmed by presence or absence of gravitational wave signals emitted by them. However, in this case the generation of relativistic plasma shocks is less certain, even though the merger of the heavy objects is a violent process in the course of which blast waves might be released that  propagate radially out, accelerate particles and terminate in the interstellar medium.

\subsubsection*{\textsf{Gamma Ray Burst Afterglows}}
GRBs are followed by long lasting afterglows covering the entire radiation spectrum over eight orders of magnitude in energy or frequency, from X-ray energies to radio wavelengths. This radiation is almost certainly emitted in the interaction with the interstellar or galactic medium of density $N$ which terminates the outflow blast from the central object \citep[for a more recent account cf., e.g.,][]{waxman06,katzea07}. This interaction results in the generation of a pair of shocks: A strong external forward shock of radius $r_s$ and a mildly relativistic reverse shock (the equivalent to a termination shock in stellar wind outflows) that, in the frame of the forward shock, propagates inward. The reverse shock is, however, short lived in this highly bursty or time-dependent case. 

One believes that within minutes of the outburst the shock Lorentz factor becomes of the order of $\Gamma \sim$ several 10$^2$ while decelerating to $\Gamma\sim$ few 10 within one day, and to $\Gamma\sim$ O(1) within few months, by then becoming non-relativistic. 

It is believed, moreover, that upstream of the shock the magnetic energy density is negligibly small compared with the kinetic energy of the flow, its upstream $\sigma$-factor (neglecting thermal energy density) probably being as small as $\sigma_1=B^2/2\mu_0 N_1m_ic^2\Gamma\approx 10^{-9}\ \mathrm{to}\ 10^{-6}$. If true, this would make the shock about non-magnetic upstream. Downstream the magnetic field is that of the ambient plasma $B\sim \mu$G and should be tangential to the shock. Since this causes problems with the condition of magnetic field continuity, such shocks necessarily must generate their own selfconsistent fields close upstream to the shock as well.  Moreover, from the observation of the emitted synchrotron radiation, $\sigma$ is believed to increase across the shock in the transition to downstream by orders of magnitude to values $\sigma_2\approx 10^{-2}\,\mathrm{to}\ \sigma_2\approx10^{-1}$. This increase is traced back to some -- still badly understood -- mechanism of generation of magnetic fields in the shock environment.

For a while, usually a few hours after the GRB, these shocks accelerate particles until the particles start radiating. Thereby the external shock decelerates (assuming proper number $\sim 4\Gamma N$ and energy $\sim 4\Gamma^2Nm_pc^2$ density conservation of the freshly shocked plasma in the plasma rest frame). The shocked plasma is contained in a comparably thin shell of width $\Delta\sim r_s/\Gamma^2$. 

The simple self-similar canonical ultra-relativistic blast-wave shock model \citep{bm76} predicts a decrease of the Lorentz factor with shock radius $r_s$ in the shock frame given approximately by
\begin{equation}
\Gamma(r_s)\simeq 150\left(\frac{{\cal E}_p}{10^{46}\,{\rm J}}\frac{1\,{\rm m}^{-3}}{N}\right)^\frac{1}{2}\left(\frac{r_s}{10^{15}\,{\rm m}}\right)^{-\frac{3}{2}}
\end{equation}
Distant observes should see the shock expanding as $r_s(t)\sim 10^{14}(t{\cal E}_p/N)^\frac{1}{4}$ m
where the proton energy ${\cal E}_p$ is measured in $10^{46}$ J, the ambient density is in m$^{-3}$, and the time $t$ is measured in seconds.  
This seems not to contradict observation of the afterglow radiation when assuming that electrons couple energetically to the protons (baryons) and obey a power law distribution d$N_e/$d$\epsilon\propto \epsilon^{-p}$ with $p>2$. Afterwards, radiation causes cooling the electrons while de-coupling them from the baryons.

However, even though this might be true in general, the details of the afterglow do often not allow for the application of the simple blast-shock model. This suggests that not only the mechanism of GRBs is not sufficiently understood today, but also the relativistic or ultra-relativistic shock theory is not yet capable of describing the afterglow satisfactorily \citep[see, e.g.,][]{nakar07,nakar10}.  

Recently it has been proposed \citep{mesz-rees10} that Population III stars might be the most probable cause for GRBs involving external shocks. Population III stars are very massive first generation stars, which may collapse already at early times and large redshifts. Once they collapse they will produce relativistic jets and splashes of radiation which could be those which appear at GRBs. This proposal can be checked by inferring about the special properties ultra-relativistic shocks should exhibit when they are involved into the generation of GRB afterglow \citep{fox-mesz06,tomaea10}. 

It is almost certain that the afterglow radiation is emitted by electrons by the synchrotron mechanism. This again requires strong downstream magnetic fields which, by the observations, should be strong enough to be close to energy equipartition on scales much larger than the nominal shock width which, for non-magnetic upstream flows is the upstream (relativistically invariant) ion inertial length $\lambda_i=c/\omega_{pi}\approx 100 /\sqrt{N_1}$ km, where $N_1$ is measured in cm$^{-3}$ and $\omega_\mathit{pi}$ is the ion-plasma frequency. Observations then require that downstream magnetic fields should be strong over downstream (index $d$) distances $L_d\approx 2\,\Gamma ct\gg \Delta_s\sim\lambda_i$, much larger than the shock width $\Delta_s$, where $t\sim 1$ day, yielding $L_d\sim 10^{10} \Delta_s$, and electrons should become accelerated by the shock to cosmic ray energies.
This implies the clarification of downstream shock electron acceleration to the observed high energies and the high inferred (near equipartition) magnetic fields. These, too, pose problems that so far cannot be considered to have been solved to any satisfaction. 

\subsection{\bf\textsf{Ultra-High Energy Cosmic Rays}}

The origin of Cosmic Rays is still insufficiently understood even today. The urgency of understanding it better can be illustrated when one notices that the mean cosmic ray energy density in the Universe is of the same order of magnitude as that of the cosmic microwave background: $u_{CR}\sim u_{CMB}$. This might be a coincidence, however, it is difficult to understand what the origin of this agreement is. 

At least part of the Cosmic Rays is believed to be accelerated in relativistic or ultra-relativistic shocks in many objects, including the sources of GRBs and starburst galaxies. A  recent HESS observation of Cosmic Rays from the latter \citep[see][]{aceroea09} with very intense gamma-ray fluxes at 2.2 GeV, implies cosmic-ray densities three orders of magnitude larger than those emitted from the centre of our galaxy. At the same time such Cosmic Rays also contribute to the structure and dynamics of those shocks. Diffusive shock acceleration seems not to work at energies higher than $10^{15}\,\mathrm{eV\ to}\ 10^{18}$\,eV \citep{meli-bier06}. Whether this limitation holds or not is unclear. 

Sub-relativistic Supernova shocks can accelerate Cosmic Rays presumably up to energies of $\lesssim10^{15}$ eV \citep{gs69,bt85,be87}. External shocks in jets of active galactic nuclei \citep{berezinsky08} and LGRBs have been suggested early on \citep[cf., e.g.,][and others]{waxman95,waxman95a,vietri95,wickea04,da06} to be sources of (galactic as well as extra-galactic) Cosmic Rays in the ultra-high energy range $\epsilon\sim 10^{18}$\,eV to $10^{20}$\,eV. This might indeed be the case with the exception of the highest cosmic ray energies $\gtrsim 10^{20}$ eV for which no reasonable shock acceleration mechanism is known \citep[cf., however, the -- debatable -- discussion in][where it was argued that external LGRB shocks could as well generate particles at such energies]{dermer06}. 
The limiting energy for shock accelerated protons has been estimated as 
\begin{equation}
\epsilon_{p,\,max}\approx 10^{\,20} \left(\!\frac{100}{\Gamma}\right)\left(\!{\frac{L_{\rm F}}{10^{\,44}\mathrm{J}}\frac{\sigma_1}{0.1}}\right)^{\!\!\frac{1}{2}} \ \ \mathrm{eV}
\end{equation}
where $L_{\rm F}$ is the undisturbed luminosity (in J) of the upstream flow and $\sigma_1=B_1^2/2\mu_0 w_1$ the ratio of upstream magnetic field energy density to the total energy density $w_1$ of the flow including rest energy \citep{waxman95}. Clearly, for the small $\sigma_1\ll 0.1$ used in the previous section, one never reaches the wanted limit of $10^{\,20}$ eV. 

Short GRBs can also accelerate protons to such high energies though at much reduced flux by internal-shock acceleration in the upstream flow, if the Lorentz factors range between $10^{\,2}\lesssim \Gamma\lesssim 10^{\,3}$ and the magnetisation is at least moderate \citep{nakar07}. For a general discussion of how to extract information on a retarding expanding external relativistic shock that can be extracted of a cosmic-ray spectrum one may consult a recent paper by \citet{katzea11}. 

In the Galaxy there is another relation for the energy density of Cosmic Rays: $u_{CR}\sim B^2/2\mu_0\sim \frac{1}{2}\rho_gV_\mathit{turb}^2$, with $\rho_g$ the density, and $V_\mathit{turb}$ the average turbulent velocity of galactic matter. This relation says that the cosmic-ray energy density is comparable to both the magnetic and turbulent gas energy densities -- a strong indication that neither the magnetic field nor the gas can be considered independent of the presence of Cosmic Rays. 

Conversely this mutual dependence also indicates that Cosmic Rays are in energy exchange with plasma and magnetic field, and as well as shocks. In particular relativistic and ultra-relativistic shocks are the most probable cosmic-ray factories and are thus responsible for this energy exchange -- in addition to general turbulence.

\section{\bf\textsf{Advances in Relativistic Shock Theory}}

Almost until the start of the third millennium, reference to relativistic shocks has been based solely on the fluid approach as given by \citet{bm76}, \citet{mko77} and others. The substantially more advanced  non-relativistic shock theory and observation in space physics \citep{tsurutani85} was mentioned only occasionally, albeit with the clear notion of not being applicable to astrophysical problems. 

The change came with the realisation that shock acceleration stagnated as long as kinetic effects and magnetic fields were not included into theory. Including these effects was  first proposed by \citet{gruzinov-wax99}, \citet{medvedev99} {and \citet{brainerd2000}} to account for the generation of the large magnetic fields which are required in the shock-radiation models of Gamma Ray Bursts \citep{piran99,piran99a}. In the following we briefly review the basic physical background of shock theory and the current state of the art with the advances in the latter coming mainly from numerical simulation.

\subsection{\bf\textsf{Fluid Dynamics of Relativistic Shocks}}

As noted above the study of relativistic and ultra-relativistic shocks began with the investigation of non-magnetised fluid dynamic shocks \citep{bm76,bm77,mko77,bo80}. Since it became clear that there are no non-magnetic shocks in the Universe, at least no visible or otherwise detectable non-magnetic shocks, we do not review these seminal works but go straight to a brief consideration of relativistic magneto-fluid dynamic (magnetohydrodynamic) shocks. 

\subsubsection*{\textsf{Relativistic Ideal {\small MHD} Rankine-Hugoniot Conditions}}
Shock jump (Rankine-Hugoniot) conditions in one-dimensional ultra-relativistic $\mathcal{E}\gg (m+\rho \mathcal{V})c^2$ non-magnetic (hydrodynamic) fluid flows ($\mathcal{V}$ is the volume, $\mathcal{E}$ the energy) for application to blast shock waves that may presumably be generated in radial outflows from astrophysical explosions were given by \citet{bm76} in their most simple form and have been widely applied. Their extension to spherical relativistic ideal {\small MHD} flows is due to \citet{ec87}. 

In terms of the velocity four-vector $u^\mu=(u^t,u^r)=(1,\mathbf{\beta}^r), \mathbf{\beta}\equiv \mathbf{V}/c$, with rest-mass density $\rho=Nm$, proper number density $N$, average rest mass per particle $m$, magnetic field $H=B_0/\sqrt{4\pi}$, with magnetic field $B_0$ and gas pressure $\mathcal{P}$ in the fluid frame, enthalpy $w=\rho c^2+g\mathcal{P}$, where as before $g=\hat\gamma/(\hat\gamma-1)$ is the ratio of the adiabatic index $\hat\gamma$, the temperature $T=\mathcal{P}/\rho$, and putting $c=1$. Working in the shock frame (index $s$) where the quantities are assumed stationary, the jump conditions (written in one line only) become
\begin{equation}\label{eq-mhd-jumps}
[\![\rho u_s^r]\!] = [\![(w+H^2) u_s^t u_s^r ]\!] = [\![(w+H^2)u_s^{r2}+\mathcal{P}+\frac{1}{2}H^2]\!] = [\![H/\rho]\!]  =  0 
\end{equation}
%
The double-brackets ${[\![\,]\!]}$ indicate the mismatches of quantities to both sides of the shock: $[\![A]\!]=A_2-A_1$ where the indices 1 and 2 refer to upstream and downstream of the shock.
In the observer's (fixed shock) frame the electromagnetic fields are given by ${\mathbf B}=u^t{\mathbf B}_0$ and ${\mathbf E}=u^r{\mathbf B}_0$. Any normal (in this case radial) magnetic field component remains continuous because of the vanishing divergence condition. In the non-magnetic limit $H\to 0$ the jump conditions become the hydrodynamic jump conditions\footnote{\!\!One should cautiously note that physically the non-magnetic (hydrodynamic) and magnetic (magneto-hydrodynamic) cases are fundamentally different. In the latter case, the gyro-magnetic motion of the charged fluid particles causes particle correlations even in the absence of any binary collisions thus justifying the collisionless ideal {\small MHD} assumption. Such correlations are missing in the hydrodynamic case which challenges the collisionless assumption, i.e. implies the presence of some kind of collisions. Moreover, applying the non-magnetic approximation to plasmas consisting of charged particles is particularly interesting. Under certain conditions (see below), such hypothetic non-magnetic plasmas act self-magnetising on very short time scales, a posteriori justifying the magnetised-shock approach. However, the validity of {\small MHD} may be questioned in principle because the large population of shock-accelerated particles around relativistic shocks \citep[cf., e.g.,][]{Gruzinov01} requires a kinetic treatment.}, and in the ultra-relativistic radially expanding symmetric case of very large bulk Lorentz factor $\Gamma\gg1$ are just another generalised implicit representation of those given by \citet{bm76}.

\subsubsection*{\textsf{Magnetised Plane Perpendicular Shock Solution}}
For a magnetised, plane and strictly perpendicular shock a set of explicit shock relations can be derived from Eq. (\ref{eq-mhd-jumps}) best when working in the downstream frame of reference \citep[cf., e.g.,][]{aa06}. The magnetic field $\mathbf{B}=B{\hat z}$ points in direction $z$ along the shock $(x,z)$-plane, the shock moves at velocity $0\ll\beta_{sh}=V_{sh}/c\lesssim 1$ upstream in direction $-x$, and the upstream flow is cold and has Lorentz factor $\Gamma$. As before with magnetisation ratio  $\sigma=B_1^2/\mu_0i mc^2N_1\Gamma$, $m=\sum_sm_s$, and  hot downstream plasma of ultra-relativistic enthalpy $w_2\approx \mathcal{P}_2{\hat\gamma}/({\hat\gamma}-1)$, one finds
\begin{eqnarray}
\frac{N_2}{N_1} = \frac{B_2}{B_1} & = &\frac{1+\beta_{sh}}{\beta_{sh}} \qquad (\equiv R_s, \ \mathrm{the\  shock\ compression\ ratio})\\
mc^2\Gamma(1+\sigma)N_1 & = & \mathcal{P}_2+\frac{B_2^2}{2\mu_0} = \frac{\beta_{sh}}{1+\beta_{sh}}\left(\frac{P_2}{{\hat\gamma}-1}+\frac{B_2^2}{2\mu_0}\right)
\end{eqnarray}
Here, the ratio of specific heats is ${\hat\gamma}=\frac{4}{3}$ in ultra-relativistic isotropic and ${\hat\gamma}=\frac{3}{2}$ two-dimensional cases, respectively. (These equations are not applicable in the strictly ultra-relativistic case for a shock with $\beta_{sh}\simeq1$, as this would require $\left.\hat\gamma< \frac{3}{2}.\right)$ After elimination of $P_2$ and $B_2$ a quadratic equation is obtained for the shock velocity $\beta_{sh}$ and its solution can be used to find the normalised downstream temperature $T_2$
\begin{eqnarray}
\hspace{-0.5cm}\beta_{sh}  =  {\frac{1}{2}}\frac{1}{1+\sigma} \Bigg\{ \Big({\hat\gamma}\left(1+\textstyle{\frac{1}{2}}\sigma\right)-1\Big) & + & \left[\Big({\hat\gamma}\left(1+\textstyle{\frac{1}{2}}\sigma\right)-1\Big)^2+4\sigma(1+\sigma)\Big(1-\textstyle{\frac{1}{2}}{\hat\gamma}\Big)\right]^\frac{1}{2} \Bigg\} \\
\frac{T_2}{mc^2} &  = & \Gamma\beta_{sh}\left(1-\frac{\sigma}{2}\frac{1-\beta_{sh}}{\beta_{sh}}\right)
\end{eqnarray}
(with temperature in energy units) showing that in this simple case the step across the shock is determined for all quantities by three parameters only: $\sigma, {\hat\gamma}$ and $\Gamma$.  Since $T_2\gtrsim0$, this also implies that $1>\beta_{sh}>\sigma/(2+\sigma)$, a condition that is easily satisfied for weakly magnetised upstream flows.

\subsubsection*{\textsf{The Case of Spherical Symmetry}}
Though a sufficiently thin shock transition can, locally, always be taken as plane, in the cases of rapidly expanding pulsar nebulae (or also ultra-relativistic jets) spherical symmetry has traditionally been assumed to be more appropriate \citep{kc84a,kc84b,ec87}. In spherical coordinates, the last of conditions (\ref{eq-mhd-jumps}) holds automatically because the radial mass continuity and azimuthal magnetic field equations assume identical form
\begin{eqnarray}
\partial_t(\rho u^t)+r^{-2}\partial_r(r^2\rho u^r)&=&0 \nonumber\\
\partial_t(r^{-1}Hu^t)+r^{-2}\partial_r\{r^2(r^{-1} H)u^r\}&=&0\nonumber
\end{eqnarray}
suggesting that $r^{-1}H$ varies like $\rho$ and thus the ratio $H/r\rho=$ const is conserved everywhere (also holding across the shock). These spherical jump conditions have been widely discussed analytically\citep[cf., e.g.,][who considered a stationary shock]{kc84a,kc84b} and have also been solved numerically for radially diverging flow under various assumptions imposed on the shock-upstream parameters \citep[cf., e.g.,][]{ec87}. 

Traditionally two different cases have been considered corresponding to $V_{sh}>V_1$ and $V_{sh}<V_1$, where $V_{sh}$ is the shock speed, and $V_1$ is the shock-upstream velocity. The former case describes so-called `forward', the latter `reverse' shocks \citep{bm76,kc84a,kc84b}.  In the first case the shocked fluid is behind the shock, in the second case it is ahead of the shock, while in both cases the shock moves outward towards increasing radii $r$.

\subsection{\bf\textsf{Self-similar shock structure}}
Self-similarity at ultra-relativistic non-magnetic astrophysical blast waves was introduced by \citet{bm76}, following \citet{sedov59} and \citet{taylor50}\footnote{It may be worth mentioning that Sir Geoffrey Taylor's paper on self-similarity solutions of blast waves was originally written as a classified paper during the Second World War in 1941.}. Below we first refer to an approximate self-similar treatment of the radially symmetric shock structure, and then treat the general self-similar case. 

\subsubsection*{\textsf{Approximate Self-similar Solutions}}
The radially symmetric, self-similar equations contain a singularity that is located at the critical velocity ratio $\eta_c=V_c/V_{sh}=\beta_c/\beta_{sh}$ with $V_c$ the critical velocity. At the shock $\eta=1$, and hence the size of the shocked region is limited between the location of the shock and the location of the singularity. For reverse shocks one has $\eta_c-1<\{\beta_{sh}\Gamma_s^2(1+\beta_{sh})\}^{-1}$. This yields for an ultra-relativistic shock with $\beta_{sh}\sim 1$ and $\Gamma_s\gg 1$ that the reverse shock-shocked region is very narrow:
\begin{equation}
\begin{array}{rcl}
\eta_c-1&<&\frac{1}{2}\Gamma_s^{-2}  \\
\end{array}
\end{equation}
Because of this narrowness the value of $V_c$ is practically identical to the velocity $\beta_{2s}$ in the shocked region adjacent to the shock. Thus in the fixed frame (after Lorentz transformation) one has for the width of the shocked region
\begin{equation}
|\eta_c-1|\sim \beta_{2s}/V_{sh}\Gamma_s^2(1\pm V_{sh}\beta_{2s})
\end{equation}
The minus sign holds for forward shocks in this expression. The value of $\beta_{2s}$ follows from solving the jump conditions. For small $\sigma$ (weak magnetic fields) the shocked-region width depends very weakly on $\sigma$. The interesting case however is that of strong magnetic fields (large $\sigma$) where $\beta_{2s}\sim 1-1/2\sigma$. This yields for forward and reverse shocks, respectively, the scalings
\begin{equation}
\begin{array}{rcl}
1-\eta_{c,\mathit{for}}&=&2\sigma(\sigma+\Gamma_s^2)^{-1},\qquad \eta_{c,\mathit{rev}}-1\sim \frac{1}{2}\Gamma^{-2}_s \\
\end{array}
\end{equation}
Hence, for reverse shocks the width shrinks with increasing $\Gamma_s$. For forward shocks, $1-\eta_{c,\mathit{for}}$ grows with $\sigma$ and becomes 1 for $\sigma\sim\Gamma^2_s$. Since for $\Gamma_s\gg 1$ one has $\beta_{2s}\sim (\Gamma^2_s-\sigma)/(\Gamma^2_s+\sigma)$, the immediate post-shock velocity becomes small (non-relativistic), so that $\eta_{c,\mathit{for}}=0$, which implies that similar to the non-relativistic case the entire region behind the shock contains shocked plasma. This is in contrast to reverse shocks where the width of the shocked plasma region shrinks as $\eta_{c,\mathit{rev}}\to 1$. Moreover, no forward shocks exist for such strong magnetic fields that $\sigma>\Gamma_s^2$, -- at least in the relativistic {\small MHD} picture. 
\begin{figure}[t!]
\centerline
{\includegraphics[width=0.8\textwidth]{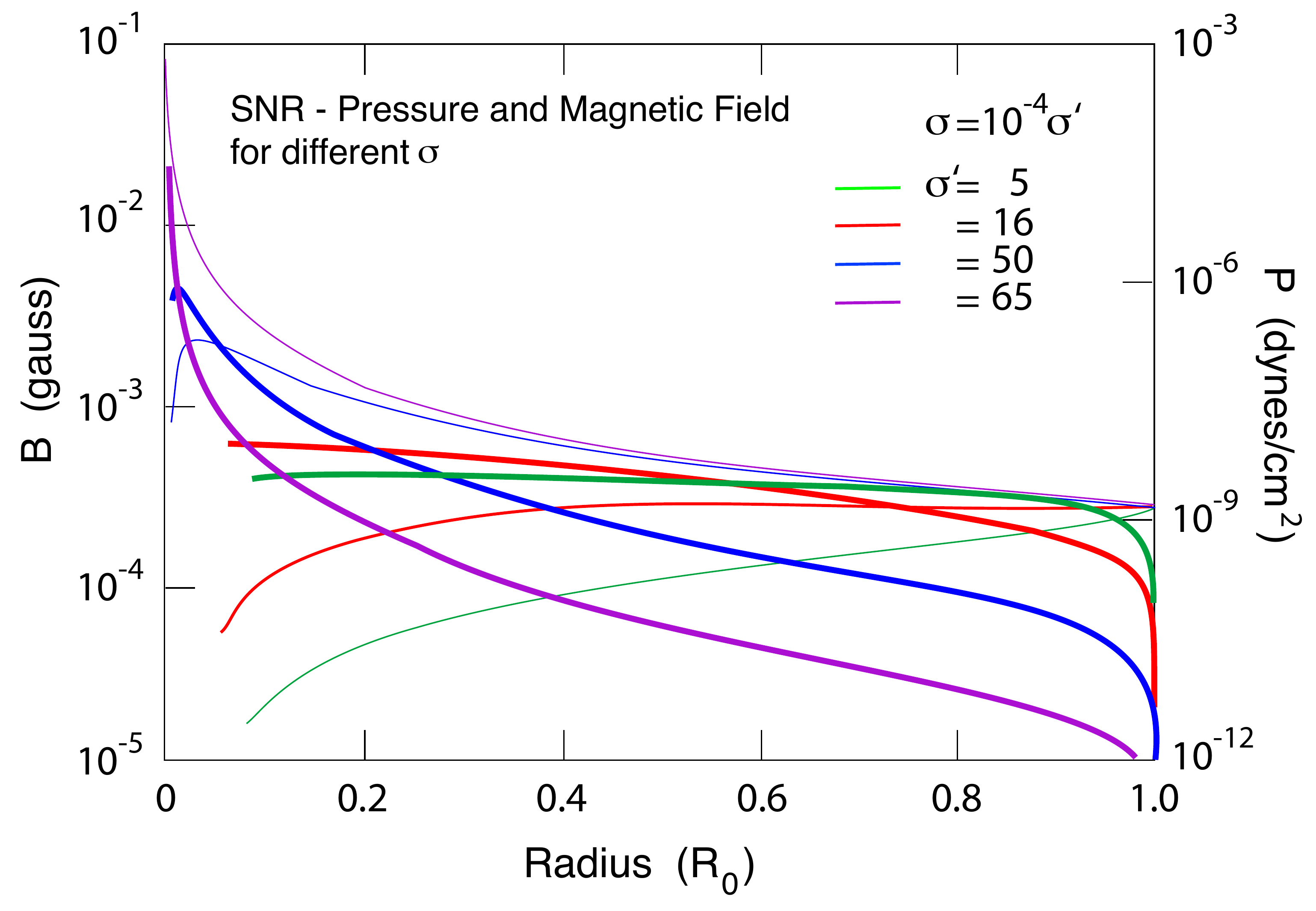} }
\caption[]
{\footnotesize The postshock pressure $P_2$ (thick lines) and magnetic field $B_2$ (thin lines) determined from the shock jump conditions  \citep[after][]{ec87} for a typical SNR like the Crab for different values of $\sigma$ and in dependence of the radius measured in SNR radii  R$_0$. The assumptions made are that $\zeta_1\sim 11$ is in the extreme relativistic range, and the velocity at $\eta_c$, the boundary of the shocked region, is 2000 km\,s$^{-1}$ corresponding the $\zeta_c=6.7\times10^{-3}$. The position of the shock front is taken at $2\eta_c^{-1}$\, pc. Here $\zeta=\tanh^{-1}(u^r/u^t), \eta=r/r_s=r/V_{sh}t $.}\label{chap1-fig2-SNR-field}
\end{figure}

The behaviour of shocked density $N_2$, pressure $P_2$, and shock-tangential (azimuthal) magnetic field $B_2$ are of particular interest. For weak magnetic fields, i.e. small $\sigma$, the density in the shocked region is about constant. An application of the self-similar jump conditions to the expanding shocked matter in the Crab nebula has been given by \citet{ec87}. 

Figure \ref{chap1-fig2-SNR-field} provides a synopsis of the evolution of the shocked magnetic field and pressure following from a simple Crab shock expansion model as function of distance and for a couple of values $\sigma$. The pressure (thick lines) drops to zero at the edge of the nebula while the magnetic field (thin lines) increases with radius toward the edge. The model which fits the observations best is that with $\sigma'\sim 16$ which corresponds to a magnetic field of $B\sim 2.6\times 10^{-4}$ G at the edge.

\subsubsection*{\textsf{Self-similar Shock Structure: General Approach}}
Recently, in view of an application to the afterglow of Gamma Ray Bursts, \citet{katzea07} developed a fully self-similar theory of collisionless shocks including (self-generated) magnetic fields. The basic assumption of this approach starts from the observation of afterglow radiation which, in order to be emitted as leptonic synchrotron radiation, requires a strong downstream magnetic field with energy density close to equipartition in a region of extension $L_d\gg \Delta_s\sim  \lambda_i$. The large observationally suggested magnitude of $L_d=\lambda_\mathit{\, corr}$, interpreting it as a shock-magnetic correlation length, is taken as evidence for self-similarity of the shock structure. This approach is based on a separation of the plasma particle distribution into two self-similar components, one non-thermal, describing the shock-accelerated high energy particle component (reflected particles, accelerated particles, a cloud of Cosmic Rays etc.), the other group are `thermal' particles belonging to the flow component and are treated as a (magneto-hydrodynamic, in the end ideal) fluid component of vanishing resistance. 

The dynamics of nonthermal particles of species $s=(e^\pm, i)$ is described by the particle distribution functions $F_s(t,{\bf p})$ which are governed by the set of (relativistic) Vlasov-Maxwell equations with current density ${\bf J}$ and single particle velocity ${\bf v}({\bf p})$ respectively defined by
\begin{equation}
{\mathbf J}=\sum\limits_s e_s\int\mathrm{d}^3{\mathbf p}_s{\mathbf v}({\mathbf p})F_s({\mathbf p}), \qquad {\mathbf v}({\mathbf p})=\frac{c{\mathbf p}_s}{\sqrt{m_s^2c^2+p_s^2}}
\end{equation}
and far upstream ($z<0$) boundary conditions 
\begin{equation}
({\bf E,B})|_{z\to -\infty}=0, \quad F_s|_{z\to -\infty}=\Gamma N_{s1}\delta^2({\bf p}_\perp)\delta(p_z-\Gamma V_1m_s)
\end{equation}

A further and crucial assumption for the development of the self-similar theory is that the self-similar shock structure is strictly stationary, i.e. the non-stationary phase has settled into a stationary final state. This assumes that the time to reach a stationary state is much shorter than the time until the shock resolves into the environment by interaction with the ambient surrounding medium, in afterglows the time of roughly one month. The further evolution of the shock within this time window is considered to be self-similar and slow. 

Under these conditions, it is argued that, for small $\sigma_1=(\lambda_i/r_{ci, th})^2<10^{-2}$, the upstream magnetic field plays no role anymore in the shock structure and the final well-developed self-similarity. This seems to be suggested by numerical simulations in pair ($e^\pm$) plasmas \citep[cf., e.g.,][who reports simulations with Lorentz factor $\Gamma\sim 30$]{spitkovsky05}. The upstream field is just important in the initial shock formation process only.\footnote{This claim is crucial as it enables to develop a stationary self-similar picture of the shock. It could, however, be questioned as it is by no means certain that a shock can reach a stationary state. The most sophisticated numerical PIC simulations of non-relativistic shocks, which do suffer much less from the problems encountered in simulating relativistic shocks, show that shocks are non-stationary on fairly long time scales of reformation \citep[cf.][]{Balogh11} {though it has also been suggested \citep{lembege2009} from  non-relativistic two-dimensional simulations that reformation might be an artefact when restricting to one dimension only}.  Shocks also generate their own upstream magnetic field in various ways by instabilities in which the shock accelerated particle component is involved. This may imply that the \emph{initial} upstream field is less important \citep{katzea07}, but in the \emph{stationary} state $\sigma_1$ cannot be neglected anymore. In addition, one may expect that relativistic shocks do also experience quasi-periodic reformation cycles. If this is the case stationarity would apply to times much longer than the typical reformation time.} Under this assumption on the `stationary shock dynamics', there will be no average large scale stationary magnetic field other than the root-mean-square magnetic field ${\bar B}(z)=\left[\sum_i B_{ii}(\Delta {\bf x}_\perp=0,\Delta t=0, z)\right]^\frac{1}{2}$ at normal distance $z$ from the shock. This field results from the shock generated magnetic fluctuations and their non-zero field correlations $B_{ij}$. The average magnetic energy density at distance $z$ is $\langle \mathcal{E}_B(z)\rangle={\bar B}^2(z)/2\mu_0$.

At large distances $|z|\gg \lambda_i$ from the shock self-similarity implies that the correlation length $\lambda_\mathit{\,corr}$ becomes the only relevant length in the system implying that the shock becomes scale invariant, if it starts diverging as $\lambda_\mathit{\,corr}/\lambda_i\to\infty$. This is assumed to be the case in self-similarity, yielding that all physical quantities become scale invariant power laws with particular power law indices $\alpha_j$ according to \citep{katzea07}
\begin{eqnarray}
{\bf B}({\bf x},t)&=&\xi^\alpha_b{\bf b}\left(\frac{{\bf x}}{\xi},\frac{t}{\xi^{\alpha_t}}\right), \qquad F_s({\bf x},{\bf p}, t)=\xi^\alpha_Ff_s\left(\frac{{\bf x}}{\xi},\frac{{\bf p}}{\xi^{\alpha_p}},\frac{t}{\xi^{\alpha_t}}\right)  \\ {\bf J}({\bf x},t)&=&\xi^{\alpha_f+3(\alpha_b+1)}{\bf j}\left(\frac{{\bf x}}{\xi},\frac{t}{\xi^{\alpha_t}}\right)
\end{eqnarray}
where $\xi=\lambda_\mathit{\,corr}/\lambda_0$, and $\lambda_0$ is any fixed reference scale as, e.g., $\lambda_i$. Denoting the non-thermal particle component by the index $s=b$, the problem then reduces to the determination of the mutual dependence of all the `similarity indices'. For the thermal particle component one has for their gyro-radius  the scaling $r_{c}=cp_{th}/eB\sim \lambda_{\,corr}^{-\alpha_b}$ and thus $r_{c}/\lambda_{\,corr}\sim \lambda_{\,corr}^{-1-\alpha_b}$. Consequently one finds that $0\geq\alpha_b>-1$ for the bounds on the magnetic field scaling exponent. 

The remaining scaling exponents can be determined when inserting the above scaling relations for the field and distribution function into the Vlasov-Maxwell equation and requiring that the different terms should have similar scaling. Then one obtains for $\alpha_t=1$, while the scaling power of the momentum should be $\alpha_p=1+\alpha_b$. This allows for the determination of the relation between $\alpha_b$ and $\alpha_f$ from the scaling of the fast particle current density $J_b$ as fixed from Amp\`ere's law. The result is $\alpha_f+3(\alpha_b+1)=\alpha_b-1$, or $\alpha_f=-2(2+\alpha_b)$. Since the average distribution function of the energetic particles in this case turns out to be a power law function, it can also be written in terms of the momentum as
\begin{equation}
\lim\limits_{p\to\infty}\langle F_b({\bf p}, \lambda_{\, corr})\rangle\simeq \left(\frac{p}{p_0}\right)^{\frac{\alpha_{f}}{\alpha_p} }= \left(\frac{p}{p_0}\right)^{-\frac{4+2\alpha_{b}}{1+\alpha_b} }
\end{equation}
where $p_0$ is a reference momentum \citep[see the discussion in][]{katzea07}. Since $\alpha_b<0$ is restricted by the above arguments, the high energy particle momentum space distribution has power $\alpha_f/\alpha_p\leq-4$, and the flattest particle distribution is obtained for $\alpha_b=0$ in a constant rms fluctuation field. Its correlation function $\langle B_i({\bf x})B_j({\bf x}+\Delta{\bf x})\rangle$ and the spectrum of energetic particles that are accelerated by the self-similar collisionless shock scale as 
\begin{equation}
{\langle} B_i({\bf x})B_j({\bf x}+\Delta{\bf x}){\rangle}\sim (\Delta{\bf x})^{2\alpha_b}, \qquad \frac{{\rm d}N_b}{{\rm d}\epsilon_b}\sim \epsilon_b^{-\frac{2}{1+\alpha_b}}
\end{equation}
with spatial distance ${\bf x}$ from the shock and particle energy $\epsilon_h$, respectively. Thus the  self-similar shock-accelerated particle energy spectrum has power $\leq -2$. 

This self-similar theory is a very important step towards an understanding of relativistic collisionless shocks. On the other hand, it is not clear whether the assumptions made can be justified other than by the heuristic arguments used by \citet{katzea07}. These assumptions are the neglect of the effects of any initial upstream magnetic field on the final self-similar state in shock evolution and, more serious, the assumption that the self-consistently generated magnetic fields can (i) indeed reach the required equi-partition amplitudes to which the self-similar theory refers, (ii) can fill the entire extended downstream and to some extent also the region upstream of the shock. Clarification of these assumption is still far from being achievable. Nevertheless, the self-similar heuristic theory provides a tool for application to shocks. It makes a clear statement about the marginal power of the high energy self-similar particle spectrum and it limits the self-similar exponent of the self-similar shock magnetic field at large distance from the shock. The range of this exponent is narrow. But it contains all the unknown physics the clarification of which is quite urgent and requires the investigation of the details of how a shock sustains itself and which processes are involved. Self-similar theory gives just the hint that the shock may sustain itself without providing any information about the underlying physical processes.

\subsection{\bf\textsf{Relativistic {\small MHD} Waves}}
Shock waves form when the linear eigenmodes of the relativistic plasma evolve non-linearly. This happens whether Waves are excited by an external driver (a piston, the central explosion in  the case of blast waves, etc.) or the shock results from wave steepening in high Mach number collisionless flows. In this respect, a relativistic plasma does not behave differently from its non-relativistic counter part. In order to infer about the nature of the shock and its evolution one thus needs to know the linear eigenmodes which are solutions of the linear dispersion relation. 

Observed astrophysical shock waves are large-scale. They evolve from large-scale low-frequency waves in plasma. One thus needs to find just the low-frequency wave branches of a relativistic plasma. 
The full theory requires a kinetic treatment based on the linearised relativistic Vlasov equation. Observations suggest that magnetic fields $\mathbf{B}$ are involved, and one is mostly dealing with magnetised plasmas. 
To first approximation (and neglecting particle creation and annihilation)  these may be treated by fluid theory based on the ideal relativistic {\small MHD} equations\footnote{\!\!In the non-magnetic case one needs to refer to kinetic theory, however, in order to learn that magnetic fields can hardly be avoided, and to conclude that relativistic collisionless shocks should almost ever be accompanied by self-generated magnetic fields. This will be discussed in more detail below.}. This is most elegantly done in the covariant formulation \citep[cf., e.g.,][]{Komissarov99}. However, for transparency and applicational  purposes the so-called 3+1 split formalism \citep[cf., e.g.,][]{km08} is more appropriate because of its similarity to non-relativistic {\small MHD}. In this formulation one adds the ideal induction equation 
\begin{equation}
\frac{\partial\mathbf{B}}{\partial t}-\nabla\times(\mathbf{v\times B})=0, \qquad \nabla\cdot\mathbf{B}=0
\end{equation}
to the set of relativistic hydrodynamic equations, which couples the magnetic field $\mathbf{B}$ to the dynamics via the velocity $\mathbf{v}$ in the Lorentz force $\mathbf{E+v\times B}$. One also adjusts the energy equation accordingly. 

Linearising around the  uniform background quantities entropy $S_0$, average mass density $\rho_0$, and magnetic field $\mathbf{B}_0$ in the plasma rest frame \citep[e.g.,][]{Komissarov99} one finds the relativistic variants of Alfv\'en and magnetosonic waves. In the 3+1 split, assuming a polytropic equation of state with adiabatic index $\hat\gamma$, the linearised Fourier transformed equations for the first order quantities (index 1) with frequency $\omega$ and wave number $\mathbf{k}$,  read
\begin{eqnarray}
\omega\mathbf{B}_1&=&\mathbf{B}_0\mathbf{k\cdot v}_1-\mathbf{v}_1\mathbf{k\cdot B}_0, \qquad\qquad \mathbf{k\cdot B}_1=0, \qquad\qquad \frac{\rho_1}{\rho_0}=\frac{\mathbf{k\cdot v}_1}{\omega} \\
\omega\mathbf{v}_1 &=&\frac{\rho_0^{\hat\gamma}S_0}{w}\left(\mathbf{k}+\frac{\mathbf{k\cdot B}_0}{\mu_0 \rho_0 h}\right)\left(\frac{\hat\gamma\, \rho_1}{\rho_0}+\frac{S_1}{S_0}\right) + \frac{1}{\mu_0 w}\Big(\mathbf{k\,B}_0\cdot\mathbf{B}_1-\mathbf{B}_1\mathbf{k\cdot B}_0 \Big)
\end{eqnarray}
Here we defined the specific enthalpy $h=c^2+gS_0\rho_0^{\hat\gamma-1}$, $w=\rho_0h+B_0^2/\mu_0$, and $g\equiv \hat\gamma/(\hat\gamma-1)$. The adiabatic index has the two limits $\hat\gamma=5/3$ for non-relativistic and $\hat\gamma=4/3$ for ultra-relativistic temperatures $T$, respectively (i.e. it depends on the internal microscopic velocities and not on the macroscopic speed of the flow). The solutions of these equations are compressible magnetosonic waves and the transverse Alfv\'en wave $\mathbf{v}_1\neq0, \mathbf{B}_1\neq0, \rho_1=S_1=\mathbf{k\cdot v}_1=(\mathbf{B,v})_1\cdot\mathbf{B}_0=0$, with refraction index $n^2\equiv k^2c^2/\omega^2=\mu_0 w/B_0^2\cos^2\theta$, and thus $V_A/c=B_0\cos\theta/\sqrt{\mu_0 w}$, where $\theta$ is the angle between wavenumber $\mathbf{k}$ and the undisturbed magnetic field $\mathbf{B}_0$. 

Shock waves evolve from the compressible modes. Introducing the sound speed $c_s^2=c^2\hat\gamma \rho_0^{\hat\gamma-1}S_0/h$, their dispersion relation is
\begin{equation}
\omega^4-\omega^2\left[\frac{k^2c^2}{w}\left(\rho_0h\frac{c_s^2}{c^2}+\frac{B^2_0}{\mu_0}\right)+c_s^2\frac{(\mathbf{k\cdot B}_0)^2}{\mu_0 w}\right]+k^2c_s^2\frac{(\mathbf{k\cdot B}_0)^2}{\mu_0 w}=0
\end{equation}
We may note that for $V_A\ll c$ one has $\rho_0 h/w=1-V_A^2/c^2$, and for propagation parallel to $\mathbf{B}_0$, i.e. $\theta=0$,  the dispersion is the same as in the non-relativistic case. 

These expressions hold in the fluid frame. The fluid moves with relativistic speed $V\lesssim c$ and bulk gamma-factor $ \Gamma >1$ with respect to the observer and shock frames. In these frames the above expressions for frequency,  wavenumber and fields appear Lorentz transformed according to
\begin{equation}
\omega=\Gamma (\omega'+\mathbf{k'\cdot V}), \quad \mathbf{k-k'}=\mathbf{V}\left[\frac{\omega'\Gamma}{c^2}+\frac{\Gamma-1}{V^2}\mathbf{k'\cdot V}\right], \quad \mathbf{B'}=\frac{\mathbf{B}}{\Gamma}+\frac{\Gamma}{\Gamma+1}\frac{\mathbf{V\cdot B}}{c^2}\mathbf{V}
\end{equation}
with $\Gamma=1/\sqrt{1-V^2/c^2}$, and $\mathbf{V}$ is the relative velocity between the two (primed and un-primed) frames. For the stationary-frame Alfv\'en velocity one obtains 
\begin{equation}
\frac{V_A'}{c}=\frac{1}{kc}\mathbf{k\cdot V'}_{A,\mathit{gr}}=\frac{\mathbf{k\cdot V}}{kc}\pm\frac{\mathbf{k\cdot V}_A/\Gamma^{2}kc}{\sqrt{1+V_A^2/\Gamma^2c^2}\pm\mathbf{V\cdot V}_A/c^2}
\end{equation}
where $\mathbf{V'}_{A,\mathit{gr}}$ is the Alfv\'enic group (wave energy flow) velocity in the stationary frame.

Instead of transforming the magnetosonic wave dispersion relation into the observer's or shock frames it is more convenient to write it in terms of the phase speed $\mathbf{v}_p=(\omega/k^2)\mathbf{k}$ which is a velocity and transforms (in terms of the transformation of the refraction index $n=kc/\omega$) according to
\begin{equation}
{n'}^{-2}\equiv \frac{{v'}_p^2}{c^2}=\frac{\Gamma^2(v_p/c-\mathbf{k\cdot V}/kc)^2}{1+\Gamma^2(v_p/c-\mathbf{k\cdot V}/kc)^2-v_p^2/c^2}\equiv\frac{\Gamma^2(n^{-1}-\mathbf{k\cdot V}/kc)^2}{1+\Gamma^2(n^{-1}-\mathbf{k\cdot V}/kc)^2-n^{-2}}
\end{equation}
In terms of the quantities in the moving frame this yields for the inverse refraction index of magnetosonic waves the quartic equation in the transformed frame
\begin{eqnarray}
\Gamma^2\left(n^{-1}-\frac{\mathbf{k\cdot V}}{kc}\right)^4&-&\frac{(1-n^{-2})}{(1-c_s^2/c^2)}\left\{\left(\frac{c_s^2}{c^2}+\frac{V_A^2}{\Gamma c^2}+\frac{(\mathbf{V\cdot V})_A^2}{c^4}\right)\left(n^{-1}-\frac{\mathbf{k\cdot V}}{kc}\right)^2\right. \nonumber \\
&-&\left.\frac{c_s^2}{c^2}\left[\frac{\mathbf{V\cdot V}_A}{c^2}\left(n^{-1}-\frac{\mathbf{k\cdot V}}{kc}\right)-\frac{\mathbf{k\cdot V}_A}{\Gamma^2 kc}\right]^2\right\}=0
\end{eqnarray}
Its solution can be given only numerically and depends heavily on the value of the Lorentz factor $\Gamma$ of the fluid. An example if given in Figure \ref{chap1-fig1-wavediag} for the group velocity phase diagram of a fast mode in a medium at oblique relativistic propagation. There the group velocity is shown in the projection(grey lines) of the three-dimensional wave-energy surface. Obviously, the transformation introduces the expected anisotropy of the wave pattern that the observer would see when having access to the magnetosonic phase and group velocities.

As in the non-relativistic case, these fast waves may steepen and develop into a shock which is a thin interface between the relativistically fast super-magnetosonic unshocked upstream flow and the slow shocked downstream flow. Once this interface has reached its approximately stationary state the properties of the downstream flow can be inferred from the known properties of the upstream flow by using the appropriate jump conditions across the shock that must be satisfied. The following section gives an example of such conditions at a spherical shock with the flow being aligned radially with a radial magnetic field.   

\begin{figure}[t!]\sidecaption\resizebox{0.54\hsize}{!}
{\includegraphics[]{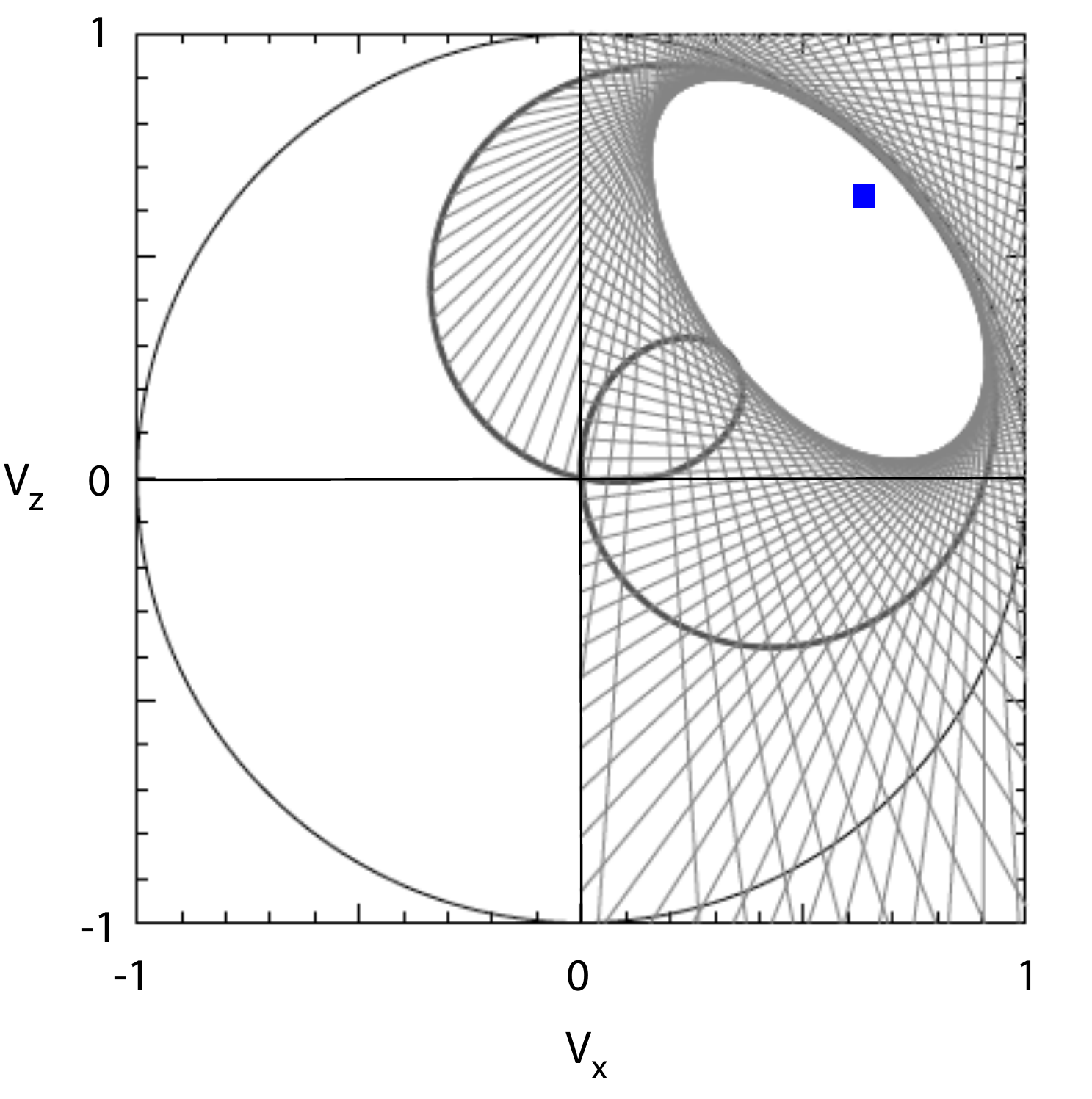} }
\caption[]
{\footnotesize The numerically calculated group velocity phase diagram \citep[after][]{km08} for the fast mode in the observer frame, projected into the $(v_x,v_z)$-plane. The cycloid line is the fast mode phase diagram; Units are normalised to $\rho_0=1, P_0=1$ and $B_0=3$. The velocity was taken as $V= 0.9c$ with direction angle $\theta=\pi/4$  (tip of velocity vector at blue dot). The wave phase is deformed by the relativistic motion.\vspace{3.2cm}}\label{chap1-fig1-wavediag}
\end{figure}

\subsection{\bf\textsf{Kinetic Wave Theory}}

The fluid theory of the previous sections is not applicable to shock formation and shock structure. It just describes the jumps of the fluid parameters across the shock under idealised conditions: when the number of shock-reflected and shock-accelerated particles does not affect the shock structure. Already for high-Mach number nonrelativistic shocks this assumption becomes critical. For relativistic (and even more ultra-relativistic) shocks the efficiency of particle reflection and acceleration is very high, and the shock structure and jump conditions become modified by the presence of a dense energetic particle component which changes wave dispersion and affects the nonlinear evolution and steepening of plasma waves. Simple fluid theory ceases to be applicable then, and relativistic shocks would require a kinetic treatment.

On the microscopic level the dynamics of collisionless plasmas  follows from the Vlasov equation for the one-particle distribution function $f(t,\mathbf{p},\mathbf{x})$ in the plasma frame
\begin{equation}
\partial_t f(t,\mathbf{p},\mathbf{x})+[\mathcal{H},f(t,\mathbf{p},\mathbf{x}]=0, \qquad \int\mathrm{d}^3\mathbf{p}f(t,\mathbf{p},\mathbf{x})=  N(t,\mathbf{x})
\end{equation}
where $\cal{H}$ is the single particle Hamiltonian, $[a,b]$ is the Poisson bracket, and $N$ the particle number density to which the distribution function is normalised. $\mathbf{p}=m\gamma\mathbf{v}$ is the relativistic momentum of the particles, and $\gamma(p)=\sqrt{1+p^2/m^2c^2}$ is the intrinsic Lorentz-factor of the plasma particles (which is a function of $\mathbf{p}$ and is not to be mixed with the external Lorentz-factor $\Gamma$ of the bulk flow). 

The one-particle phase space distribution $f(t,\mathbf{p},\mathbf{x})$ is Lorentz-invariant \citep[see, e.g.,][]{landau1975,VanKampen69}\footnote{The invariance of the distribution can be trivially seen if one recalls that the distribution is the probability of finding a particle in its phase space volume element. When correctly (covariantly) Lorentz transforming the latter from one frame into another frame this probability does of course not change.}  as long as particle creation or  annihilation are absent. Similarly, the pressure tensor $\mathsf{P}$ is an invariant as it is defined as the second differential moment of $f(\mathbf{p})$ in terms of the difference $\mathbf{p-P}$ between the single particle momenta and the bulk momentum $\mathbf{P}=m\Gamma\mathbf{V}$, which is the first moment of the moving distribution function being non-zero only in a streaming plasma. The (isotropic) scalar pressure\footnote{Thermally anisotropic but otherwise equilibrium plasmas require a different definition of temperature which is still debated upon. The correct definition is analogous to Boltzmann's inverse definition and defines the inverse temperature as a four vector \citep[cf.][]{Kampen68}.} is given by $P=\frac{1}{3}\mathrm{trace}\mathsf{\,P}\,(=NT$ for an ideal plasma). 

Since shocks are not in thermal equilibrium but require continuous driving by the upstream flow and the interaction with kind of obstacles, they generate a large number of plasma waves in all frequency regimes from the plasma frequency $\omega_{pe}=\sqrt{4\pi e^2 N/m_e}$ down to zero-frequency plasma modes.  They also emit radiation at frequencies $\omega>\omega_{pe}$ well above the plasma frequency by nonlinear processes and wave-particle interaction. Linear theory can be used to infer about the possible eigenmodes and their growth rates. However, the situation is complicated by the inhomogeneity and variability of the plasma in the shock making the application of the straight linear kinetic approach obsolete.  The basic picture drawn by \citet{sagdeev66} long ago for non-relativistic shocks is still valid in its basic lines and holds also for relativistic shocks what concerns shock formation and the responsibility of low frequency plasma modes.  

The complete spectrum of linear kinetic electromagnetic eigenmodes $\omega(\mathbf{k})=\omega_\mathbf{\,k}$ excited in the shock and shock environment follows from the linear dispersion relation 
\begin{equation}\label{eq:disp}
n^2\equiv\frac{k^2c^2}{\omega^2}=\mathrm{Det}\,\left[n^2\frac{\mathbf{kk}}{k^2}+\mathbf{\epsilon}(\mathbf{k},\omega_\mathbf{\,k})\right], \qquad \mathbf{\epsilon}(\mathbf{k},\omega_\mathbf{\,k})=\mathsf{I}+\frac{i}{\mathbf{\epsilon_0\omega_\mathbf{\,k}}}\mathbf{\sigma}(\mathbf{k},\omega_{\mathbf{\,k}})
\end{equation}
where $\mathsf{I}$ is the unit tensor, and $\mathbf{\epsilon,\sigma}$ are the dielectric  and plasma conductivity tensors, respectively. The latter contain the entire plasma response and thus depend on the assumptions on the plasma model, including the various plasma components and sources of free energy. 

For purely transverse electromagnetic waves the longitudinal first term in the dispersion relation drops out, and the dispersion relation simplifies to just $n^2=\mathrm{Det}\,\mathbf{\epsilon}(\mathbf{k},\omega_\mathbf{\,k})$. In this latter form it applies to the generation of the lowest frequency magnetic oscillations that constitute the shock, in particular to the description of the self-excited shock magnetic field by the \citet{Weibel59}-zero-frequency magnetic instability that is claimed to take responsibility for a much higher shock compression ratio $R_s$ than provided by {\small MHD} shock theory. The latter applies to shock formation in the form of the \citet{Fried59}-`filamentation' instability which describes the head-on interaction of two (collisionless) plasma beams, a model that applies to the spatially limited shock interaction region where the cold fast upstream flow mixes into the hot slow downstream plasma.

The above expressions can also be written in covariant form \citep[cf., e.g.][]{Dewar77} which in the relativistic case is sometimes more convenient for explicit calculation of the dielectric tensor expression. This makes use of the four-dimensional field tensor $F^{\mu\nu}=\partial^\mu A^\nu -\partial^\nu A^\mu$, with $A^\mu=(\phi,\mathbf{A})$ the 4-vector potential, and $J^\mu$ the current 4-vector\footnote{We use the flat metric tensor in the form $\eta_{\mu\nu}=\eta^{\mu\nu}= \mathrm{diag}(1,-1,-1,-1)$.}. In linear approximation in Fourier-space $k\equiv k^\mu=(\omega/c,\mathbf{k})$ with scalar 4-vector product $k\cdot x=\eta_{\mu\nu}k^\mu x^\nu=\omega t-\mathbf{k\cdot x}$ we have $-iF^{\mu\nu}(k)=k^\mu A^\nu-k^\nu A^\mu$, and the connection between the current and potential of species $s$ is given through $4\pi\ ^{(s)\!\!}J^\mu=\, ^{(s)\!}\lambda^\mu_{\ ~\nu}(k)A^\nu(k)$, where $^{(s)\!}\lambda^{\mu\nu}(k)$ is the polarisation tensor contribution of species $s$ which has the symmetry property $k_\mu\, ^{(s)}\lambda^{\mu\nu}= \, ^{(s)}\lambda^{\mu\nu}k_\nu=0$. The linear 4-current is the linear response of the plasma to a perturbation of the plasma distribution $f(p,x,t)$, where $p\equiv p^\mu=m_su_s$ is the 4-momentum, and $u\equiv u^\mu= {\dot x^\mu}$ the 4-velocity, the overdot indicating differentiation with respect to  proper time. This yields for the linear current $^{(s)}J(k)$ from the Vlasov equation
\begin{equation}
^{(s)\!}J^\mu(k)=\frac{\epsilon_0\omega_s^2}{N_s}\int\mathrm{d}^4p\frac{\partial f_{s0}(p)}{\partial p^\nu}\,p^4\left(A^\nu-\frac{\eta_{\alpha\beta}A^\alpha p^\beta}{\eta_{\alpha\beta}k^\alpha p^\beta}k^\nu\right) = \frac{^{(s)\!}\lambda^{\mu\nu}(k)}{4\pi}A^\nu
\end{equation}
where $\omega_s^2=4\pi e^2N_s/m_s$ is the square of the $s$-species plasma frequency, and $f_{s0}(p)$ the (known) undisturbed relativistic $s$-species distribution function. Thus the polarisation tensor becomes
\begin{equation}
^{(s)\!}\lambda^\mu_{\ ~\nu}(k) = -\frac{\epsilon_0\omega_s^2}{N_s}\int\mathrm{d}^4pf_{s0}(p)\left[\delta^\mu_\nu-\frac{k^\mu p_\nu+p^\mu k_\nu}{\eta_{\alpha\beta}k^\alpha p^\beta}+ \frac{\eta_{\alpha\beta}k^\alpha k^\beta p^\mu p_\nu}{(\eta_{\alpha\beta}k^\alpha p^\beta)^2} \right]
\end{equation}
The set of equations for the linear 4-vector potential $A^\nu(k)$ then becomes $D^{\mu\nu}(k)A_\nu(k)=0$ with $D^{\mu\nu}(k)\equiv (\eta_{\alpha\beta}k^\alpha k^\beta) \eta^{\mu\nu} -k^\mu k^\nu +\lambda^{\mu\nu}(k)$, where now $\lambda^{\mu\nu}(k)=\sum_s\, ^{(s)\!}\lambda^{\mu\nu}(k)$ is the full polarisation tensor. 

The covariantly written but otherwise identical to the above non-covariant dispersion relation can directly be read from these expressions as
\begin{equation}
\mathrm{Det}\, \left[(\eta_{\alpha\beta}k^\alpha k^\beta)\eta^{\mu\nu} -k^\mu k^\nu+\lambda^{\mu\nu}(k)\right]=0
\end{equation}
In this form it is explicitly formulated in terms of the undisturbed properly prescribed relativistic plasma distribution function $f_{s0}(p)$, holding in a homogeneous plasma, i.e. for wavelengths substantially shorter than any inhomogeneity thus imposing restrictions on its applicability to long wavelength perturbations of the shock transition layer.

The above theory has been formulated in terms of the 4-vector potential $A^\mu$ which is subject to proper gauging. The most appropriate choice is the Lorentz-gauge $k^\mu A_\mu=0$. In addition one may fix the direction of the wave vector $\mathbf{k}=(k_x,0,0)$. Then $k^\mu=(\omega/c,\mathbf{k})$. This reduces the problem to the three independent polarisations: one longitudinal and two transversal polarisations with unit vectors $\mathit{e_\ell, e}_{\perp 1},e_{\perp 2}$. The explicit expressions for the polarisation unit vectors are
\begin{equation}
\sqrt{| \eta_{\alpha\beta} k^\alpha k^\beta | }\ e_\ell^\mu =(k_x,\omega/c,0,0), \quad e_{\perp 1}=(0,0,1,0), \quad e_{\perp 2}=(0,0,0,1)
\end{equation}
In this orthogonal system the 4$\times$4 tensors $D^{\mu\nu}, \lambda^{\mu\nu}$ reduce to 3$\times$3 tensors $D_{ij}=e_i^\mu D_{\mu\nu}e_j^\nu$, $\lambda_{i,j}=e_i^\mu \lambda_{\mu\nu}e_j^\nu$, and $\eta_{ij}=e_i\cdot e_j$, with $i,j=1,2,3$. As a result the dispersion relation becomes just a different version of Eq.\,(\ref{eq:disp})
\begin{equation}
\mathrm{Det}[D_{ij}(k)] \equiv \mathrm{Det} [\eta_{\mu\nu} k^\mu k^\nu\eta_{ij}-k_ik_j+\lambda_{ij}(k)]=0 
\end{equation}
In order to calculate the elements of $D_{ij}$ it is convenient to normalise the undisturbed distribution function taking into account that the phase space volume is covariant: $f_{s0}(p)\to {\hat f}_{s0}(\mathbf{p})[N_s/\gamma(\mathbf{p})]\delta[p^0-m_sc\gamma(\mathbf{p})]$, where $p^\mu =(m_sc \gamma(\mathbf{p}),\mathbf{p})$, $p_i=p^\mu\cdot e_i$, and ${\hat f}_{s0}(\mathbf{p})$ is the ordinary one-particle phase space distribution, which depends on the 3-momentum $\mathbf{p}$, $\int\mathrm{d}\mathbf{p} {\hat f}_{s0}$ $(\mathbf{p}) = 1$, and $\gamma(\mathbf{p})=\sqrt{1+|\mathbf{p}|^2/m_s^2c^2}$ is the \emph{internal} gamma-factor. Then the reduced polarisation tensor reads
\begin{equation}\label{lambdab}
^{(s)}\lambda_{ij}=-\omega_s^2\int\frac{\mathrm{d}\mathbf{p}\,{\hat f}_{s0}(\mathbf{p})}{\gamma(\mathbf{p})}\left[\eta_{ij}+\frac{(k\cdot k)p_ip_j}{(k\cdot k)^2}\right]
\end{equation}
\citep[A more conventional form of this general expression can be found, e.g., in Eq. (8) of ][]{bretea10}. Calculation of its components is straightforward but the result depends on the model. The non-zero components for a moving species $s$ are (dropping the superscript $s$) $\lambda_{\ell,\ell}$, $\lambda_{\perp 1,\perp 1}$, $\lambda_{\perp 2,\perp 2}$, $\lambda_{\perp 2,\ell}=\lambda_{\ell,\perp 2}^*$. For counterstreaming cold beams and electromagnetic (transverse) modes they have been given explicitly \citep[e.g.][]{silvaea02,aw07}.

\subsubsection*{\textsf{Shock-relevant Wave Instabilities: The Fluid Picture}}
For several reasons only the unstable solutions of the dispersion relation are of interest. They yield instability (growth, interaction, wave cascade, turbulence, and absorption) of plasma waves as being fundamental to shock physics. The astrophysical interest in all these plasma waves and instabilities which may generate them is rather limited; it is dictated by the observation of Cosmic Rays and radiation, with shocks themselves considered just  mediators of both. Astrophysical interest in the very shock structure and formation (taken in this order) has been reluctantly arisen from the as well with reluctance accepted insight that probably neither particle acceleration nor its consequence, the generation of radiation, will be understood without at least some understanding of the principles of shock formation. 

Shock formation on its own is a consequence of instability. Firstly, shock waves result from instability of waves in collisionless plasma when these waves steepen by nonlinear processes, whether growing in the absence of dissipation or being pushed to grow by external drivers. Secondly, these waves are involved in the generation of dissipation, shock thermalisation, and reflection and acceleration of particles with the latter two a collisionless way of dissipating large amounts of the upstream energy supply. Finally and, for relativistic shocks most important, a number of plasma waves seem to sign responsibility for amplification or even generation of the shock-required strong downstream magnetic fields by generating current flow in the shocked plasma environment of the shock, i.e. the apparently very high compression ratios $R_s$ which the contemporary diffusive acceleration theories require, much higher than any gasdynamic and magneto-hydrodynamic theory can provide, even when pushed to its extremes.

Because of this reasons, relativistic shock theories in the astrophysical context have in the recent years turned to investigate some shock-related linear instabilities. These are, in the first place, electromagnetic instabilities driven in the presence of the high energy particle component. It is commonly assumed that the upstream flow is cold, an assumption that is good for at most mildly-relativistic temperatures $T\ll m_ic^2$ and high Mach numbers with $\Gamma\gg 1$. In this case the distribution function degenerates to become a Dirac-delta function, and the fluid dynamical picture can to first approximation be used to investigate the relevant linearly unstable plasma waves. One distinguishes electrostatic (longitudinal, ${\bf k} \| {\bf E}$, wave vector parallel to wave electric field) and electromagnetic (transverse, including mixed) modes. The last decade has seen high activity in analytical and numerical investigation of these linearly unstable relativistic modes \citep[sometimes also claims on their non-linear evolution; as for an incomplete list one may refer to][]{achterberg83,yd87,califanoea97,kazimuraea98,zweibel02,bell04,ba08,bretea10,lp10,lp11}.

Since the relativistic treatment of the full kinetic theory of wave growth is complex, the main analytical efforts have focussed on the approximate fluid picture which to some degree is equivalent to assuming a `waterbag model' for the unstable set of particle distributions \citep[as first used by][in an investigation of the non-relativistic Weibel-filamentation mode]{yd87}. There have been claims that the `waterbag model'  introduced spurious effects \citep{bret09} but these are physically less important subtleties. The main problem with the fluid picture is that it applies only to a subclass of modes; it yields only `reactive' and `non-resonant' instabilities and ignores thermal spreads of the resonances. 

In the cold non-magnetic ($\sigma_1=0$) beam approximation the beam distribution is $F_b({\bf p})\propto \delta(p_x-\Gamma^2mc^2)\delta(p_\perp)$ where $\Gamma^2\sim\gamma_b$ is the beam (index $b$) Lorentz factor, and the beam propagates in $x$-direction with $p_\perp$ the transverse (cold beam) momentum ${\bf p}=(p_x,p_\perp), p_x=mc\gamma_b\beta_x^b$, $c\beta_x^b=\sqrt{1-\gamma_b^{-2}}$. The beam-fluid polarisation (susceptibility) tensor then becomes
\begin{equation}
^b\!\lambda_{ij}=-\frac{e^2}{\epsilon_0m\omega^2}\int\frac{{\rm d}^3{\bf p}\,F_b({\bf p})}{\gamma_b}\left[\delta_{ij}+\frac{(k_i\beta^b_j+k_j\beta^b_i)c}{\omega-ck_i\beta^b_i}-\frac{(\omega^2-k^2c^2)\beta^b_i\beta^b_j}{(\omega-ck_i\beta^b_i)^2}\right]
\end{equation}
For the cold-beam distribution the integral including the factor in front on the right-hand side then simply becomes a factor of $-(\omega_b/\omega)^2$ before the expression in the brackets, where $\omega_b$ is the beam plasma frequency. This susceptibility is to be added to the cold fluid susceptibility $-(\omega_p/\omega)^2$, with $\omega_p^2=\omega^2_e(1+\mu)$, and $\mu=m_e/m_i$. The resulting dispersion relation has longitudinal electrostatic solutions, as Langmuir waves, and one short wavelength, small phase velocity $\omega/k\ll c$ electromagnetic very low frequency solution, the transverse Weibel (`filamentation') mode for $k_x\to 0, k_z\neq 0$ with electric field along and magnetic field perpendicular to the beam (and $k_z$, therefore called filamentation). The growth rate is
\begin{equation}
\mathrm{Im}(\omega)=\omega_b/\sqrt{1+\omega_p^2/k_z^2c^2}\rightarrow \omega_b \quad\mathrm{for}\quad k_z\gg\omega_p/c
\end{equation}
Since the initial (upstream) magnetic field was zero, this instability generates a weak about zero-frequency magnetic field by filamenting of the microscopic beam currents \citep[{for the first astrophysical applications cf.}][]{medvedev99,gw99,Gruzinov01,wa04,le06,aw07,achterbergea07}. {This is a well-known process since the pioneering work of \citet{Fried59}. It has been elaborated in early analytical \citep{davidson1972a} and numerical \citep{lee1973} work and was observed also in laboratory laser-plasma experiments \citep{tatarakis2003}. } Clearly, in the zero-temperature approximation, any temperature effects {are excluded}, and the {original} Weibel instability \citep{Weibel59} caused by temperature anisotropy is not recovered here. Moreover, this instability is of very short wavelength and thus can generate magnetic fields only on a short distance of the order of the existence of the particle beam. Hence when the upstream beam penetrates the shock and is thermalised, the field is created just inside the thermalisation length, i.e. inside the shock ramp transition, and the already mentioned problem arises of how a magnetic field can be transferred to the far downstream. On the other hand, if the shock accelerated (cosmic ray) energetic particles in the downstream can be assumed to represent a beam then the magnetic field would exist all over their confinement scale.

Among the longitudinal electrostatic instabilities, Langmuir modes are of little importance compared with the electron Buneman mode {\citep{buneman1958,buneman1959} which is the most active inside the shock ramp current region to which it is restricted. The Buneman mode causes violent thermalisation of the electrons and has profound nonlinear effects  resulting in the production of large numbers of electron holes with signatures in real space as well as in phase space,\footnote{{Electron holes have been predicted first by \citet{bernstein1957} as kinetic structures both in real and momentum space. Their spatial scales are between a few 10 and a few 100 Debye-lengths, identifying them as micro-scale features which move at speeds very close to the bulk current flow velocity. They cause non-gaussian organisation of the electron distribution in phase space \citep[for textbook accounts see also, e.g.,][]{davidson1972,treumann1997}. The statistical theory based on coarse graining of the phase space has been given by \citet{dupree1972,dupree1982}. Electron holes resulting from current instability  have been investigated analytically and numerically by \citep[cf., e.g.,][and many others]{berk1967,schamel1972,schamel1975,schamel1983,schamel2005,luque2005}. Their most important properties are (i) that they contain strong localised electric potentials of the order of the available plasma-kinetic energy, (ii) reflect and accelerate electrons to energies much larger than thermal, and (iii) trap the low energy component in the hole potential well. More recent numerical simulations \citep[for the most sophisticated examples see][]{newman2001,newman2007,main2006,goldman2007} have investigated their configuration space and phase space dynamics showing that electron holes are an inevitable consequence of plasmas which carry strong electric currents in which case the holes completely determine the micro-scale dynamics of the plasma.}} subsequent electron heating in the flow direction and electron acceleration. The latter is of interest because the phase space holes accelerate a substantial fraction of electrons into a very cold electron beam while the bulk electron component is substantially heated up to temperatures of the electrostatic hole-potential drop, which is a large fraction of the initial energy stored in the current flow \citep[cf., e.g.,][]{newman2001}.}

The secondary cold electron beam escapes along the Weibel-magnetic field into the downstream plasma (note that it cannot escape far upstream because of the relativistic shock speed and the retarding shock potential which inhibits upstream acceleration of electrons) where it serves to generate secondary electron holes via the Buneman instability. The beam continuously regenerates itself in the holes while it heats the surrounding electrons along its path in the magnetised downstream matter up to a distance where its energy becomes insufficient to further feed the generation of electron holes.
The Buneman instability (or its magnetised equivalent, the modified two-stream instability) is probably the main responsible for the generation of a hot electron gas surrounding the shock transition which becomes visible in the emitted high energy radiation spectrum. It might also be the main responsible in driving the Weibel mode via the electron temperature anisotropy it causes and which will grow when the filamentation instability ceases. This whole process is continuously fed by the relativistic or ultra-relativistic upstream flow. Thus the grossly unexplored combination of a Weibel-like instability and a Buneman-like instability seems to be of vital importance in the physics of relativistic shocks.

This picture requires understanding how the magnetic field can be amplified or generated near the shock over distances much larger than the shock ramp transition. Since it is believed that the Weibel instability just generates short-scale magnetic fields in the region of the shock transition only, one is on the search for mechanisms capable of producing magnetic fields on scales larger than the electron or ion inertial lengths on which the Weibel instability grows.. 

Steps in this direction have been proposed by \citet{bell04} who assumed that the shock-accelerated Cosmic Rays signed responsible for the generation of return currents in the ambient flow. Once these undergo fluid instability they can non-resonantly grow and as well generate about zero-frequency magnetic fields. This mechanism has become subject to intense investigation {\citep[cf., e.g.,][and others]{reville2006,reville2008,ba08,ab09,rs09,ohira2009, bykovea11,ss11}} both analytically and numerically.

In an ambient magnetic field ${\bf B}=B_z{\hat z}$, in the ion-inertial (Hall) regime $\omega_{ci}\ll\omega\ll \omega_{ce}$ of unmagnetised ions and magnetised electrons, the Weibel-filamentation growth rate of the wave with wave vector perpendicular to both the beam velocity and magnetic field becomes
\begin{equation}
\mathrm{Im}(\omega)=\omega_b\cos\thetabn/\sqrt{1+\omega_p^2/k_z^2c^2}\rightarrow \omega_b\cos\thetabn \quad\mathrm{for}\quad k_z\gg\omega_p/c
\end{equation}
In the limit of $\cos\theta_{vB}=0, k\to k_z$, with $n=kc/\omega$,  the dispersion relation reads
\begin{equation}
\left(\epsilon_1-n^2+ {^b\!\lambda}_{xx}\right)\left(\epsilon_1-n^2+ {^b\!\lambda}_{yy}\right)-\epsilon_2=0, \quad \epsilon_1\simeq1-\frac{\omega_i^2}{\omega^2}+\frac{\omega_e^2}{\omega_{ce}^2}, \quad \epsilon_2\simeq \frac{\omega_e^2}{\omega\omega_{ce}}
\end{equation}
and has  as solutions the growing ordinary Weibel filamentation and a whistler mode. These waves grow on the electron inertial scale $\lambda_e=c/\omega_e$ thus being of rather short wavelength. We will brief on the Weibel mechanism in the context of numerical simulations.

\citet{lp10,lp11} discussed a number of other instabilities of which they assume that it will be possible to excite them in the shock-upstream region: whistlers, Alfv\'en modes, Bell modes and their relative mutual limitations in dependence on the assumed ratio $\mu$ of electron-to-ion mass. From their investigation it seems that other waves grow faster than the filamentation instability and thus would outrun it being of greater importance  (even though not generating magnetic fields implying that one of the big problems in ultra-relativistic shock physics remains unsolved). 
However, a wave dominates over others not necessarily if its growth rate is larger. Wave growth requires in addition (i) that the source of free energy exists long enough for the wave to grow to sufficient amplitude, (ii) that the growing wave starts from a sufficiently large thermal fluctuation level, and (iii) that the growing wave does not start decaying into other wave modes at an early stage. Neither of these conditions has so far been checked properly. For instance, in a cold plasma of zero temperature the thermal fluctuation level of electrostatic low frequency modes is just on the quantum level. On the other hand, the magnetic fluctuation level might be substantial such that the Weibel mode may grow to large amplitude already before the electrostatic modes have started. In non-relativistic reconnection it has recently been shown \citep{tnb10} that the Weibel instability selects the largest fluctuation level for growing. 

Whether the conclusion that the Weibel filamentation mode does not grow is correct or not cannot be decided easily. Our previous discussion showed that the upstream foreshock (precursor) region of the relativistic shock is extremely narrow and, therefore, any unstably excited waves will be spatially restricted to the vicinity of or very close to the shock ramp itself. Speaking about superluminal waves is relevant only for the phase speed which does not transport energy. Since the group speeds of the waves will always be less than $c$, the energy of the unstable waves will also be confined to the shock ramp and must accumulate there. This implies that the waves very rapidly reach their nonlinear state, and linear analysis breaks down. All kinds of wave-wave and wave-particle interactions take place after this happens, and it cannot be said that one or the other mode does not grow. The most probable case is that the accumulation of wave and particle energy at the shock ramp will provide a very rigid barrier for the upstream flow and cause rapid thermalisation inside the ramp with possible broadening of the shock ramp in the downstream direction such that the shock becomes a broad downstream transition region that will differ substantially from the narrow ramp obtained when investigating low-Mach number or non-relativistic shocks.

\subsubsection*{\textsf{Shock-relevant Wave Instabilities: Kinetic Theory}}
It has been recognised that a pure fluid approach to the unstable modes around relativistic shocks cannot properly account for the main problems. This has stimulated  the investigation of the kinetic theory of unstable waves with the emphasis on long wavelength waves. Because of the complexity of such investigations one should in principle turn to numerical simulations. In this section we briefly review the most recent analytical and numerical work only, restricting to the main most recent proposals \citep{rabinakea10,bretea10}. As for the case of the fluid theory these focus on waves in the \emph{upstream} (foreshock or `precursor') region of relativistic shocks with the main interest in the generation of large-scale magnetic fields as the result of cosmic ray beam-upstream flow interaction \citep[being of most interest in cosmic ray acceleration theory, cf.,e.g.,][and others]{bell04,bell05,katzea07,keshetea09,bykovea11}, assuming that the Cosmic Rays have Lorentz factors $\gamma\gg\Gamma$ and can run far ahead of the ultra-relativistic shock. Hence, this theory is subject to the same restrictive comments as the above fluid approaches.

The basic kinetic theory of relativistic beams passing through a cold plasma has been given long ago \citep{akhiezer75}. Such streaming can excite magnetic waves both resonantly and non-resonantly. The long-wavelengths resonant waves are Alfv\'en waves which grow when the particle streaming velocity in weakly magnetised  plasma exceeds the background Alfv\'en speed $v_b>V_{1A}$. The growth rate in this case is well known \citep[e.g.,][]{be87}: Im$(\omega)_\mathit{res}\approx \omega_{c,cr}[(kv_b/\omega) -1]N_b/N$, with $N=\sum_iN_i$. 

Some general conclusion on {long wavelength} purely magnetic Weibel modes can be drawn \citep{rabinakea10} including a drifting cosmic ray component with distribution function $F_i(p)$ as function of the relativistic particle momentum $p$ of species $i$. The \emph{total} plasma frequency is given as the sum 
\begin{equation}
\omega_p^2\equiv\sum_i\omega_{i}^2=\sum_i \frac{e^2N_i}{\epsilon_0m_i}, \qquad N_i=\int\frac{{\rm d}^3p}{\gamma_i(p)}F_i(p)
\end{equation}
including the integral over the relativistic distribution function $F_i$ (which is a somewhat unusual definition of the partial density $N_i$). It includes an average over the inverse (internal cosmic ray beam) Lorentz factor $\gamma_i(\mathbf{p})$ which results from the relativistic expression of the mass in the denominators of the partial plasma frequencies.\footnote{This formula results from Eq. (\ref{lambdab}). It assumes implicitly that the phase-space volume element d$x^3$d$p^3$ is a Lorentz invariant.  Accounting for Lorentz contraction, the spatial volume element transforms from the co-moving (primed) frame to the observer's frame as d$x^3=$d$x'^3/\gamma(p),\, \gamma(p)=\sqrt{1+p^2/m^2c^2}$ \citep[cf., e.g.,][Section 10]{landau1975}. The momentum-space element, on the other hand, can be shown to transform as d$p^3=\gamma(p)$d$p'^3$ which cancels the extra proper (internal) Lorentz factor in the spatial volume element, thus yielding d$x^3$d$p^3=$\ d$x'^3$d$p'^3$. Indeed,  the phase-space volume element turns out to be a Lorentz scalar.
} 

The cold plasma dispersion relation
with  background flow and cosmic ray beam included becomes a sixth-order polynomial. Only two of its solutions yielding instability (indices $\|,\perp$ refer to the direction of the cosmic ray `beam' ${\bf v}_b$, and $\langle\cdots\rangle$ means averaging over $F_i$)
\begin{equation}
\mathrm{Im}(\omega)=+\omega_b\left[(1+\omega_p^2/k_\perp^2c^2)^{-1}+(k_\|c/\omega_b)^2\langle\gamma^{-2}\rangle\right]^\frac{1}{2}, \qquad k_\|\langle\gamma^{-2}\rangle\ll k_\perp\ll \omega_p
\end{equation}
When a hypothetical beam spread of angle $\Delta\theta$ is taken into account, \citet{rabinakea10} find that instability exists below a critical spreading angle $\Delta\theta<\Delta\theta_\mathit{crit}$ for waves of frequency $\omega\simeq \beta k_\|c$ under the condition $\beta\Delta\theta<\omega_b/\omega_p$, with $\beta=v_b/c$. This has been shown to hold for all axi-symmetric cosmic ray beams. \citet{rabinakea10} note that these long wavelength modes are \emph{not the fastest growing} modes. Shorter wavelength Weibel-filamentation modes will grow faster, as has been known since \citet{Weibel59}.

In application to relativistic shocks, one has $\omega_b/\omega_p\sim\alpha_b^\frac{1}{2}(\gamma_b/\Gamma)^{-\frac{s}{2}}$, with $\alpha_b$ the fraction of post-shock energy carried by the cosmic ray beam, and $s\sim 2$ the approximate power law index of the cosmic ray spectrum. `Far upstream' it is assumed that $\gamma_b\gg\Gamma$. Then, according to the above estimates, the long wavelength modes grow at rate $\mathrm{Im}(\omega)\simeq(\omega_b/\omega_p)k_\perp c$. Since the wave should grow on time scales shorter than the deflection time of cosmic ray particles $\tau_\mathit{def}\sim\gamma m/eB$ by an angle $\sim\Gamma^{-1}$, the ratio of growth to deflection time determines that waves can grow for transverse wave numbers (in shock frame quantities)
\begin{equation}
k_\perp c/\omega_p\gtrsim(\sigma_1/\alpha_b)^\frac{1}{2}\Gamma^{-1}, \quad \mathrm{for} \quad \gamma_b(x)\lesssim \Gamma\left(\alpha_b^\frac{1}{2}\Gamma\right)^\frac{2}{s}
\end{equation}
In electron-proton plasma $\omega_b/\omega_p\sim\sqrt{\alpha_b\mu}(\gamma/\Gamma)^{-s/2}$, and for growing long wavelength modes it is required \citep{le06} that $\Gamma\gtrsim 10^2\sqrt{\alpha_b/0.1}$. Smaller Lorentz factors suppress the growth of long-wavelength waves. In addition the minimum shock-frame energy fraction of the cosmic ray beam as function of distance must be less than $\alpha_b\lesssim\alpha_b^\mathrm{min}=m_i\gamma_b(x)\mu^{1/s}$.

These quite general results for cold beams of a certain angular spread have been generalised to a more complete plasma model including a thermal spread of the particle distribution \citep[cf.][where a  table is given compiling the maximum growth rates of the two-stream, filamentation, and so-called `oblique' instabilities in a cold beam-plasma system for different parameter choices]{bretea10}. This can be achieved by referring to a relativistic thermal Maxwell--J\"uttner distribution function $F_\mathit{MJ}({\bf p})\propto \exp\,\{-\xi[\gamma({\bf p})-\beta_bp_x]\}$ instead of the waterbag cold plasma delta-distribution. This is an isotropic-temperature thermal-equilibrium distribution function which holds in the relativistic isothermal case.\footnote{For a discussion of its validity see \citet{dunkel2007}.} It is normalised to one if using the factor of proportionality $\xi/4\pi\gamma^2({\bf p})K_2[\xi/\gamma({\bf p})]$, where $\xi\equiv mc^2\!/T$, and $K_2(x)$ is the modified Bessel function of second order. Generally,  relativistic flows with non-relativistic temperatures $T< mc^2$ which are of interest in relativistic-shock physics, will exhibit negligible temperature effects. The parameter space of the two-stream and Weibel modes has been investigated for the J\"uttner distribution in order to elucidate any residual temperature effects and to determine the transition between two-stream and Weibel/filamentation instabilities with the known result that at non-relativistic flows the system is dominated by two-stream modes. For relativistic and ultra-relativistic flows the known and expected result is that the Weibel/filamentation instability family takes over. Though, since many parameters are involved (beam density ratio, beam spread, temperatures, Lorentz factors, mass ratio) the system of solutions becomes complex and non-transparent in the thermal J\"uttner  case. Clearly, in the bulk moving frame one has $\Gamma=1$. Hence, temperature effects may, in this frame, affect the evolution of other instabilities which, when growing in the direction perpendicular to the bulk flow, will survive and might also influence the conditions at the shock transition. Effects of this kind have not yet been investigated properly \citep[see, however,][]{dieckmann09}.

In view of the application to particle acceleration and the generation of radiation it is of particular interest to estimate the maximum achievable (in the above sense of the general self-similar theory, i.e. the rms) amplitude of the stationary magnetic field that is generated by the Weibel-filamentation instability (here we do not enter into a discussion of the electrostatic modes even though these might be necessary in preparing the conditions for growth of the Weibel instability). Such an estimate requires the precise knowledge of the saturation (stabilisation) mechanism of the instability which by itself requires the solution of the full nonlinear dynamics of the wave-particle interaction. Since this is an impossible task, a particular mechanism has to be assumed. Quasi-linear relaxation (i.e. the random-phase mean field approach) is invalid in this context because no depletion of the beam takes place. 
\begin{figure}[t!]
\centerline{{\includegraphics[width=1.0\textwidth,clip=]{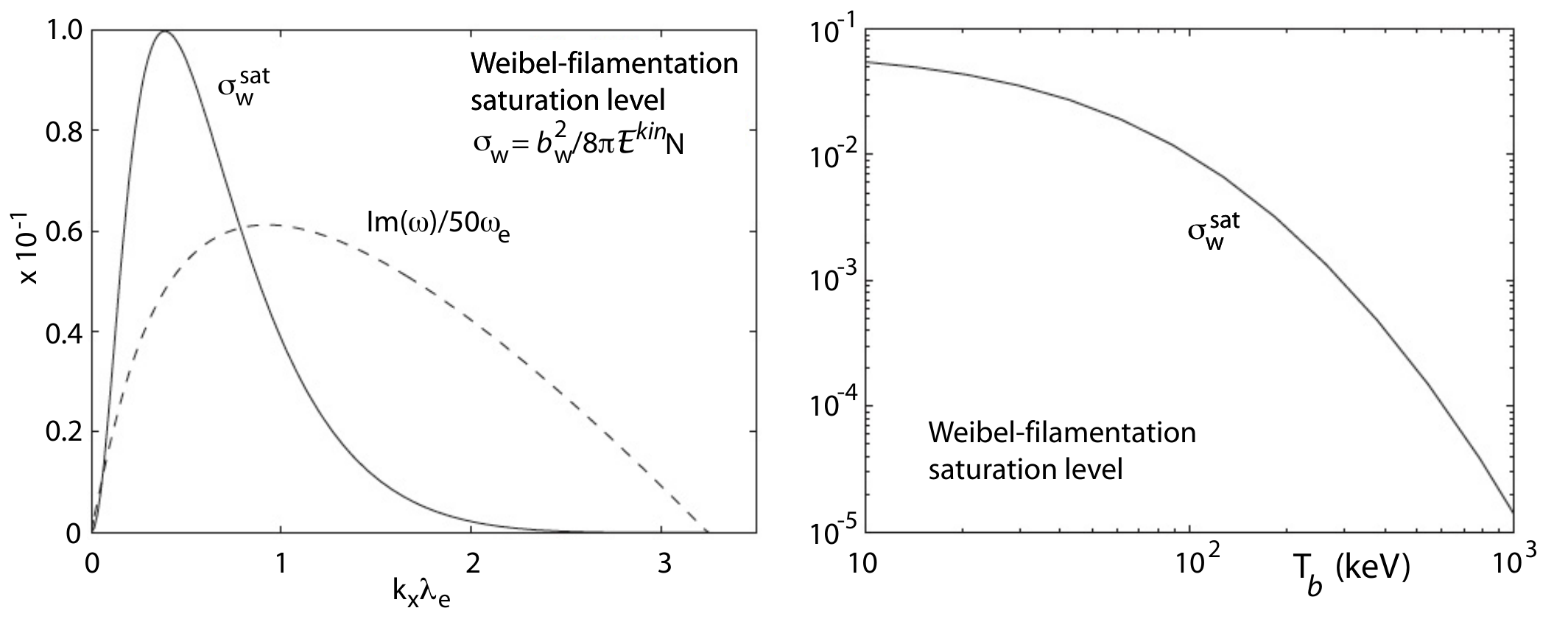}
}}
\caption[]
{\footnotesize The electron Weibel filamentation instability magnetic saturation energy density $\sigma_w^{sat}$ normalised to the total kinetic energy density \citep[after][]{bretea10}. Calculated for $\gamma_b=3, N_b/N=0.8, T_1=5$ keV. {\it Left}: Dependence on wave number $kc/\omega_e$ for $T_b=100$ keV. {\it Right}: Dependence on beam temperature.}\label{relshock-fig-bsat}
\end{figure}

The usual assumption is that the magnetic wave saturates by trapping the beam particles in the magnetic mirrors of the wave. This is the simplest choice from a large variety of possibilities like wave scattering, cascade, wave decay and other parametric interactions including non-linear wave-particle interactions. It supposes that one single wave number $k$ grows fastest and dominates the growth of all other waves, clearly contradicting the self-similar theory assumption. Sticking to this mechanism, the trapping (bounce) frequency in a sinusoidal wave-mirror in the wave frame is $\omega_\mathit{tr}=\sqrt{ev_bb_w/m\gamma_{b,tr}c}$, where $b_w$ is the wave magnetic amplitude. Hence, the bounce frequency in the wave well increases with wave amplitude $b_w$, beam velocity $v_b$ and wave number $k$ and decreases with trapped beam Lorentz factor $\gamma_{b,tr}$.  Assuming that the wave stops growing once the bounce frequency $\omega_\mathit{tr}\sim \mathrm{Im}(\omega)$ becomes comparable to the growth rate \citep[cf., e.g.,][]{silvaea02}, which yields for the saturation amplitude of the filamentation instability
\begin{equation}
b_{w}^\mathit{sat}\simeq\frac{m}{ek}\{\mathrm{Im}[\omega(k)]\}^2\left\langle{\gamma_{b,tr}}/{\beta_{b,tr}}\right\rangle_\mathit{tr}
\end{equation}
The average indicated by the angular brackets $\langle\cdots\rangle$ has to be taken solely over the trapped (bouncing) particle distribution $F_{b,tr}$, and the growth rate is given by any of the above expressions for the filamentation instability, depending on the chosen model. An example of the Weibel saturation field based on the magnetic trapping saturation assumption is given in Figure \ref{relshock-fig-bsat}. It must, however, be stressed again that the magnetic trapping is just one apparently obvious mechanism of saturation based on the assumption that no other nonlinear effects set on earlier and the wave grows linearly until trapping in the wave wells cuts it off. Whether this proposition is realistic is not known. Trapping requires quite large wave amplitudes, a fact that may suggest that long before trapping sets on other processes like cascading may have taken over and either wave saturation ends up with smaller rms amplitude or the spectral energy density is distributed over a larger range in wave number and frequency space.

\subsubsection*{\textsf{CR Modified Shocks: Bell-like Instabilities}}
One particular subclass of electromagnetic instabilities is worth to be singled out. These are the Bell-like instabilities  (originally anticipated though not worked out) by \citet{achterberg83}. They were recognised in their importance and proposed as a fundamental mechanism for generation of magnetic fields near shocks in the presence of Cosmic Rays only more recently by \citet{bell04,bell05}. They differ from the Weibel-filamentation mechanism which is based on beam-beam interaction (or temperature anisotropy). Bell-like instabilities are \emph{non-resonant} electromagnetic very low frequency instabilities which rely on the presence of  cosmic-ray `return currents' flowing in the shock background plasma environment and closing the presumable  current that is believed to be carried by the shock-accelerated high-energy Cosmic Rays into upstream. Of course, they are in principle  already contained in the general theory given above and {have more recently been investigated based on general kinetic theory \citep{reville2006,reville2007,reville2008,amatoea08,ba08,ab09} and also with particle-in-cell (PIC) numerical  simulations \citep{ohira2009}}. 

The original proposal by \citet{bell04}, following numerical simulation work by \citet{lb00},  was based on a fluid ({\small MHD}) model including a hypothetical cosmic ray current density ${\bf J}_{cr}=eN_{cr}{\bf v}_b$ (with $v_b$ the cosmic ray beam velocity) which the background plasma compensates (on its frame of reference) by an electron return current of ${\bf J}_\mathit{ret}=-eN_{cr}{\bf V}$, assuming that for reasons of charge neutrality this current transports the same charge density as the Cosmic Rays, and its velocity is the background plasma velocity induced by the Cosmic Rays and yielding the additional Lorentz-force term $-({\bf J}_\mathit{cr}-{\bf J}_\mathit{ret})\times{\bf B}$ in the {\small MHD}-momentum equation of the upstream magnetised background flow in its proper frame. \citet{bell04}, working in the upstream frame, assumed a parallel shock such that the original upstream magnetic field is parallel to the shock normal.  $J_\|$ is the free input parameter generating a first order perpendicular magnetic field component ${\bf B}_\perp$ and perpendicular current ${\bf J}_\perp=a(J_\|/B_\|){\bf B}_\perp$.  The quantity $a$ is calculated to first order as function of $\lambda=(kr_{c,cr})^{-1}$, the inverse product of wave number and upstream cosmic ray gyro-radius. This yields a linear upstream frame dispersion relation $\omega^2-k^2V_{A1}^2-R[1-a(\omega, \lambda)-\omega/kv_b]=0$ with $R=kB_\|J_\|/m_iN$ which is a cosmic ray modified upstream dispersion relation containing Alfv\`en waves. For large cosmic ray velocities $v_b$ the last term in the brackets can be dropped, in particular as the wave frequency is small, $\omega\sim 0$. The Cosmic Rays turn out to drive a magnetic  wave field in the upstream plasma resulting from a non-resonant instability. For a power law cosmic ray momentum distribution $F(p)\propto p^{-4}$ in the interval $p_1<p<p_2$ the quantity $a$ can be calculated. \citet{bell04} derived the dispersion relation 
\begin{equation}
\omega^2-k^2V_{A1}^2\pm\zeta v_b^2(1-a_1)k/r_{c,cr1}=0, \quad \mathrm{where}\quad \zeta=r_{c,cr1}|BJ_\||/v_b^2m_iN
\end{equation}
and $a_1\to 1$ for $\lambda_1\equiv (kr_{c,cr1})^{-1}\gg 1$ while $a_1\to 0$ for $\lambda_1\ll 1$. For $\lambda_1=1$ a stable Alfv\`en wave is obtained in the uninteresting case $\zeta v_b^2 \ll V_{A1}^2$, when the Cosmic Rays do not carry currents. In the opposite case the dispersion relation is  $\omega^2\approx \pm\zeta (v_b/4r_{c,cr1})^2(1-\frac{3}{4}\pi i)$ and instability arises. These are purely current driven non-resonant {non-firehose} modes, very low frequency magnetic oscillations driven by the cosmic-ray current of particles which on the unstable short scales are non-magnetised. Conditions under which this theory is applicable have been discussed in depth by \citet{bykovea11a}.

The most relevant instability is found for moderately short wave lengths $1< kr_{c,cr1}<(v_b/V_{A1})^2\zeta$. The maximum growth rate becomes $\mathrm{Im}(\omega)|_m\approx\frac{1}{2}\zeta\mathcal{M}_{bA}\omega_{c1,cr}$, where $\mathcal{M}_{bA}= v_b/V_{A1}$, with maximum unstable wave number $k_m\approx (v_b/V_{A1})^2/2r_{c,cr1}$, and the predicted saturation wave amplitude $\delta B/B_1\propto \mathcal{M}_{bA}P_{cr}/m_iN_1V_{sh}^2\approx 300$ is quite large. Note that the assumption is that the initial upstream field is not zero. Nonetheless, in a very weak initial upstream field $B_1$ the saturation level may still be quite low. In favour of the Bell instability one should, however, note that the saturation level is only weakly dependent on the background field; it is determined by the cosmic ray pressure at the shock surface: $\delta B^2/8\pi\sim (V_{sh}/2c)P_\mathit{cr,1}$. {On the other hand, \citet{reville2006} have shown that plasma temperature effects lead to further reduction of the growth rate and also affect the most unstable wavelengths. Still, Cosmic Ray diffusion seems to be amplified substantially, coming close to the Bohm limit, as {\small MHD} simulations by the same authors suggest \citep{reville2008}. Moreover}, numerical PIC simulations performed recently \citep{rs09} seem to indicate that due to early nonlinear saturation of the instability  the amplification which can be achieved is much less than Bell's above estimate, barely exceeding a factor of 10. {Related numerical simulations by \citet{niemiecea08} with a rather different setting had already previously suggested even lower saturation levels, barely exceeding the background magnetic field value.}

There are additional reservations concerning the Bell mechanism with return current in the upstream region. The validity of the magnetohydrodynamic approach in the Bell model (where the return current is carried by background electrons that move together with the Cosmic Rays, which is possible only when the shocks are parallel) has nevertheless been confirmed by kinetic calculations \citep{reville2006,ab09} and by the PIC simulation of \citet{ohira2009}. The flow of the currents in the upstream region implies that the (highly relativistic) shock approaches with light velocity such that it remains questionable whether the comparably slow waves will have time to grow fast enough to reach any substantial amplitudes for saturation before being overcome and digested by the shock ramp.  

\begin{figure}[t!]\sidecaption\resizebox{0.54\hsize}{!}
{\includegraphics[]{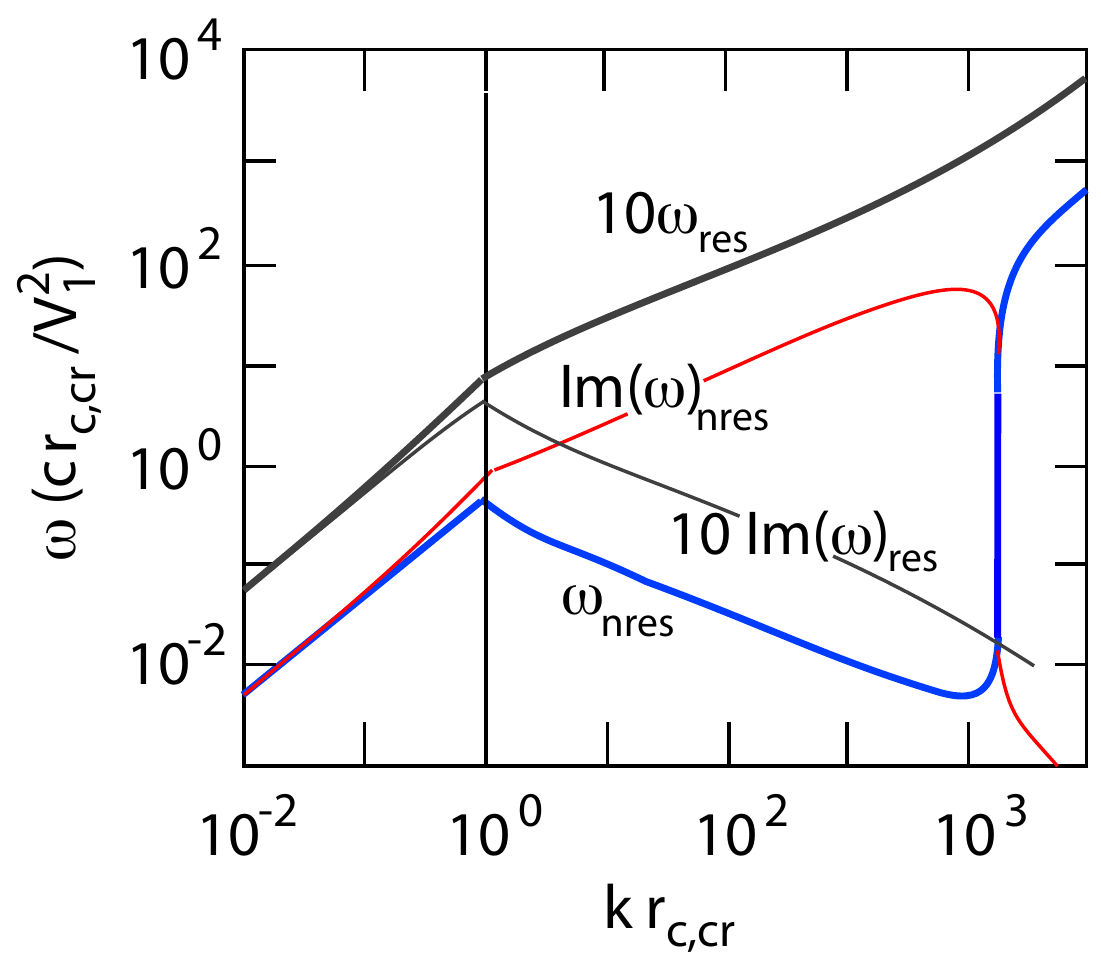} }
\caption[]
{\footnotesize Comparison between the kinetic resonant and non-resonant cosmic ray return current driven magnetic upstream modes for the case when the non-resonant modes is the more important zero frequency oscillation, i.e. in the parameter range $Z\gg 1$ \citep[after][]{ab09}. The numbers chosen are $V_1=10^4$ km\,s$^{-1}$, $B_0=1\mu$G, $N_i=1$\,cm$^{-3}$. The calculation assumes a cosmic ray acceleration efficiency of $\eta=P_{cr}/m_iN_iV_1^2=10^{-1}$ and a maximum cosmic ray momentum $p_{\max}=10^5 m_ic$. Frequency $\omega$ and growth rate Im${\omega}$ are given as functions of wave number $kr_{c,cr}$. Long wave length grow at similar rates for both modes while the non-resonant Bell-like mode dominates for short wavelengths.\vspace{0.5cm}}\label{ab09-fig}
\end{figure}

Also, the maximum growing wave numbers remain to be large since the cosmic ray Mach number $\mathcal{M}_{bA}\gg 1$ is large and, thus, the unstable waves are of quite short wavelengths. This was believed to be in contrast to the resonant streaming instability modes  \citep[cf., e.g.,][]{lc83}; however,  \citet{ab09} have shown that both types of waves have maximum growing wave numbers $kr_\mathit{c,cr}\sim 1$ and comparable though small growth rates at long wavelengths (see Figure \ref{ab09-fig}){;  \citet{reville2006} performed similar calculations for the relativistic kinetic case including temperature effects which yield essentially a similar result thus confirming that growth will occur for wavelength below the gyro-radius of the resonant cosmic ray particles though closer to the gyro-radius than in the non-relativistic case. This result was also confirmed in the simulations of \citet{ohira2009}.} One would not only like to have long wavelengths but also a theory for mostly perpendicular shocks because in fast streams with weak magnetic fields, i.e. $\sigma_1\ll 1$ the shocks are very close to being perpendicular. 

\citet{ab09} have checked the competition between the non-resonant (Bell-like) and resonant modes in a fully kinetic approach for the Bell settings and for the \citet{achterberg83} case when the cosmic ray current generates a compensatory drift in the upstream background, i.e. for the density $N_e=N_i+N_\mathit{cr}$ and zero current $N_iV_1=N_ev_e$ conditions.  In the latter case the dispersion relation agrees with Bell's only for $N_\mathit{cr}\ll N_i$. They extended their calculation to a cosmic ray power law spectrum $\propto p^{-s}$ with $4\leq s<5$. It turns out that the relative importance of the modes is determined by the parameter $Z=\zeta (v_b/V_A)^2$. For $Z\gg 1$ the non-resonant Bell-like mode grows almost non-oscillatory and at a fast rate, dominating the resonant mode. Conversely, for $Z\ll 1$ the resonant modes dominate. Figure \ref{ab09-fig} shows a comparison of the two modes, resonant and non-resonant for the case when $Z\gg 1$. The growth rates of resonant waves maximise at $kr_{c,cr}\approx 1$ while the non-resonant growth increases deep into the short wavelength domain, and in agreement with Bell's prediction the non-resonant mode takes over at short wavelengths $kr_{c,cr}> 1$ and the wave becomes about zero frequency. For long wavelengths $kr_{c,cr}< 1$ there is no difference between the growth rates and frequencies of the two modes.

\subsubsection*{\textsf{The Oblique Case}}
It has already been noted that Bell's case, even though supported by kinetic theory, is marginal as it does not include any obliquity neither for the magnetic field nor for wave propagation. It holds just for a strictly parallel shock while it is suggestive that ultra-relativistic shocks will rather be either magnetically oblique or perpendicular. 

Obliquely propagating waves should be important in shock physics and shock-particle acceleration. This is known from non-relativistic shock theory and simulation \citep[cf., e.g., the comprehensive account of the various plasma waves and their importance in all aspects of nonrelativistic shocks given in][]{Balogh11}. In relativistic shock theory the importance of obliquity of waves has been noted as well \citep[see the wordy but non-transparent discussion in][]{bret09}. However, in view of the very large number of  sources of free energy in the interaction of relativistic streaming inhomogeneous magnetised/non-magnetised, current-carrying/current-compensated plasmas with their environment -- as in the case of external shocks -- with inclusion of high energy accelerated particle populations, any simple listing of the possible instabilities of all kinds must necessarily be incomplete and erratic. It just contributes to confusion. 

In non-magnetised flows (upstream) obliquity refers to the angle between upstream flow and wave vector $\mathbf{k}$. In the presence of a background magnetic field however weak or once a magnetic field $\mathbf{B}$ is generated (as, e.g. in the cases of Weibel's and Bell's instabilities) it refers to the angle between $\mathbf{B}$ and $\mathbf{k}$. Growing magnetic modes in this case should not have  purely magnetic properties like non-resonant Bell modes; because of the implied obliquely directed electric field component, obliqueness rather introduces density fluctuations. for large-amplitude Alfv\'en or the wanted Bell modes (at least in second order) and, thus, the waves become compressible, belonging to the family of magnetosonic waves. 

In initially even weakly magnetised upstream flows the main currents carried by the energetic particle component in their gyration motion. These currents flow perpendicular to the magnetic field; they cannot be compensated in the manner proposed in Bell's original mechanism \citep{bell04} because the magnetised electrons do not follow the gyrating energetic nucleons. They close via field-aligned currents carried by electrons which are accelerated along the magnetic field out of the inflow-plasma background. While the gyrating nucleonic-cosmic ray-current rings excite both electrostatic modified-two stream instabilities and, through the resonant ion-cyclotron maser mechanism, upstream low-frequency electromagnetic cyclotron instabilities \citep[cf., e.g.,][]{aa06}, the ionic equivalent to the celebrated resonant electron-cyclotron \citep[cf., e.g.,][]{treumann06} maser, the field-aligned electron closure currents are distributed over the entire cosmic ray cyclotron orbit. They can undergo other instabilities both electrostatic and electromagnetic.  \citet{bell05} gave a first inclusion of transverse currents, and \citet{bykovea09,bykovea11} included a magnetised cosmic ray current, showing that short wavelength perturbations affected the longer wavelength instability. First numerical PIC simulations performed showed that the amplification of magnetic field in the parallel current Bell instability is lower than analytically estimated \citep{rs10}. Under the assumption that some cosmic ray particles remain magnetised and drive a perpendicular current flowing closer to the shock they however suggest an additional short-wavelength amplification of the Bell-like magnetic field. This will be discussed below in the simulation section.
\begin{figure}[t!]\sidecaption\resizebox{0.54\hsize}{!}
{\includegraphics[]{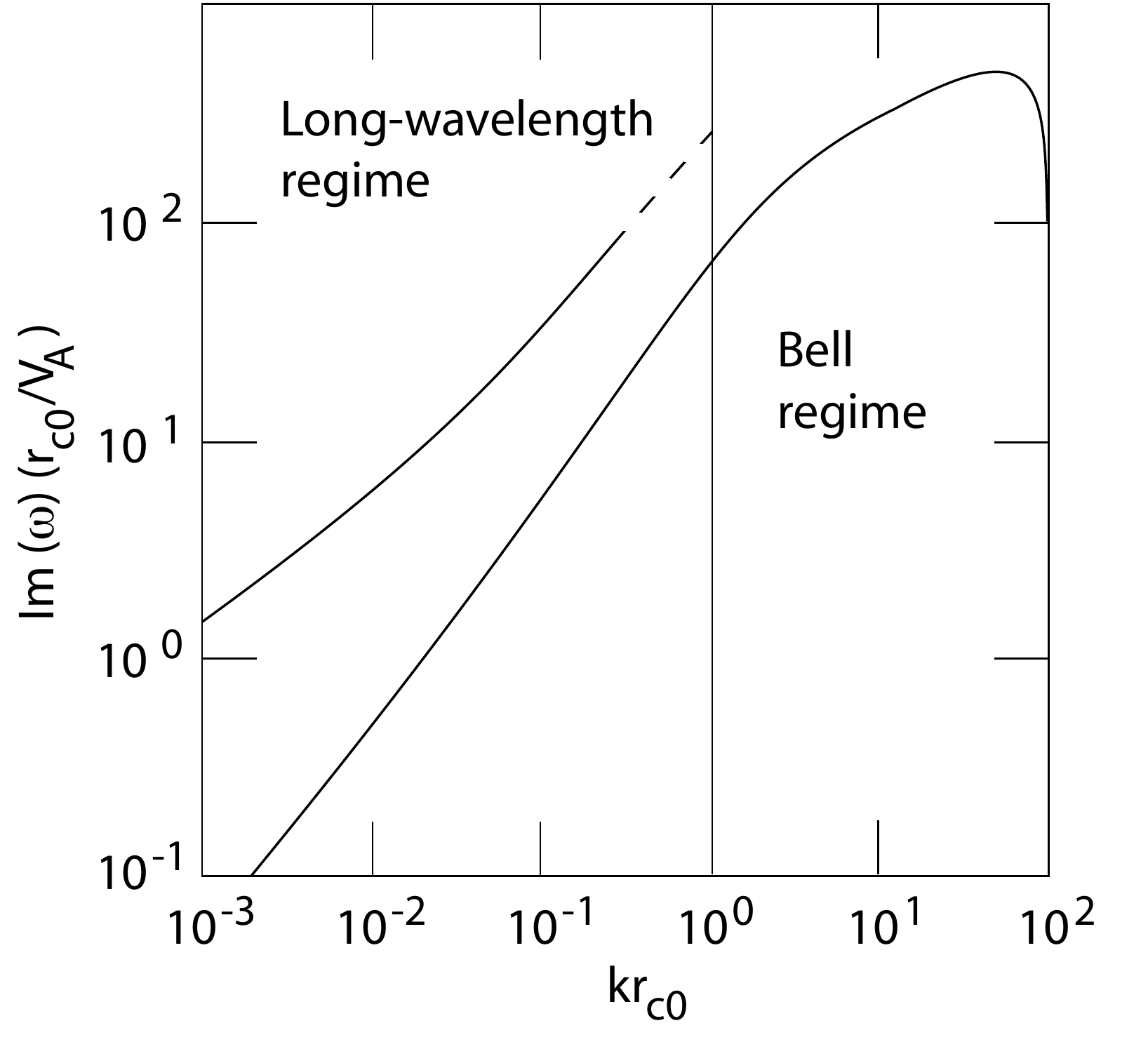} }
\caption[]
{\footnotesize Growth rates of the parallel Bell mode (same as in Figure \ref{ab09-fig}) compared to the growth rate of the long-wavelength mode that is excited in the presence of short wavelength Bell-mode turbulence. The calculation uses the following initial parameters: Bell-mode wavenumber of maximum growth $k_0r_{c0}=100$, cosmic ray spectral power law index $\alpha=4$, mixing length paramter $\xi=5$, and mean free path parameter $\eta=10$ \citep[after][]{bykovea11}. Shown is the mode with the largest growth rate. Clearly the turbulence driven long wavelength mode has increasingly larger growth rate than the Bell mode in the long wavelength regime.\vspace{1.0cm}}\label{relsh-fig-byk11}
\end{figure}

\subsubsection*{\textsf{Transition to Long Wavelengths}}
The point which interests us here is how Bell's mechanism extends into the long-wavelength domain obtaining amplified magnetic fields on the larger scale, as required by the diffusive shock acceleration mechanism. A step into that direction has been gone by \citet{bykovea11a,bykovea11}. These authors propose that the short scale magnetic field perturbations generated by the Bell instability provide a turbulent background for large-scale mean fields to grow. The reason is that the current carried by the cosmic ray component responds to the presence of the short scale turbulence, shorter than the cosmic ray gyroradius which, in the quasilinear (mean field) approach, scatters the Cosmic Rays diffusively and so diffuses the cosmic ray current. In the last  consequence, this leads to damping of the short-scale $kr_{c,cr}>1$ modes and growth of long-scale $kr_{c,cr}<1$ modes with propagation parallel to the initial magnetic field on the expense of the short scale modes.  This dominance of growth of the new (diffusively excited) long-scale modes over the Bell modes is shown in Figure \ref{relsh-fig-byk11}. 

The response in the mean field equations is modelled by assuming diffusive scattering of the current carrying Cosmic Rays at the short Bell fluctuations which impose a finite (anomalous) mean free path on the Cosmic Rays. The latter heuristically modelled as $\ell_\mathit{mfp}= \eta r_{c0}$, being proportional to $r_{c0}$, the initial gyroradius of the Cosmic Rays in the undisturbed field $B_0$, assuming $\eta> 1$. The extreme Bohm limit (fastest diffusion) would correspond to  $\eta=1$. 
\citet{bykovea11} use the Bell velocity fluctuation spectrum  $4\pi\rho|{\bf v}({\bf k}|^2=|{\bf b}({\bf k)}|^2k_1/|k_\||$ as function of the magnetic fluctuation spectrum $|{\bf b}({\bf k})|^2$ in Bell's modes. This fluctuation spectrum is understood as a given background turbulence. Averaging the dynamic equations and the induction equation properly in its presence produces a prefactor $(1+\kappa_t)$ on the mean Lorentz (`ponderomotive') force term in the dynamic equations, where $\kappa_t$ is a turbulent coefficient that is responsible for the transfer of energy into the long wavelength regime. It can be expressed through the average Bell-mode magnetic fluctuation spectral density $\langle{{\bf b}^2}\rangle$ and $\tau_\mathit{corr}\approx 2\pi\xi/k_0\sqrt{\langle|{\bf v}^2|\rangle}$, the correlation time, as
\begin{equation}
\kappa_t=\frac{\tau_\mathit{corr}}{B_0}\int_0^\infty{\rm d}k_z\left(\frac{k_1|k_z|}{4\pi\rho}\right)^\frac{1}{2}\langle{\bf b}^2(k_z)\rangle
\end{equation}
with $z$ along the undisturbed magnetic field ${\bf B}_0$, $k_0=4\pi J_{cr}/cB_0$ the wave number of maximum growth of Bell modes, mixing length parameter $\xi\sim 5$ for short wavelength Bell modes, and $k_1$ the (mean) threshold wavenumber $1<kr_{c0}<k_1r_{c0}$ of Bell's modes, defined by $k_1\langle B\rangle=(4\pi /c)\langle J_{cr}\rangle$ where averaging is over the short modes, and $J_{cr}\approx ecN_{cr}\sigma_{cr}$, with $\sigma_{cr}>B^2/4\pi\epsilon_{cr}$ referring to Cosmic Rays only.

 \citet{bykovea11} obtain a general dispersion relation which in the `intermediate' regime, where the wavelength of the unstable modes is longer than that of Bell modes but still shorter than the mean free path, i.e. $\eta^{-1}< kr_{c,cr}< 1$, yields a parallel propagating wave with right-hand polarisation and growth rate
\begin{equation}
{\bf b}_\mathit{int}=b_\mathit{int}({\hat x}+i{\hat y}), \qquad \mathrm{Im}(\omega)_\mathit{int}\approx 4\pi kV_A\left(\xi\langle{\bf b}^2/B_0^2\rangle\right)^\frac{1}{2}
\end{equation}
The polarisation of this mode is opposite to that of the left-hand polarised Bell modes (and thus offers its possible capacity of partially balancing the magnetic helicity of the generated magnetic fields). For even longer wavelengths in the `hydrodynamic' regime $kr_{c,cr}<\eta^{-1}$ when the mean free path is shorter than the unstable mode wavelength, the system becomes diffusively collisional and can be described by appropriately defined magneto-hydrodynamic equations. Their linear unstable solution s are both, left and right-hand polarised, waves which grow at the same rate
\begin{equation}
\mathrm{Im}(\omega)_\mathit{hyd}\approx \left(\frac{\pi}{2}\frac{k_0}{\eta k}\frac{\langle{\bf b^2}\rangle}{B_0^2}\right)^\frac{1}{2}kV_A, \qquad \frac{k_\|}{k}\equiv\cos\theta>\cos\theta_\mathit{max}\approx \frac{1}{\eta}
\end{equation}
These very long wavelength waves are definitely oblique, and the maximum growing modes, for large $\eta$, propagate about perpendicular to the magnetic field. 

The critical dependence of the unstably excited long wavelength magnetic fields on the finiteness of the mean free path $\lambda_\mathit{mfp}$ through the parameter $\eta$ implies that field energy is dissipated. This dissipation goes on the cost of the driver, i.e. the Cosmic Rays via Bell's instability, but the dependence on $\eta$ implies that some part of the cosmic ray energy is also transferred to the background plasma, predominantly the ions in this case. \citet{bykovea08} have reviewed ion heating and heat transfer to electrons in shocks for the case when it can be  modelled as kind of Coulomb collisions (justified approximately by the assumption of  a finite mean free path $\lambda_\mathit{mfp}$). 

The work of \citet{bykovea11} ist the first quasi-linear (in the literal meaning of the word) attempt of describing the self-consistent generation of low frequency quasi-stationary magnetic fields in the shock environment -- actually just in the upstream pre-shock region. It shows that a gyrating cosmic ray gas that by the Bell mechanism excites short-scale magnetic fields in the lowest non-linear (i.e. quasi-linear or mean field) approximation is diffused by its self-generated short-scale wave spectrum, and the response of the upstream plasma to this diffusive partial depletion of the cosmic ray beam is to excite another branch of wave spectrum the scales of which are much longer than those of Bell modes. These waves escape scattering and can grow on the expense of the cosmic ray and Bell modes.This theory is interesting and promising in providing (still short though somewhat) larger-scale waves than obtained by simple linear Bell theory. It is a first and obviously efficient step toward a theory of inverse wave cascading from short into long scales as known in some models of turbulence where, as in this model, the wave energy-injection source is at the short scales, the scales of Bell's modes shorter than the cosmic ray gyro-radius. 

\section{\bf\textsf{Simulation studies}}
Most progress in genuine collisionless shock theory is attributed to recent numerical simulation studies  based on full particle, either PIC or Vlasov, codes. This is probably the only way of self-consistently treating the arising highly nonlinear problems when dealing with high-Mach number collisionless shocks to which fluid theory is an insufficient approximation. It should, however, be stressed that the astrophysical interest is by no means in the physics of shock formation and shock structure; its focus is on the capability of shocks to accelerate particles -- on the one hand for reproducing the observed (or inferred) energy spectrum of Cosmic Rays, on the other hand for understanding the observed energetic photons which are generated in those systems and which are believed to necessarily result from particles which have been accelerated at the shock. In this view, the relativistic shock structure itself is just a mediator. The model underlying these expectations \citep[cf., e.g.,][]{drury83,be87,md01} is the diffusive shock particle acceleration model (DSA) which, for being effective, requires strong magnetic fields in the shock environment. We noted already that this implies that the shock self-consistently generates magnetic fields. Since the need for field amplification was realised, the main interest in relativistic shock simulations focussed on magnetic field generation and amplification up to magnetic field strengths which substantially increase the classical {\small MHD} compression ratio. Following the initial work by \citet{yd87}, this happened roughly one decade ago \citep{nishikawaea97,califanoea97,califanoea98,califanoea98a,kazimuraea98,medvedev99,gw99,bu99,nishikawaea03}. These authors concentrated on the effects of the Weibel-filamentation instability. Their work caused a sudden inflation of activity in theory and simulation on the Weibel-filamentation mode which has not yet ceased at the time of writing \citep[cf., e.g.,][]{dieckmann07,dieckmann09}. It was later joined by Bell's instability, which however requires the presence of already otherwise produced Cosmic Rays. In the following we review recent numerical work in view of how relativistic shocks form and how they  contribute to the needs of astrophysics  -- with the exception of high energy laser fusion plasmas, the only place in the Universe where relativistic shocks exist.   

\subsubsection*{\textsf{Limitations}}
Before proceeding, we note that all numerical simulations encounter a number of serious limitations. These should be kept in mind prior to naively applying the simulations to a given astrophysical situation. Published simulations are frequently not transparent enough such that their astrophysical validity can from time to time hardly be inferred. 

The first of these limitations is related to the number of particles that can be accessed in a simulation. This number is limited by the capability of following the paths of all particles in a given simulation box. Available computing powers allow treating only a comparably small number of particles of say (in the very best case) $<10^{13}$ in total. This number determines the physical size of the simulation box, cells, and resolution. It can be compared to the number $N_D=\frac{4\pi}{3}N\lambda_D^3$ in the Debye sphere which must be large, $N_D\gg 1$, in order for the plasma picture to be valid. 

In relativistic shock plasmas of $T\sim 10$ keV $\ll m_ec^2$ and $N\sim 10^{-7}$ m$^{-3}$ for instance, one has $\lambda_D\sim5\times10^3$ m, $N_D\sim 10^{15}$ particles per Debye sphere, implying that not even the number of particles in one single Debye-volume could be simulated by present techniques. Thus, each `simulation particle' is a macro-particle which represents a huge number of real particles; this supposes that some microscopic mechanisms are implicit to the simulation which cause the particles to be correlated (or clumped) together on the micro-scale for the entire simulation time $\tau_\mathit{sim}$ such that they obey a common dynamic. One may assume that these are just the particles in the Debye sphere which are held together by electrostatic fields. Then the number of macro-particles contained in the simulation box is in fact the number of Debye-spheres which is also suggested by the requirement of charge neutrality. For sufficient statistics each simulation cell of linear size $d_\mathit{cells}$ should contain a substantial number of such macro-particles; in the best PIC simulations these are about $\sim100$ macro-particles per cell corresponding to a linear cell size of $d_\mathit{cells}\sim 5\times 10^5$ m. Managing the calculation of $\sim10^{12}$ macro-particle paths then allows designing a linear 1D-simulation of $N_\mathit{cells}\sim10^{10}$ cells corresponding to a physical scale of $L_\mathrm{1D}\lesssim5\times10^{15}$ m $\sim0.2$ pc. In higher dimension these numbers (10 particles per cell in 2D and 3 in 3D)  for reasons of reliable statistics require a reduction of cell number (to $10^5$ in 2D, and $10^3$ in 3D) and thus apply to shorter linear physical scales.  

Relativistic shocks move at high speed $V_{sh}\sim c$ across the box. They quickly encounter the boundaries where the results become affected by boundary effects. This restricts the astrophysically useful simulation time. 
On the expense of the box size, the boundary problem can be softened by producing the shock in a collision of two counterstreaming plasmas instead of a reflecting wall and leading to the generation of two shocks;  in the case of external shocks, however, this method suffers from the necessity of considering two media of different physical properties: the stream and the environment. Another possibility is filling the box with a medium of artificially high dielectric constant $\epsilon\gg 1$. This, on the other hand, affects the growth of instabilities and wave propagation, thus causing other unwanted effects. {In addition, \citet{vay2008} recently noted that the standard method in numerical simulations used to advance the particle momentum in time, the so-called Boris pusher, may break down at relativistic particle velocities because the split of the electric and magnetic components of the acceleration inherent to it causes a spurious force on the relativistic particles. Though the effect of this spurious force is not easily quantifiable in highly variable (turbulent) fields, the interpretation of particle motion and the ultra-relativistic part of the distribution function require some care. This effect becomes crucial  in strong electromagnetic fields; in the rather weak magnetic fields related to relativistic shocks it might be less important. } 
 
Finally, the ion-to-electron mass ratio $\mu_{ie}=m_i\gamma_i/m_e\gamma_e$ causes troubles as it sets the scales of electrons and ions far apart. Since relativistic electrons must be assumed to be cooler than ions (because of their high radiation efficiency), the presence of the internal Lorentz factors in the above ratio make the problem even worse. For this reason, simulations in the relativistic domain are usually performed either for pair plasmas (and thus for mass ratio $\mu=1$), or using small mass ratios $\mu_{sim}\ll \mu_{ie}$ of the order of at most a few times 100. Simulations of non-relativistic shocks \citep[cf., the review of shock theory in][]{Balogh11}  unambiguously demonstrated that the behaviour of collisionless shocks changes substantially when the mass ratio becomes realistic and that low-mass ratio simulations can be quite misleading on what concerns shock formation and shock structure. Whether this also holds for relativistic shocks is not known. One may suspect that similar changes occur when it will become possible to increase the mass ratio also in relativistic simulations.  

\begin{figure}[t!]\sidecaption\resizebox{0.54\hsize}{!}
{\includegraphics[]{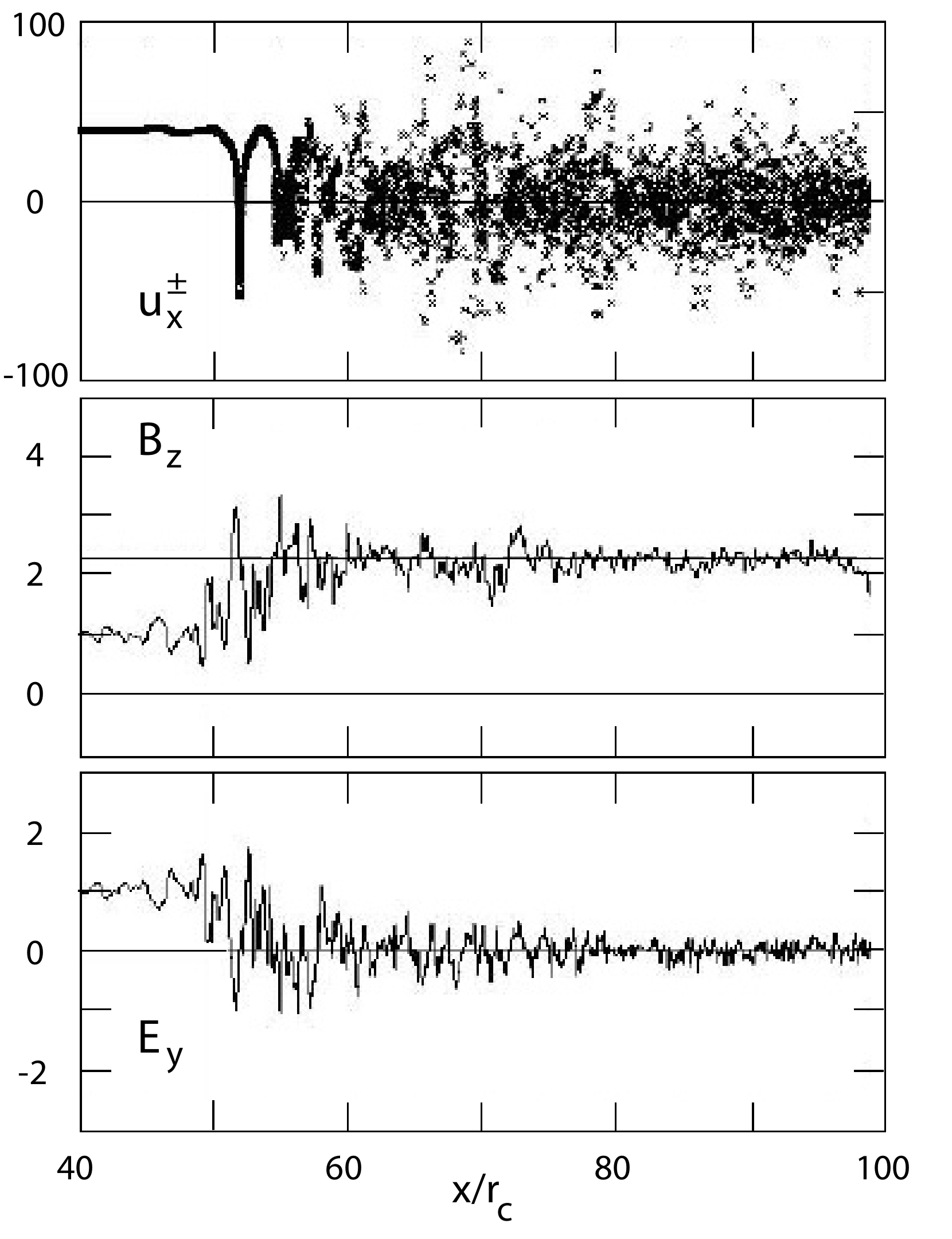} }
\caption[]
{\footnotesize Relativistic perpendicular shock formation in 1D-simulation of a cold charge neutralised pair plasma of density $N_+=N_-$ and upstream Lorentz factor $\Gamma=40$. The shock is produced in the interaction between the inflow beam (coming in from left flowing in $+x$ direction) and the beam that is reflected from the impermeable boundary at the right and flows in $-x$ direction. Shown is the 4-velocity $u_x$ (top), which initially is $u\approx\Gamma$, magnetic field $B_z$, and electric field $E_y$. The fields are normalised to $B_0$. The spatial coordinate is in terms of the initial gyro-radius $r_{c0}$. The shock propagates to the left. It causes electromagnetic precursor waves in the field components which slightly distort the incoming beam in front of the shock (the large turnover in $u_x$ to negative values) and subsequent beam plasma thermalisation with downstream turbulence and production of a small number of energetic particles only up to twice the upstream 4-velocity (i.e. up to $\Gamma\sim80$). This acceleration is not seen upstream; it occurs several gyro-radii downstream. Compression ratio is $\approx2$ \citep[data taken from][]{aa06}. \vspace{0.1cm}}\label{relshock-fig-sim}
\end{figure}

\subsection{\bf\textsf{Pair Plasmas}}
The simplest collisionless particle simulation models use electron-positron pairs $\pm e$ in 1D spatially. A typical yet highly idealised simulation of this kind \citep{aa06} is shown in Figure \ref{relshock-fig-sim} for a strictly perpendicular relativistic shock of moderate upstream bulk Lorentz factor $\Gamma=40$ and $\sigma=1$ (i.e., $\sigma^-=\sigma^+=2$), with a box of $N_\mathit{cells}=1024$ cells, $d_\mathit{cells}=0.1r_{c0}$ and time step  $\Delta t=d_\mathit{cells}/c$. The simulation uses a reflecting impermeable wall, and the shock is caused by the interaction of the two counter streaming (cold) plasmas in $\pm x$ directions. As expected, one observes an electromagnetic precursor wave running ahead of the shock at the speed of light. 

The shock occurs where the incoming beam velocity is turned over into upstream direction and is accompanied by strong field oscillation and magnetic overshoot. Further downstream the plasma becomes thermalised with turbulent field  and compression ratio $\approx 2$. Interestingly, a small number of energetic particles is generated \emph{downstream} reaching a Lorentz factor $\Gamma\approx80$. These particles do not reach the upstream region, at least not in the simulation time as their upstream directed velocity is not high enough to overcome the shock, and there is not the slightest sign of upstream particle acceleration. \citet{aa06} argue that these are simply thermalized particles, estimating that the effective downstream temperature reached is $T_2\approx 15\, mc^2$, which implies that the downstream heating is strong, and converts the downstream region into a region with relativistic temperatures. This, in turn, would imply  that processes like particle creation/annihilation would ultimately come into play. It is nevertheless interesting that the tail of the downstream `quasi-thermal' distribution is long enough to show the presence of quite energetic particles which have been accelerated in the downstream turbulence and not at the shock proper.

\subsubsection*{\textsf{Simulations in Higher Dimensions}}

The first higher dimensional PIC simulations with astrophysical applications have been performed by \citet{silvaea03}. These were 3D fully kinetic electromagnetic relativistic simulations of shock waves produced in the collision of two interpenetrating shells of electron-positron pair plasmas. The total number of particles whose paths could be followed amounted to just $\approx10^{\,8}$. Since the focus was on the generation of magnetic fields via the Weibel/filamentation mode, the shells were assumed to be initially non-magnetised with $\sigma=0$, and the simulation was performed on a box of $25.6^2\times10\lambda_e^3$ cells for a simulation time of $\tau\omega_e=150$ and periodic boundary conditions. Here, as before, $\lambda_e=c/\omega_e$ and $\omega_e$ are the electron-skin length (electron inertial length) and electron plasma frequency, respectively. The inital Lorentz factors used were $\Gamma\approx1.17$ (weakly relativistic) and $10.05$ with thermal spreads $\Delta\beta_\mathit{th}\approx 0.085$ and $0.01$. Under these conditions the Weibel-filamentation mode grows, the current decayed into narrow filaments, and in the interaction region (which was on the size of a few $\lambda_e$) a quasi-stationary magnetic field evolved in both cases, in the relativistic case within times $\approx15\omega_e^{-1}$, afterwards saturating on an about constant level. These qualitative observations confirm the Weibel linear and nonlinear theories of the previous sections, but quantitative application to astrophysics was hardly possible.   
\begin{figure}[t!]\sidecaption\resizebox{0.54\hsize}{!}
{\includegraphics[]{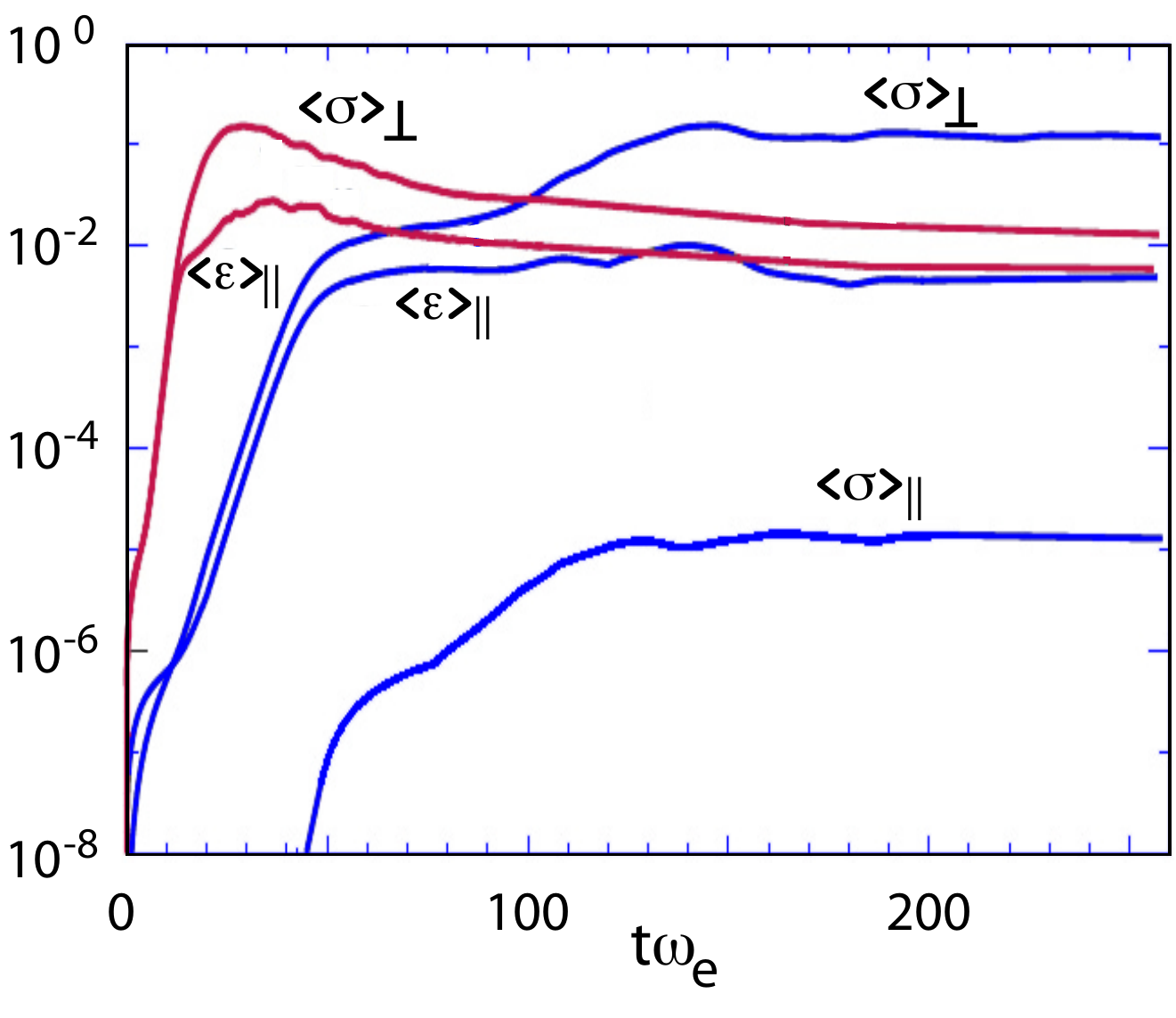} }
\caption[]
{\footnotesize Average magnetic $\langle\sigma\rangle$ and electric $\langle\epsilon\rangle$ energy density ratios in the collision of two pair shells for two different Lorentz factors $\Gamma=10$ (red curves) and $\Gamma=100$ (blue curves) as function of time.  Due to the stronger coupling of the two-stream and Weibel modes the weakly relativistic case saturates earlier and at a lower level than the ultra-relativistic case. The Weibel perpendicular magnetic level is considerably higher in the latter case. Moreover, in saturation the transverse magnetic field energy density is five orders of magnitude higher than the longitudinal \citep[data taken from][]{jaroschekea04}. \vspace{0.9cm}}\label{relshock-fig-jaro-a}
\end{figure}

A similar 3D-model of counter-streaming pair-plasma shells of equal density and bulk momentum being initially charge and current free has been used by \citet{jaroschekea04,jaroschekea05} again with emphasis on the Weibel-filamentation mode using a similar but slightly bigger grid with $5\times10^8$ particles. The simulations use Lorentz factors $\Gamma=10$ for internal shocks and $\Gamma=100$, the latter corresponding to the ultra-relativistic case, and thermal spreads $\beta_\mathit{th}=10^{-5}$ and $10^{-6}$, respectively. In the ultra-relativistic case, which models the external shock scenario relevant in GRB afterglows, simulation times were $\tau\omega_e=260$, corresponding to roughly $5\times10^4$ time steps. 
In the mixing region of the two plasma shells the shock structure evolves into current filaments surrounded by toroidal magnetic fields. Gradually, neighbouring parallel filaments merge until saturation when the filaments are large and at sufficient distance separated by their proper toroidal fields. 

The instability is an oblique combination between the Two-stream and Weibel filamentation instabilities but, for increasing Lorentz factor, evolves into a quasi-perpendicular Weibel mode with propagation perpendicular to the current-magnetic field structure, implying that the system becomes quasi-2D though some obliquity is necessary to keep the global system current-free. It is important to note that strict two-dimensionality would require the closure of currents somewhere `outside' along the magnetic field and would thus cause different effects related to field-aligned current flow.

Figure \ref{relshock-fig-jaro-a} shows the time-evolution of the normalised energy densities in parallel and perpendicular directions to the counter-streaming shells. Here the volume-averaged energy is defined as $\langle\sigma\rangle= V^{-1}\int\mathrm{d}x^3(B^2/2\mu_0)/4(\Gamma-1)Nm_ec^2$. The weakly relativistic case (red curves) saturates early and at a lower level than the ultra-relativistic case. These fields are confined to the interaction region which is narrow, and the simulations do also suffer from the small boxes and the limited simulation time. Saturation is due to particle trapping and thus is diffusion limited. This yields life-times of trapped particles and magnetic field decay of $\tau_\mathit{diff}\sim 10^9\omega_e^{-1}$, a comparably short though reasonable time of $\sim10^5$ s, i.e. of the order of one or few days in real times. This time is not far from the afterglow time. One may notice that nucleonic components will shift the action to the ions thus extending all times by a factor which is a function of the mass ratio $\mu=m_i/m_e$. The electrons, after initially having contributed to seed Weibel fields become readily magnetised and behave passively while the ions will take over letting the electrons radiate in a magnetic field which will further increase, this time now on the ion time scale.    
\begin{figure}[t!]\sidecaption\resizebox{0.34\hsize}{!}
{\includegraphics[]{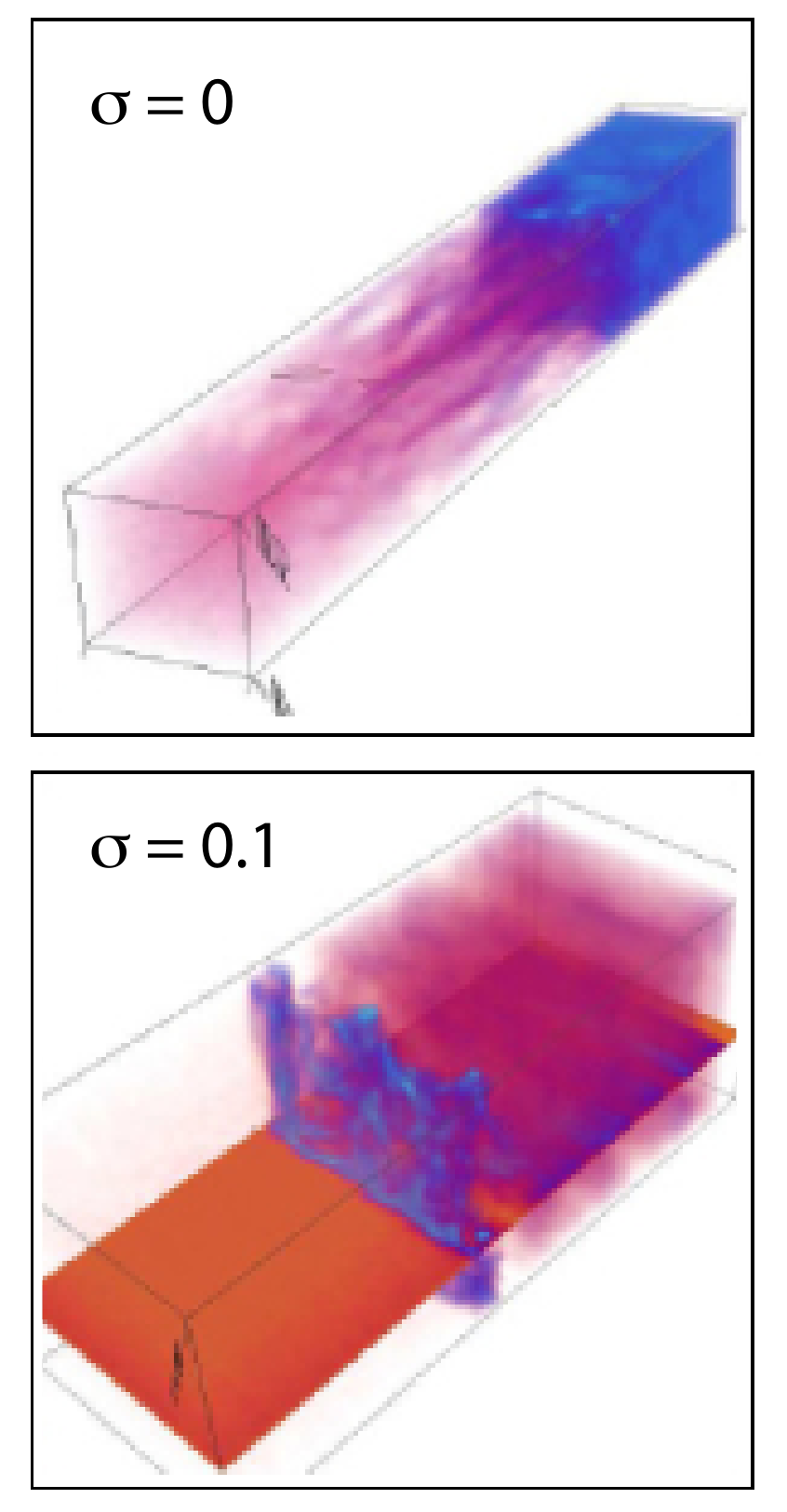} }
\caption[]
{\footnotesize Plasma density structure in two 3D PIC simulations with different $\sigma=0$ and $\sigma=0.1$ for two colliding pair plasmas with $\Gamma=15$ \citep[data taken from][]{spitkovsky05}. {\it Top}: In the absence of initial magnetic fields the Weibel filamentation instability generates stretched current and density filaments and the corresponding transverse (toroidal) magnetic fields. This causes a comparably broad shock transition region ultimately yielding a shock of compression ratio $N_2/N_1\approx{\hat\gamma}/({\hat\gamma}-1)$ and a compressed field which is essentially restricted to the shock, whose width equals the skin depth of $\approx10^2\lambda_e$. Downstream the plasma is thermal but particles of gyroradius larger than the shock width can escape upstream because they move at speed $c$, while the shock moves upstream with speed $c/3$ only. {\it Bottom}: In the magnetised case, the upstream flow runs into the shock front, is retarded and causes a narrow transition by the diamagnetic current of stopped and gyrating particles as well as a stationary precursor wave which is radiated from the shock. The ramp current increases the magnetic field over a  length of $\lesssim10\lambda_e$ or a few gyro-radii in the compressed field, leaving it enhanced and turbulent over some downstream distance. Hence, non-magnetic and magnetic shocks obey a completely different physics and have a completely different structure; they are controlled by different instabilities. Since, however, the magnetic field is generated in a very short time, any ultra-relativistic shock will readily become magnetic after a short time. \vspace{0.1cm}}\label{relshock-fig-spit-a}
\end{figure}

\subsubsection*{\textsf{Shock Structure}}
Systematic investigations of the shock structure have been initiated by \citet{nishikawa05} and \citet{spitkovsky05} who performed 3D PIC simulations on perpendicular shocks in pair plasma. \citet{nishikawa05} used a comparably small simulation box with an $85^2\times160$ grid, $\lambda_e=4.8d_\mathit{cell}$, and performed several independent simulations with a moderate total particle number varying between $5\times10^7$ and $18\times10^7$ particles (a number which distributed over the available simulation cells this corresponded to 27 particles per cell per species); they also applied periodic boundary conditions. The main difference their and other simulations as, for instance, those of \citet{spitkovsky05} which we are going to discuss in more depth, is that they separately injected fast spatially limited jets of different width into a given plasma set up at rest. They used both very narrow and wide jets with a moderate Lorentz factor $\Gamma=5$, corresponding to relativistic though not ultra-relativistic jets, and followed the space-time evolution of the plasma-jet configuration. These simulations are well suited to investigate the mixing of fast jet material with the environment by effects which are caused by the presence of the lateral jet boundaries. What concerns the formation of the relativistic shock which is generated at the head of the jet so the main result was the generation of the Weibel-filamentation instability which was observed in all cases and found to be strongest for an initial $\sigma=0$. 

\paragraph{Effects of Magnetisation.} The $\sigma$-dependence of the relativistic shock properties have been investigated more closely by \citet{spitkovsky05} in a number of simulations of pair plasmas in head-on collision of two counter-straming quasi-neutral plasmas. This arrangement  is perfectly suited for the investigation of relativistic shock formation and monitoring the evolution of shock structure that is independent of the effects of the distant lateral plasma boundaries. 

The most important finding that could be extracted from these simulations was that the shock structure indeed sensitively depends on $\sigma$. At low magnetisations $\sigma\lesssim 10^{-2}$ the shock turns out to be indeed mediated by the Weibel instability. However, at larger magnetisation, i.e. in the presence of external fields or also after the Weibel fields have run nonlinear, the shock couples with other waves (such as Bell modes) and settles in high magnetic field saturation. The shock is then completely controlled by magnetic reflection of particles and substantially less by the Weibel instability. Apparently the transition is gradual with increasing $\sigma$; the newly generated magnetic field gradually takes over and starts dominating shock dynamics via the Lorentz force that it exerts on the particles.

 What concerns the downstream particle distributions, \citet{spitkovsky05} argues that these distributions are always thermal, apparently for all $\sigma$, though this claim is not yet proved by the simulations themselves and might in its stringency possibly hold only for pair shocks. In fact, the simulations by \citet{spitkovsky05} show indeed no direct sign of any shock-particle acceleration. But we may recall that the 1D-simulations by \citet{aa06} as well yielded a thermalised downstream distribution, but they also indicated the presence of a small number of quite energetic particles in both downstream and upstream directions. After a sufficiently long time those among the latter having upstream directed velocities might in nature outrun the slightly slower shock and reach the shock front, possibly even escaping to upstream to be further accelerated. We might note that any positively charged downstream particle that makes it up to the shock front experiences the upstream accelerated shock potential which may kick it out into the upstream shock vicinity. Nevertheless, these simulations of relativistic shocks do not unambiguously support the view that shock acceleration is acting and is indeed the canonical acceleration mechanism in the Universe. \citet{spitkovsky05} argued that there are probably no shocks with low-magnetic field with dimensions $>1D$ in nature. 

\begin{figure}[t!]
\hspace{-1mm}\centerline{\includegraphics[width=\textwidth,height=6cm]{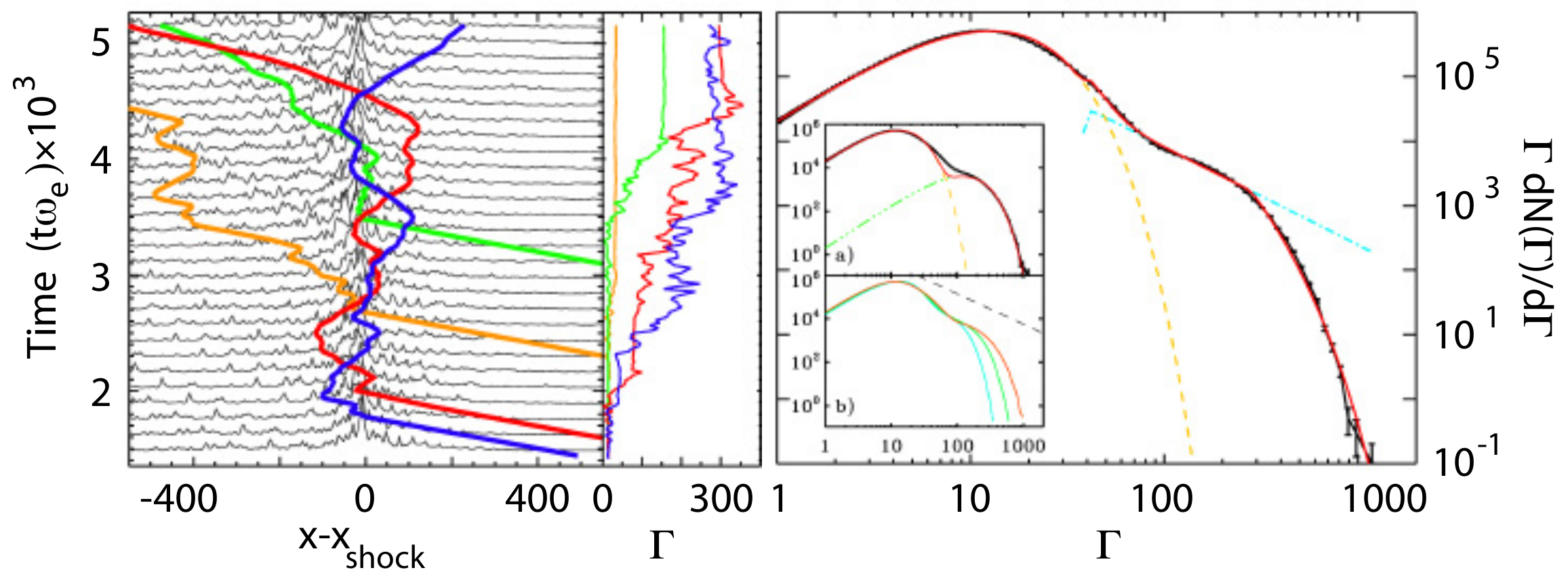} }
\caption[]
{\footnotesize \emph{Left}: The paths of four selected particles (in colour) out of the nonthermal tail of the downstream particle distribution drawn on top of the magnetic field as function of time. The particles manage to pass the shock ramp being reflected and accelerated. The gain in energy is shown in the right part of this panel for all four particles. Inspection shows that the main energy gain is accumulated in resting at the shock ramp and moving back and forth around the ramp on small oscillations only. \emph{Right}: The particle spectrum downstream in a 100$\lambda_e$ wide slice at downstream distance 500$\lambda_e$. Maxwellian fits are shown, and a power law fit to the flat region is indicated. \citep[data taken from][]{spitkovsky08}.  }\label{relshock-fig-spit-b}
\end{figure}

\paragraph{Long-term Behaviour.} In order to check the long-term behaviour of shocks, \citet{spitkovsky08} also performed 2D PIC simulations in unmagnetized pair plasma and found that the particle distribution downstream of such shocks consists of two components: a relativistic Maxwellian, with a characteristic temperature set by the upstream kinetic energy of the flow, and a high-energy tail that extends to energies $>100$ times that of the thermal peak. 

This high-energy tail is best fitted by a negative power law in energy of index $2.4 \pm0.1$, modified by an exponential cutoff owing to the finite size of the simulation box \citep{bu99}. The cutoff moves to higher energies with simulation time, causing an increase of the power-law range. The number of particles in the tail is approximately estimated as $N_\mathit{tail}\sim0.01N_2$. However, these particles carry a substantial fraction of $\approx10$\% of the downstream kinetic energy. The idea is that particles from the hot thermal downstream background above a certain energy with velocity directed upstream are in the long term indeed fast enough to outrun the comparably slow shock which moves with nominal speed $\beta_\mathit{sh}\sim c/3$ upstream. Once in the upstream region, these particles are believed to be back -scattered and to undergo Fermi-type acceleration. 

\citet{spitkovsky08} analysed the trajectories of some of these particles in a similar way as \citet{jaroschekea04,jaroschekea05}  suggesting that the particles indeed may have escaped upstream and were back-scattered in the magnetic fields generated self-consistently by the Weibel instability. If a large enough number of such particles can accumulate with time  in the tail of the downstream distribution they could in addition enhance the backscattering efficiency by exciting Bell modes and their oblique relatives. Subsequently the shock would become strongly magnetised, possess a magnetised foreshock and would become mediated by the self-generated energetic particle component. 

\begin{figure}[t!]
\hspace{1mm}\centerline{\includegraphics[width=\textwidth,height=7cm]{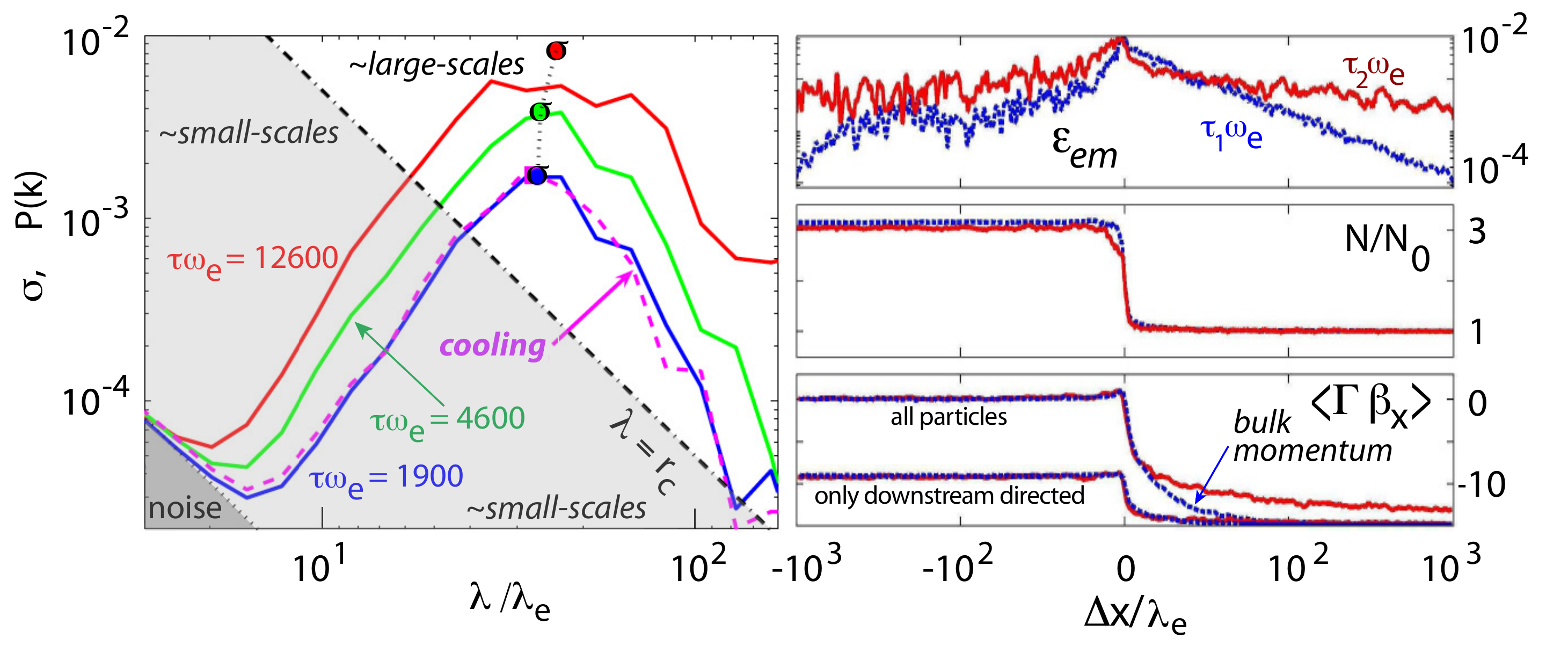} }
\caption[]
{\footnotesize \emph{Left}: Simulated spectrum of  the 2D-magnetic power $P(k)$ and of the evolution of the $\sigma$ parameter taken in the downstream interval $-1000\lambda_e<\Delta x<-200\lambda_e$ at three progressing times, with $k=2\pi/\lambda$ wave number, $\Delta\,x$ downstream distance from the shock ramp, and $\lambda_e=c\omega_e$ the electron skin depth. The increase of the integrated magnetic energy fraction $\sigma(\tau)$ with time $\tau$ is shown as big dots. The diagonal corresponds approximately to the demarcation line $\lambda=r_c$  between small-scale (magnetised, wavelength $<$ rms-gyroradius $r_c\simeq \sqrt{2P(k)}\lambda_e$) and large-scale (non-magnetised) fluctuations. The maximum of the growing power is for all times above this line indicating maximum growth of downstream fluctuations at small wave numbers-long wavelengths (i.e. at larger scales). Artificially cooling the system by replacing particles above $\Gamma=80$ with thermal particles stops the wave growth as indicated by the pink (`cooling') curve. \emph{Right top}: Normalised electromagnetic energy density $\epsilon_\mathit{em}$ for two simulation times $\tau_1\omega_e=2250, \tau_2\omega_e=11925$, early (blue colour) and late (red colour) times, respectively. Note the increase towards higher fluctuation levels at larger distance to both sides of the shock. \emph{Right middle}: Normalised density $N/N_0$. With time the shock profile flattens a bit towards downstream.  \emph{Right bottom}: Average particle momentum taking all particles and only downward streaming particles, respectively. Note the substantial increase in upstream energetic particles at late time. The upstream flow starts retards substantially closer to the shock ramp than one observes the upstream energetic particles. Near and behind the shock the main momentum contribution comes from the energetic particles \citep[after][]{keshetea09}. These simulations show the presence of an over-abundant population of high-energy accelerated particles upstream of the shock.}\label{relshock-fig-keshet}
\end{figure}

Because of their small boxes and comparably short simulation times,  all these simulations, even those of \citet{spitkovsky08}, are incapable of detecting any massive shock acceleration of particles, though we already noted some downstream indications for the potential presence of accelerated particles whose origin still remains unclear; it might probably rather to be searched in a process of downstream turbulent acceleration than in a Fermi-like mechanism. 

Recently \citet{keshetea09} extended both the length $L$ of the box and the simulation time $\tau$ up to a combined number $(L/\lambda_e)^2\tau\omega_e$ in non-magnetic (initial $\sigma=0$) pair-plasma shock simulations with $\Gamma=15$. They used the same code as \citet{spitkovsky05} but replaced the two interacting plasma streams by one stream only and applied a reflecting-wall set-up in a box of total scale $6300^2\times1024=4\times10^{10}$ cells. The total number of particles was increased to $\approx2\times10^{10}$. 

What interests us here is the long-term evolution of the shock (i.e. its state close to the end of the simulation time $\tau\omega_e\gtrsim 10^3$) as the short term evolution is identical to that described above. The important findings in these numerical experiments can be summarised in four main points: (i) The gradual generation of `larger scale' and `larger amplitude' magnetic field fluctuations surrounding the shock transition -- the term `larger' will be defined more precisely below -- than in the former short-term simulations.  They also extend far into upstream, possibly (though not necessarily) suggesting that energetic particles manage to pass the shock front from downstream to upstream, where they contribute to the generation of magnetic fluctuations. One should, however, note that these authors do not follow single particle orbits and thus cannot infer about the origin of the high-energy particles. (ii) An increase of the total normalised electromagnetic fluctuation energy $\epsilon_\mathit{em}\equiv \sigma + \langle |{\bf E}|^2\rangle/m_eN_1c^2(\Gamma_0-1)$ occurs both downstream and upstream of the shock ramp.(iii) The density profile of the shock is modified by the presence of energetic accelerated particles, and the shock slightly accelerates. This is an effect of the presence of the increasingly hotter (more energetic) downstream plasma. (iv) The upstream energetic particles increase the average upstream momentum far ahead of the shock at distances $\sim10^3\lambda_e$ while the bulk momentum of the upstream flow starts decreasing only at upstream shock-distances $\lesssim 10^2\lambda_e$. 

These results can be read from Figure \ref{relshock-fig-keshet}. The left panel shows the evolution of the magnetic power spectrum and average $\sigma$. One finds that the maximum magnetic power is generated at `non-magnetised' scales (larger than the gyro-radius) of the order of $\lambda\sim 3\times10^2\lambda_e$ which is still very short, yet one order of magnitude larger than the typical Weibel wavelengths, and the wavelength does slowly increase with time. Still, the simulations stop after a comparably short time and  practically no saturation is reached. Also, though the wave power increases by somewhat more than two orders of magnitude, its peak in these simulations  remains about constant at $\sigma_\mathit{max}\sim 7$\% at the shock front. What concerns the high-energy particles that possibly escape to upstream one should, however, carefully take into account that the simulations by \citet{spitkovsky08} indicate that this happens only in the unmagnetised case while the magnetisation inhibits such an escape for at least a while. Figure  \ref{relshock-fig-keshet} shows that there is substantial magnetisation caused. Hence, it cannot be concluded that the upstream particles are those which have escaped from downstream. For this to know the particle orbits should have been followed. also, one would wish to have magnetised simulations for long enough times to see whether then sufficiently hot particles can be generated downstream for escape to upstream as well and how long this will take. In any case, the simulation times are short compared with any natural time scale for direct comparison.

Similar long-term  simulations in unmagnetised electron positron plasma were recently performed in 3D by \citet{nishikawaea10} who injected $\Gamma=15$ and $\Gamma=100$ electron-positron jets into a long though narrow box (on a $4005\times131^2$ grid). Their goal was to examine the nonlinear stage of the Weibel instability and its capacity for particle acceleration. These authors found cold jet and ambient electrons  to become thermalized in the resulting shocks. In these simulations the pair jet interacts with the background pair plasma and causes an external bow shock. Large amplitude electromagnetic fields generated behind this shock lead to turbulent downstream magnetic fields. The Weibel instability is indeed identified; it evolves  nonlinearly to large amplitudes. The authors argue that  the field would  be strong enough for causing the synchrotron afterglow emission in Gamma Ray Bursts.

\subsubsection*{\textsf{Magnetic Field Obliqueness and Acceleration}}
\paragraph{Dependence on Magnetic Inclination.} So far either initially unmagnetised or magnetised  perpendicular pair shocks have been discussed.  An attempt to include oblique shock angles has been undertaken by \citet{ss09a} who performed 2.5D and 3D PIC simulations in order to investigate the dependence on $\thetabn$, the angle between the upstream magnetic field ${\bf B}_1$ and the shock normal ${\bf n}$. A set up with Lorentz factor $\Gamma=15$ is used with cell number 50000. The plasma skin depth chosen to be $\lambda_e=10$ cells, each cell containing initially 2+2 particles.  Emphasis is put on particle escape to upstream and particle acceleration.  In order to control the Weibel instability, the initial magnetisation is set to $\sigma=0.1\gtrsim 10^{-3}$ \citep[cf.][for discussion]{spitkovsky05}. In doing so the simulations can directly be applied to Pulsar Wind Nebulae (PWNe) and internal shocks in GRBs and AGN jets. 

\paragraph{Sub-luminal versus Super-luminal Shocks.} Particles can escape upstream only when their upstream speed component along the magnetic field is larger than the sub-luminal speed of the shock. Otherwise the shock is super-luminal with respect to the particles, and they will be swept downstream. In the reflecting-wall frame the critical angle between sub-luminal and super-luminal shocks is $\tan\theta_\mathit{Bn,crit}=\Gamma\tan(\cos^{-1}\beta_\mathit{sh})$, with $\beta_\mathit{sh}$ the shock speed in the upstream frame; this yields $\theta_\mathit{Bn,crit}\approx 34^\circ$ in the wall frame \citep{ss09a}. Such quasi-parallel relativistic shocks thus will behave similarly to unmagnetised shocks, possessing foreshocks and are subject to mediation by the Weibel filamentation instability which is the counterpart of the resonant magnetic-pulsation instability which is excited in the foreshocks of quasi-parallel  non-relativistic shocks. 

\begin{figure}[t!]\sidecaption\resizebox{0.65\hsize}{!}
{\includegraphics[]{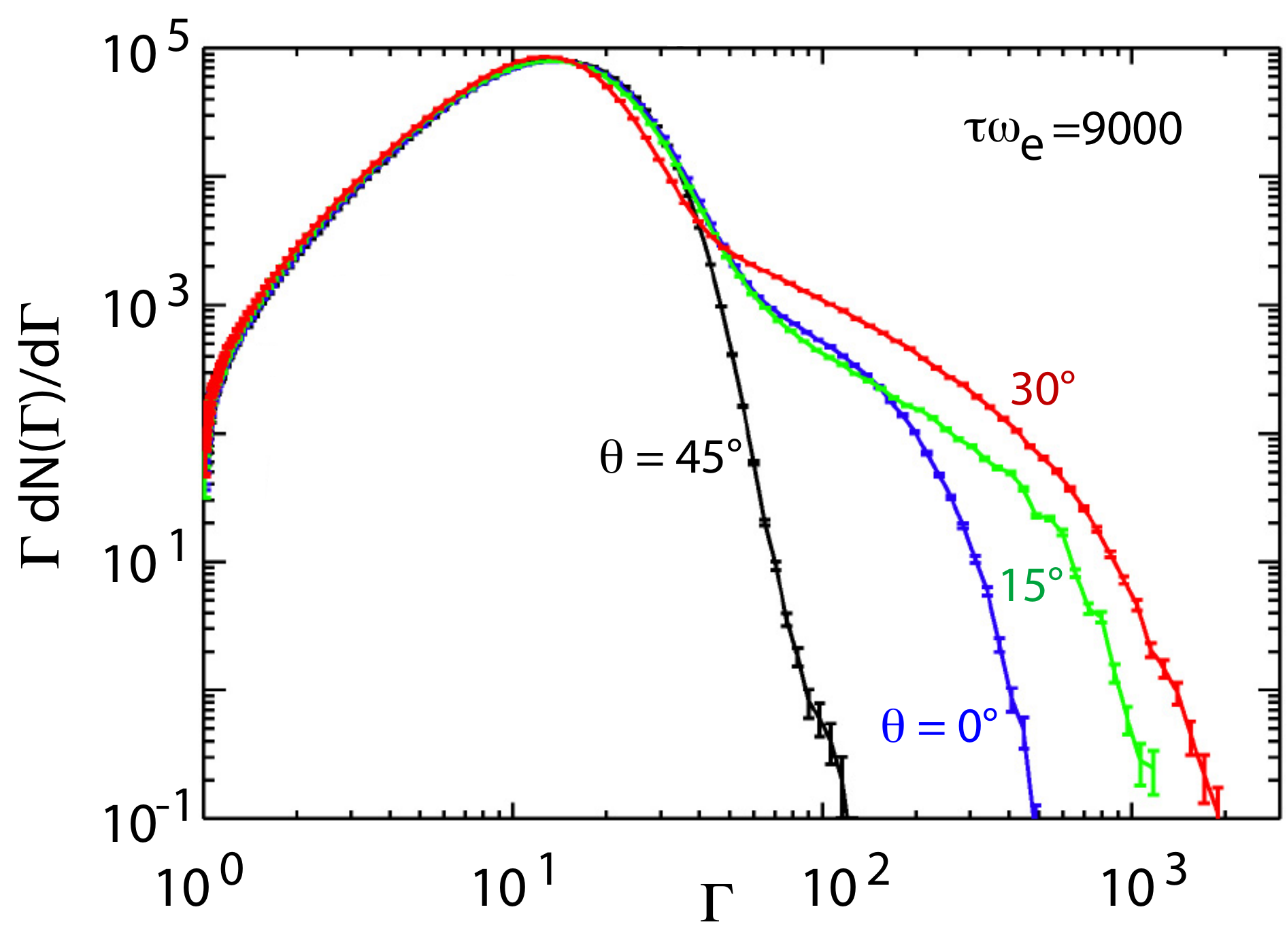} }
\caption[]
{\footnotesize The evolution of the downstream particle-energy spectrum as function of the upstream magnetic inclination angle $\theta_\mathit{Bn}$ for $\tau\omega_e=9000$ and $\Gamma$ running from 1 to $2\times 10^3$. Below the critical inclination angle $\theta\approx 34^\circ$ particles can escape to the upstream region, undergo SDA and return to downstream, where they contribute to a nonthermal tail on the particle energy spectrum. Above the critical inclination, no particles  escape to upstream and no super-thermal tail evolves  
\citep[after][]{ss09a}. \vspace{0.5cm}}\label{relshock-fig-sido}
\end{figure}

Figure \ref{relshock-fig-sido} shows the simulated downstream particle-energy spectra as function of $\Gamma$ and in dependence on the magnetic inclination angle $\theta$ taken at identical simulation time $\tau\omega_e=9000$. The simulated energy spectra confirm indeed that only sub-luminal shocks let downstream particles return to upstream and become accelerated. Such particles, in addition, contribute to the generation of upstream waves which, in the relativistic domain, parallels what is known from non-relativistic oblique shocks \citep[cf.][]{Balogh11}. In contrast, `super-luminal' shocks\footnote{To avoid confusion, one should keep in mind that the term `super-luminal' does not mean that the shock is moving at velocity faster than the speed of light. It is just the geometric point of intersection of the inclined upstream magnetic field and the shock front which moves at an apparently super-luminal speed along the shock.} do not permit particles to escape upstream and therefore lack any upstream waves. One may thus conclude that the presence of an oblique upstream magnetic field at angles below the critical  angle suppresses the escape of downstream particles to upstream and thus also suppresses the self-consistent generation of upstream waves and further shock acceleration of the escaping particles to higher energies either by the diffusive shock (DSA) or shock drift (SDA) acceleration mechanisms. This conclusion holds for the self-consistent acceleration and upstream injection of particles from the background pair plasma and should be valid until the downstream turbulence accelerates particles to energies such high that their gyroradii become so large that they effectively demagnetise in the downstream plasma. 

In the sub-luminal case the downstream  particles which are believed to escape upstream become accelerated by the SDA mechanism, finally return to downstream and contribute to the non-thermal tail of the downstream particle energy spectrum. Like in Figure \ref{relshock-fig-spit-b} the downstream particle spectrum consists of a relativistic thermal background plus a non-thermal tail with power index lying between $2.8\pm 0.1\lesssim \alpha\lesssim3\pm0.1$ and having a high-energy exponential cut-off. The evolution of the particle spectra for different magnetic inclination angles is shown in Figure \ref{relshock-fig-sido}. The self-consistent DSA mechanism with shock generated particle injection works only for nearly parallel shocks.

\subsection{\bf\textsf{Electron-Ion Plasmas}}
Electron-positron (i.e. pair) plasmas may not be rare in astrophysics. They are found in various objects, particularly in pulsar winds, pulsar wind nebulae and possibly also in Gamma Ray Bursts, where they may be created by various processes. The choice to use pair plasmas in simulations has not been guided by their abundance but by the advantage the mass ratio $\mu=1$ offers. 

Real astrophysical shocks, in particular external shocks, involve ions and, because of this reason, are believed to be much stronger cosmic-ray accelerators than pair shocks which, however, accelerate leptons and thus are directly subject to synchrotron emission. The main interest in external shocks should really be on treating electron-nucleon plasmas with, possibly, an admixture of an additional dilute pair component. Such plasmas have $\mu\gg 1$, in case of a pure electron-proton plasma $\mu=1836$. This mass ratio puts the time scales on which electrons and nucleons evolve far apart yet does not allow to treat the ions as either immobile or as a simple fluid. 

In the following we briefly review a few contemporary attempts to simulate relativistic collisionless shocks that include nucleons. Basically two different types of simulations have been performed: (i) those where the focus is on the generation (and injection) of energetic particles by the relativistic shock-production process as well as on the production of magnetic fields; this resembles the pair plasma simulations; (ii) simulations where a high-energy cosmic ray component is assumed to be present and the focus is on its further acceleration by considering a beam-plasma interaction which produces strong magnetic fields and scatters the Cosmic Rays. In addition, cosmic-ray-modified and also radiation-modified shocks have been treated. However, in collisionless shock theory on which we focus, these modifications are of secondary importance. Though of interest in cases where the cosmic ray component is externally given, we may spare any efforts to look into such modifications until a genuine understanding of relativistic shock formation, shock structure and shock properties under collisionless conditions has been obtained, a state from which we are still far away.

\subsubsection*{\textsf{No Cosmic Rays}}
\paragraph{Initially Non-magnetic Shocks.}
The first successful PIC simulational attempts by \citet{nishikawaea03,nishikawaea06}, \citet{frederiksenea04}, \citet{hededalea04,hededalea-NC05}, and \citet{aa06} to infer the 3D-shock structure, magnetic field generation and particle acceleration in nonmagnetised relativistic electron-proton plasmas still used a small mass ratio of $\mu=16$. \citet{frederiksenea04} worked with a  Lorentz factor $\Gamma=3$ and performed simulations on a small grid $200^2\times800$ with 25 particles per cell, a total of $8\times10^8$ particles and box sizes $(10\lambda_i,10\lambda_i,40\lambda_i)$. The shock was generated in counter-streaming (along $z$) plasma interaction. One of the plasma streams was taken to have 3 times higher density, and the simulations were performed with periodic boundary conditions in ($x,y$) and open conditions in $z$. The simulation time $\tau\omega_e=480$ (corresponding to $\tau\omega_i=120$) was long enough to allow the relativistic particles to cross the box along $z$ roughly $\sim 3$ times. As expected, this simulations allows for the electron Weibel filamentation mode which grows nonlinearly and deflects ions. At later times the ions react with the ion-Weibel mode. Electrons are heated by mixing the two streams. The power spectrum of the magnetic field is a power law toward smaller scales, and cascades inversely in the transition region from electron to ion Weibel modes. 

\begin{figure}[t!]
\centerline{\includegraphics[width=\textwidth]{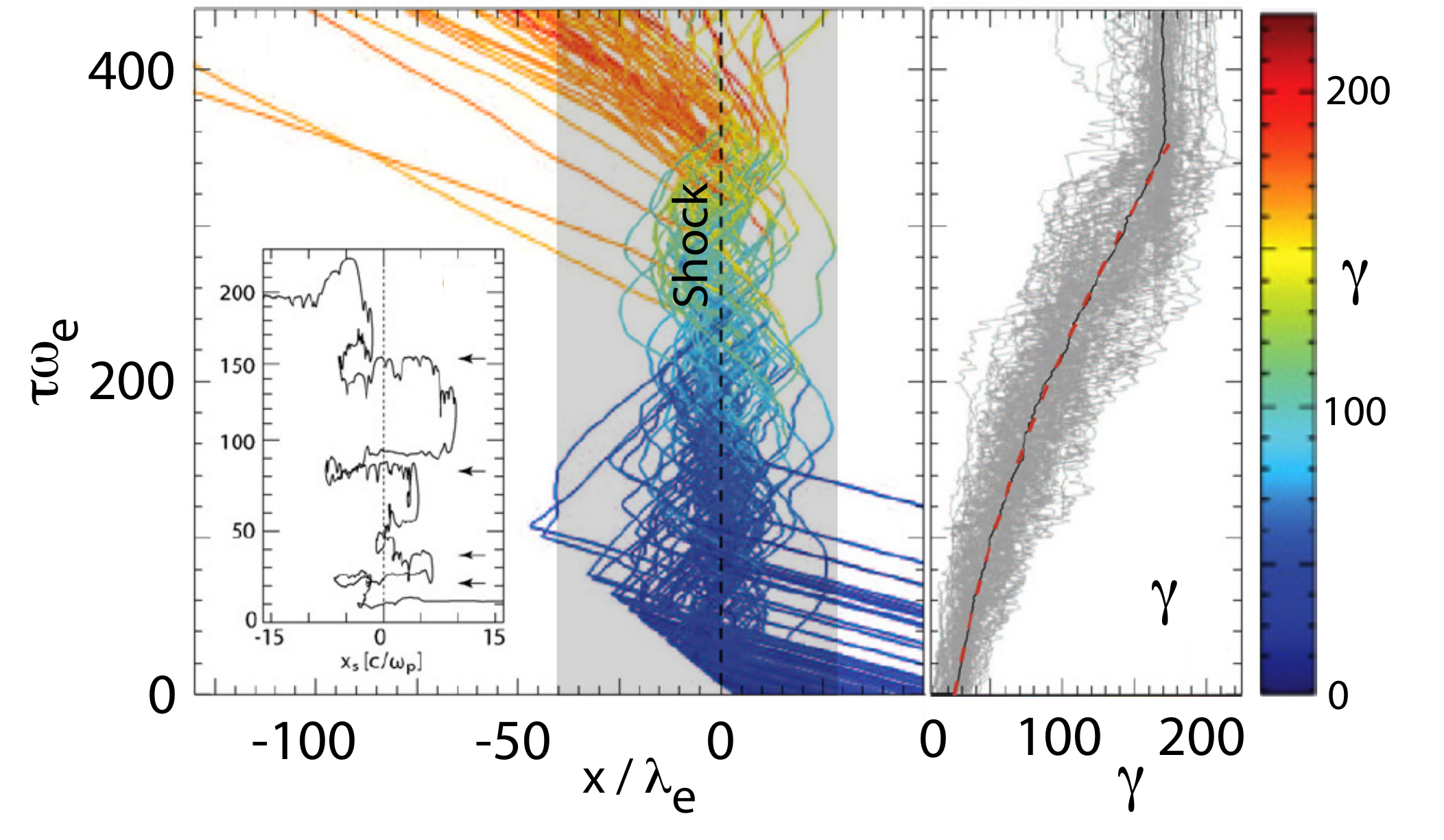} }
\caption[]
{\footnotesize \emph{Left}: Time-evolution of the orbits of the 80 most energetic ions in a non-magnetised  relativistic shock simulation with $\Gamma=20$. The particles are coming from the upstream flow, are back-scattered in the magnetic turbulence in the shock transition, staying within the distance of an ion inertial length $\lambda_i\approx 50 \lambda_e$, and are locally fast enough to traverse the shock a short distance back upstream. They do not make it far into upstream but oscillate around the shock thereby picking up energy as shown in the inset for one particle. One should note that all particles during their shock-drift phases remain within the ion-inertial length shock-transition region of $\Delta x \sim50\lambda_e$ cross-shock extension, as indicated by the shading thus moving together with the shock upstream at the shock velocity. The colour code along the orbits of the particles indicates the increase in particle energy. Once the particles have gained sufficient energy to overcome the reflecting shock potential and the scattering upstream turbulence, they escape with the flow into downstream direction where they produce the high energy tail of the particle spectrum. \emph{Right}: The increase in the proper ion Lorentz factor along the particle orbits. The energies along the orbits form a cloud showing the average increase of the Lorentz factor for all 80 ions up to saturation at about $\gamma\sim150$  \citep[after][]{martinsea09}. }\label{relshock-fig-martins}
\end{figure}

Larger Lorentz factors $\Gamma=15$ which are more suitable for ultra-relativistic shocks, and different numbers and sizes of cells ($125^2\times2000$) were used by  \citet{hededalea04,hededalea-NC05}. These allow to examine in detail the energy evolution of turbulently heated electrons. It turns out that their spectrum is quasi-thermal yet contains a very short power law region just before the exponential cut-off  produced by  escape-losses due the finite length of the simulation box. Though this power law range (being shorter than one decade in energy) might not be overwhelmingly convincing, the observed substantial heating of the electrons in the ion-electron Weibel-mode interaction indicates that relativistic shocks may indeed generate high electron temperatures. In nature these temperatures might possibly  reach the high values needed to explain the observed synchrotron radiation.

Larger mass ratios between $\mu=16$ and $\mu=1000$ have recently been simulated by \citet{spitkovsky08a} in 2.5D dimensions at $\Gamma=15$. He produced non-magnetised shocks in his previously used reflecting-wall PIC code. This time the simulations were run until a steady final state was reached. The code uses 10 cells per $\lambda_e$ and thus resolves the ion-inertial length scale $\lambda_i$. The typical stationary shock thickness in this case becomes $\approx 50\lambda_i$ and seems to be practically independent on the mass ratio up to  value of $\mu\approx 1000$. 

As is expected, the dominant mechanism of magnetic field generation can indeed be attributed to the ion-Weibel instability. As is also expected, the magnetic energy peaks are found right at the shock transition. It comes close to equi-partition with very high final magnetisation ratio $\sigma\approx 1$ locally, but the average magnetic energy remains $\lesssim15\%$ of equipartition at the shock. The high local value of $\sigma$ is caused by the large overshoot in the magnetic field in the shock ramp. 

The downstream region contains isotropised flows and highly structured magnetic islands. Upstream of the shock the Weibel filaments are non-stationary. Stationarity is reached by electron heating which stops the ion-Weibel instability from growing. Electron heating is large with average downstream-electron energy ratio $\epsilon_e\sim 0.5$ as needed in GRB afterglows, but the electrons are mainly thermal. Hence, it seems that electron-proton shocks behave rather similarly to relativistic pair shocks.

Similar results have been obtained by \citet{martinsea09} for a relativistic electron-ion stationary shock in a PIC simulation with initial $\Gamma\approx20$ but small $\mu=32$ and with a resolution of 10 cells per $\lambda_e$. This high resolution was chosen in order to follow the single particle dynamics. As before, the magnetisation evolved up to $\sim 20\%$ of equipartition near the shock, somewhat less than in the above described simulations by \citet{spitkovsky08a} which is probably due to the smaller mass ratio. Interestingly, both electron and ion energy distributions evolved into high energy tails up to proper Lorentz factors $\gamma\gtrsim 100$. These authors picked out the 80 most energetic particles in their simulation and followed their orbits. Those energetic ions who became most strongly accelerated  started hanging around at the shock front for several 100 plasma times in which they picked up energy until being push to their energy limits and being able to overcome the reflecting shock potential and make it to downstream. This process is very similar to shock-foot formation in quasi-perpendicular non-relativistic shocks with the main difference that in the relativistic case the magnetic shock field is self-consistently generated by the Weibel-mode instability. This is shown in Figure \ref{relshock-fig-martins}. It is most interesting that their is no indication for the ions to run the shock out. Instead, all the high energy ions seem to originate in the upstream region and do for quite long times not escape from the shock transition staying within an ion skin-depth distance of the shock ramp.

\paragraph{Magnetised Shocks.} Since astrophysical shocks in PWNe, GRBs and AGNs are magnetised (as is obvious from observation of synchrotron radiation), it is important to investigate the effect of magnetic fields on shock formation. In non-relativistic shocks magnetic fields are the most important ingredient as well, because no collisionless non-relativistic shocks would exist without them (we do not speak of the microscopic electrostatic shocks here). 

The dependence of PWNe termination shocks and internal shocks in GRBs on the presence of an ambient magnetic field has recently been subject to 2.5D PIC simulations in electron-proton plasmas \citep{ss10} for small mass ratios $1<\mu\leq 100$. It turns out that, like in the pair plasma case, relativistic ions (i.e. protons)  can return along the magnetic field and enter the shock  upstream region only when the shock stays `subluminal'. This implies that the shock must be nearly parallel with magnetic  inclination angle $\theta_{Bn}<30^\circ$ for $\Gamma>5$ and $\sigma>0.03$. Such ions are further accelerated by `surfing' along the shock. Their downstream energy spectrum evolves into a thermal spectrum with an  energetic-ion power-law tail containing $\sim3\%$ of the downstream ions while consuming  $\sim10\%$ of flow energy. Electrons are accelerated much less strongly. This is due to their stronger magnetisation. Their number is one order of magnitude smaller than that of the energetic ions. These findings are not particularly encouraging concerning the ultra-relativistic shock particle (by either SDA or DSA) acceleration, for it must be expected that most magnetised collisionless shocks are very close to being perpendicular with $\thetabn\gtrsim 80^\circ$.

These simulations are supported by an independent 3D PIC simulations \citep{hn05} where a relativistic plasma jet was injected into a plasma immersed into an ambient magnetic field either parallel or perpendicular to the jet. The direction of the ambient field affects the evolution of the electron Weibel instability and the associated shock in simulations with (low) mass ratio $\mu=20$ when the jet has Lorentz factor $\Gamma=5$. Like in a  pure pair plasma (as discussed above), the Weibel two-stream instability  grows in the parallel and weakly oblique case below the threshold magnetic field strength $\sigma\sim 0.03$ where it gives rise to a transverse magnetic field. 

If the ambient field is perpendicular,  the picture in the jet case becomes very complex. Behind the jet front (i.e. the shock) inertial separation between electrons and ions causes charge-separation fields in which both ambient and jet electrons become strongly accelerated.This kind of electrostatic acceleration, or heating of the electrons, transfers excess energy stored in the ions to the electrons and thus provides a way of electron acceleration which is, however, restricted just to the shock front. In an extended jet-shock front the electrons in this field remain magnetised and will in addition perform an ${\bf E\times B}$ drift along the shock front, which gives rise to an electron sheet current, whose magnetic field  is high while being  restricted to the shock where it generates the shock overshoot field. This current can become quite strong, probably even strong enough to lead to the excitation of the Buneman instability (or to the modified-two stream instability), which both result in the formation of electron holes in the shock transition region and cause further electron heating and non-thermal acceleration. These mechanisms have not yet been sufficiently explored for relativistic shock though being well-known in non-relativistic shock dynamics \citep{Balogh11}. {First investigations in this direction have only recently been published \citep{dieckmann09,dieckmannea2010,murphy2010}. In particular, \citet{murphy2010} in weakly relativistic numerical simulations of a shock, found that the shock evolved into a quasi-perpendicular shock just because of the presence of this drift current sheet which became unstable and formed magnetic structures, i.e. magnetic flux tubes of fairly homogeneous strong magnetic field. Relativistic electrons gyrating in these field may contribute to the generation of synchrotron radiation.}

\subsubsection*{\textsf{Cosmic Ray Beams and External Shocks}}
Simulations including Cosmic Rays put emphasis on the generation/amplification of magnetic fields either by the Weibel filamentation or Bell-like modes \citep[cf., e.g.,][]{aa06,amatoea08}, the modification of the shock, increase of the compression factor and further acceleration of Cosmic Rays, boosting their energies  up as much as possible. This kind of approach is less suited for understanding shock formation than for purposes of application (mostly to external SNR shocks) where the shock does already exist. The cosmic-ray particles are then either generated in the self-consistent and still badly understood evolution of the shock or -- and more probably -- originate from  external sources, encounter the shock and are accelerated to higher energies, modify the shock and contributinge to amplification of the shock magnetic field. In the following we only consider recent simulations where cosmic ray beam plasma interactions are investigated in view of generation of magnetic fields (Weibel and Bell modes).

Typical approaches of this kind are found in \citet{niemiecea08} who, in view of application to the external shocks found in SNRs, inject a weakly relativistic $\Gamma=2$ cold cosmic ray beam of current $J_\mathit{cr}=eN_\mathit{cr}c\beta_\mathit{sh}$ into the plasma which is assumed to drift with the shock along the magnetic field $B_{0\|}$ into the thermal upstream medium. $\beta_\mathit{sh}$ is the normalised relativistic shock speed. The shock is assumed as quasi-parallel. The set-up is charge neutral, and the current is compensated by the return current $J_\mathit{ret}=-eN_eV_d$ carried by the background electron component. Periodic boundary conditions and various 3D and 2D grids  have been used. Mass ratios range from $\mu=3$ to $\mu=500$, and the angle of the ambient magnetic field is varied from $\theta=40^\circ$ to $80^\circ$. The main result of these simulations is that the growth of the magnetic field is slower, strongly affected by the obliqueness of the field, and the saturation level substantially smaller than analytically predicted by linear and non-linear theories and by the assumed saturation through magnetic particle trapping and scattering (as has been discussed in earlier sections), which would just yield $\langle|{\bf b}^2|\rangle/B_{0\|}^2\approx 1$. Regarding the effect of the magnetic field turbulence on the particle distribution it is conjectured that in all simulations alignment of the bulk and cosmic ray flows is obtained. 

Additional simulations have recently been performed by the same authors with cosmic ray beam-plasma interactions at much higher Lorentz factors up to $\Gamma=300$ in 2.5D with mass ratio $\mu=20$ and cosmic-ray flow along the ambient magnetic field \citep{niemiecea10}. Referring to the linear growth rate of the Weibel or Bell instabilities, the simulation time allows for 20 to 30 e-foldings of the growing waves. These simulations essentially confirm the above result of generation of magnetic fields bei either instability while reaching much higher field amplifications up to $\langle|{\bf b}^2_\perp|\rangle/B_{0\|}^2\approx 5\times 10^2$ in the highest ultra-relativistic case and maximum-growing wavelengths with $\lambda_\mathit{max}/\lambda_i$ approximately 10 to 30. It is, however, not entirely clear what this large amplification factor actually means if the simulation starts with initially zero magnetic field. The amplification is roughly proportional to $\Gamma$, with the parallel field being less amplified by about one order of magnitude. The reason for the magnetic field amplification is the non-resonant interaction between the cosmic ray beam and the plasma, and the Cosmic Rays respond by being scattered in the wave field approximately corresponding to Bohm diffusion. 

\subsection{\bf\textsf{Radiation}}

\subsubsection*{\textsf{Self-consistent Photon Spectra from Simulations.} }
The ultimate link (and proof of astrophysical validity) between the theoretical and simulation results is the reproduction of observed radiation spectra by radiation spectra inferred from simulated shocks. Until recently, such attempts have not been particularly successful, because the `synthetic' spectra from simulations could not be measured directly. With the available grid size sufficiently short wavelengths, which occur in real space, cannot be resolved. Spectra were inferred subsequently from the particle distribution by applying radiation theory, which is a reasonable approach as long as radiative losses and the shock response to the radiated photons can be neglected. 

Recently a different approach has been developed which is based on post-simulation processing codes which have become available. With their help radiation spectra can be calculated directly from the particle motion by considering the sub-gyro-orbit `jitter'-acceleration-deceleration sequences of electrons and positrons when following the particle orbits and by use of the far field representation of the radiation field \citep{hededal05,martinsea09a}. A theoretical account for this radiation including simulation results has been given recently  \citep{medvedev09,medvedevea10}. 
The new methods allow for resolution up to 1\% of the cell size and have been applied to both  collisionless pair and electron-proton shocks in order to determine the  shape of the emitted photon spectrum  \citep{ss09b,nishikawaea10,nishikawaea10a,frederiksenea10}. 

\citet{ss09b} exploit $\sim10^4$ energetic electrons moving in the self-consistently produced fields of their 2D PIC shock simulations and were followed over 135 plasma times. Such spectra must be regarded as local and instantaneous. The photon spectra obtained in these simulations obey the expected shape of synchrotron spectra of electrons moving in a strong magnetic field. Toward low frequencies all calculated spectra consistently decay like a $\sim\omega^{-\frac{2}{3}}$ power law,  in perfect agreement with the expectation of a 2D synchrotron spectrum emitted in magnetised media. Qualitatively though not quantitatively (yet not surprisingly) this result confirms that the downstream electrons, which have been self-consistently accelerated in the downstream shock turbulence, are the sources of synchrotron radiation that is emitted by relativistic shocks.  

\citet{frederiksenea10}, in contrast, in a similar investigation found a number of additional results in 2D and 3D PIC simulations of pair and electron-proton plasmas, the most prominent one is that the spectral peak increases first as $\omega_\mathit{peak}\propto \Gamma^2$ and then as $\propto\Gamma^3$. Moreover, the spectral shape and peak depends on angle. The spectra are anisotropic, indicating beaming, which seems more pronounced in electron-proton than in pair plasma. Their spectra, in contrast to \citet{ss09b} are interpreted as non-synchrotron and become synchrotron only if the magnetic field is taken as static and not self-consistent. In addition, their spectra differ substantially in 2D and 3D simulations.  

These methods, as ingenious as they are and as intriguing and seducing as their results may be, should however be taken with substantial care. Not only that the two above mentioned calculations produced different and partly contradicting spectra. The spectra derived from the simulations by these methods are by no means real particle spectra but spectra extracted from macro-particles which do not account for the emission of each single electron. They, instead, refer to a very large coherent clump of (non-physical) particles with identical or hidden internal dynamics, and it is not clear what the radiation spectra which are determined in the above described way really mean. Their uncritical application to real problems must therefore be cautioned.

\subsubsection*{\textsf{Radiation Mediated Shocks.}} 
Another broad research field in relativistic shocks is centred on the question to what extent shocks are modified by their own self-consistently produced radiation. Radiation may indeed modify shocks, but the effect is different from that caused by the production of Cosmic Rays. Radiation results in energy loss and, if  reabsorbed, causes redistribution of energy in the shock environment. This problem becomes urgent in hot relativistic shocks which emit very intense radiation. Such shocks have so far been excluded from our discussion.  Here, we just note that \citet{budnikea10} have recently investigated some of the consequences that intense self-generated radiation has on the relativistic shock structure for Lorentz factors in the interval $6<\Gamma<30$ under the assumption of coupling of electrons and protons through plasma processes.  These authors  provide the relevant equations and boundary conditions for later application. Shocks of this kind become radiation-mediated. Simulation of the related problems would involve the self-generated radiation field in the equations of motion of the simulated macro-particles, which also includes pair-production in the intense radiation fields. So far the radiation produced in simulations is, however, undetectable. In order to observe its effect its intensity would need to be up-scaled to the expected local strengths. This is still far from any practical application.

\subsubsection*{\textsf{Relevance for Modelling Gamma Ray Bursts.} }
{Following the original suggestion  by \citet{remington1999},} \citet{ms09} have addressed the important question, whether such 3D PIC simulations {\citep[and also related laboratory laser plasma experiments like those performed by][]{kuramitsu2011}} are just qualitative and, quantitatively, will or will not in the near future become comparable with astrophysical observations, as for instance the radiation from GRBs. In order to answer this question, one needs to compare the radiative cooling time of relativistic electrons in the self-consistently generated shock magnetic field and the microscopic dynamical time of the evolution of collisionless relativistic shocks, i.e. the inverse proton plasma time $\tau_\mathit{cool}\omega_i$. In current simulations in 3D, the result of such a comparison is that, for $\tau_\mathit{cool}\omega_i\lesssim 10^3$, electron cooling in the vicinity of the shock becomes efficient such that the electrons contribute to a substantial loss of kinetic energy in the form of emitted radiation. Current 3D PIC simulations can, indeed, resolve this type of relativistic shocks, and thus permit  the detailed study of their micro-scale structure. 

With modern methods (and the caution noted above) it might become possible to infer the radiation spectrum of the electrons, and even to determining their spectral slopes and peak or integrated powers. The radiation is, however, restricted to that emitted from the close vicinity of the shock corresponding to the motion of the electrons in the simulation. Such spectra can, with caution, be compared with observed radiation spectra to provide information about the particle population, energy, and the type of shock  responsible for the generation of the radiating particle distribution. Shocks of this kind are found as internal radiative shocks in baryon dominated GRBs. On the other hand, radiative cooling limits any Fermi acceleration (if present) of the electrons, and this implies that the electron spectrum is thermal, i.e. lacking a high energy tail, and should peak in the multi-MeV range. In the opposite case when the radiation efficiency of the electrons  is less than that predicted by the simulations, the electron distributions may possess energetic tails. External shocks, on the other hand, like those believed to be responsible for  the extremely early afterglow phase, will probably be cooled only when they are still highly relativistic.

\section{Conclusions}
Despite intense and increasing research over the last decade, which is reflected in the vastly growing number of recent publications on this subject, understanding of relativistic collisionless shocks is still far from completion or from a situation, where derivation from first principles becomes feasible. The reason for this lack of understanding lies in the practical non-availability of collisionless relativistic shocks, neither on Earth nor in the near-Earth environment {though this might change \citep{remington1999}} when laboratory experiments will manage to generate highly relativistic shocks in laser fusion plasmas \citep[cf. the comments and references given in][]{ms09}. In particular the three questions of interest in astrophysics, {(i)} the generation of strong large-scale magnetic fields in shock interaction, {(ii)} the related shock acceleration of Cosmic Rays to very high energies, and {(iii)} the self-consistent generation of radiation still lack any sufficient understanding and mapping to observations. Yet progress cannot be denied and has been accounted for in this review \citep[and also the in-depth discussion given by][]{medvedev09a}. 

In the following we very briefly summarise what, as seen by us, is the current state of the art with respect to a final answer to these questions.

\subsubsection*{\textsf{Magnetic Field Generation}}
Generation of magnetic fields in relativistic shocks is believed necessary and unavoidable in  increasing the shock compression ratio, as required by the DSA mechanism (in case it will be confirmed that this is the only, unavoidable, or the dominant mechanism of particle acceleration), as well as for the generation of synchrotron radiation. 

The main processes of magnetic-field generation in relativistic shocks that have been put forward during the last two decades are based either on the Weibel filamentation instability, which has been confirmed by the numerical simulations in pair and electron-ion plasmas, or the generation of Bell-like modes which has also been confirmed in numerical simulations. 

The Weibel instability works best in initially non-magnetised (or very weakly magnetised) low-temperature relativistic flows or in flows about parallel to the ambient magnetic field. Saturation levels are, however, relatively low, and are restricted to distances of few (ion-nucleon) skin depths, not far downstream of the shock. However, the Weibel modes re-magnetise the shock transition such that the shock becomes about magnetically perpendicular. {In fact, being a transverse mode, the Weibel filamentation instability generates quasi-toroidal transverse magnetic fields by splitting the original homogeneous beam up it into beamlets or beam filaments on the fastest growing transverse scale. This implies a highly structure magnetic field along the shock surface different from the model of a homogeneous quasi-perpendicular shock plane. Since the filamentation instability, however, provides an entire spectrum of scales, the structure of the magnetic field along the shock surface becomes complex. }

The finding in numerical simulations that the downstream region carries magnetic fields far more distant from the shock transition than the Weibel mechanism predicts is not well understood yet. These fields form vortices {\citep[as first identified in most recent simulations of][]{murphy2010}}, which might suggest that in 3D the Weibel fields in the shock transition evolve nonlinearly into current loops and (possibly reconnected) field vortices{\footnote{When this happens, it would, on the other hand, put into question the common assumption that such field vortices stabilise the shock transition by particle trapping. At the contrary, one expects that the shock transition becomes highly structured and complex in the tangential direction \citep[as suggested by][]{murphy2010}, with saturation and stabilisation of the Weibel fields provided by non-linear effects quite different from trapping, an effect whose consequences have not been investigated yet.}} which are advected far downstream by the shocked turbulent flow and thereby possibly fill an extended downstream region with small-scale magnetic fields.

Bell-like modes require the presence of a cosmic-ray population upstream of quasi-parallel shocks. This population is either generated self-consistently by the shock (i.e. raising the injection problem) or it is imposed from the outside. The Bell-like fields occupy  the shock-foreshock, causing the foreshock to become turbulent on short scales yet offering the possibility of diffusive interaction which drives a secondary wave population possessing a longer-wavelength spectrum of scales up to several cosmic-ray gyro-radii. 

The turbulence caused by these waves retards the upstream flow though possibly only slightly. Since the turbulent waves have both a very low frequency and a very low upstream phase speed, the relativistic upstream flow must advect them toward the shock transition, where they  accumulate. Depending on how much time this takes, the waves with longer wavelength will grow to non-linear amplitudes and contribute to a nonstationarity of the shock, thereby forcing the shock to reform on irregular time-scales. This reformation will be similar to what is known from upstream magnetic pulsations in non-relativistic quasi-parallel shocks \citep{Balogh11}, a process which has not been given any attention so far in relativistic shock theory!

{Once this happens, the shock will continually radiate Bell-like (and also Weibel) magnetic fields to the downstream region. Being frozen into the flow like quasi-stationary structures, magnetic vortices and flux tubes of different scales, these fields become advected downstream by the flow with very little turbulent attenuation only thereby contributing to and extended region of downstream turbulence. They will gradually occupy almost the entire downstream domain up to distances very far away from the narrow shock ramp and shock transition. Since the plasma on the scales of these structure is manifestly collisionless, they indeed experience only very little attenuation. This process might provide the main mechanism of generating highly turbulent magnetic fields in the shocked medium far behind relativistic collisionless shocks.}

\subsubsection*{\textsf{Particle Acceleration and Radiation}}
There can be no doubt that relativistic and, in particular, ultra-relativistic shocks accelerate charged particles to high energies. In the present review we have not dealt with particle acceleration. There is a wide literature on this subject, mainly favouring the DSA mechanism the consultation of which the reader is referred to. However, it remains unclear whether the DSA mechanism is involved in the acceleration process at ultra-relativistic shocks. Available simulations on relativistic shock formation in the absence of external Cosmic Rays provide little or at least no unambiguous indication for the DSA mechanism and would restrict it just to the case of highly magnetised nearly parallel shocks. 

The problem is buried in the following: in very fast shocks it is difficult for upstream directed particles of the thermal downstream particle distribution to outrun the shock. These particles must originate in the downstream thermal tail and must have been pre-accelerated by some turbulent downstream mechanism to Lorentz factors substantially larger than the shock's proper Lorentz factor such that their upstream speed exceeds the shock velocity.\footnote{{It is sometimes argued that in ultra-relativistic shocks also the downstream particles move at velocity of light $c$ which is independent of any coordinate frame and, thus, the particles would also move at light velocity with respect to the shock system. This, however, holds only for photons. The velocity of any massive particle must be transformed according to the rules of relativity, and so, relative to an ultra-relativistic shock, they will have difficulties to catch up with the shock unless their $\Gamma$ factor becomes very high, i.e. unless they have already been accelerated to very high energies in the region downstream of the shock.}}  In addition, for magnetised shocks this outrunning will be made possible only if the shock is nearly parallel with magnetic inclination angle against the shock normal, i.e. $\thetabn\lesssim30^\circ$. This excludes a large class of relativistic and ultra-relativistic shocks, in particular external shocks, which have always been believed to be close to perpendicular. 

If a particle makes it finally up to the shock, simulations seem to indicate that it will oscillate for a while around the shock front, experiencing acceleration until it ultimately escapes downstream and contributes to a power law tail. This kind of acceleration resembles SDA where the particles during the phases when they enter into upstream close to the shock experience the upstream convection electric field and become accelerated in a way that is similar to pick-up ion acceleration in the shock-foot region over the part of their upstream gyro-orbit exposed to the upstream convection electric field ${\bf E}=-{\bf \beta\times B}_1$. This may happen many times for, when the particles enter the shock from upstream, the strong shock potential reflects the particles back upstream and the shock ramp sweeps them along until they have picked up sufficient energy to overcome the shock potential and leave from the shock to downstream. For this process it is even not at all needed that downstream particles reach the shock front. In Figure \ref{relshock-fig-martins}, for instance, all ultimately accelerated ions have come solely from the upstream flow while becoming accelerated to large Lorentz factors $\Gamma$ in this process and, in principle, the measured final Lorentz factors $\Gamma_\mathit{fin}\sim 150$ of the particles in this simulation can be taken as mapping the cross-shock potential $U_\mathit{sh}\lesssim m_ic^2\Gamma_\mathit{fin}/e$ ($\approx 150$ GV in this particular case{; accounting for the reduced mass used in the simulations this amounts to $\approx 4.5$ GV in a proton plasma. The simulations use a $\Gamma=20$ which implies protons of energy $\sim 20$ GeV, thus the cross-shock potential consumes roughly 25\% of the kinetic energy of the upstream flow}) which the accelerated particles must overcome.  In non-relativistic shocks, these particles would be considered as shock-reflected ions, while in the ultra-relativistic case with high shock velocities $\beta_\mathit{sh}$ they cannot escape far upstream from the shock. The shock sweeps them in front of it upstream over a distance until they can escape downstream from the shock. Possibly particle acceleration at ultra-relativistic shocks is just this kind of reflection-surfing  SDA acceleration. 

It remains unclear, however, how in electron-nucleon plasmas the energy is transferred from shock accelerated nucleons to electrons so that the ultimately observed photon-radiation spectra are generated. The mechanism of electron acceleration is still badly understood. Recently, referring to the process of reconnection and generation of strong electric fields known from  near-Earth space physics, it has been suggested in view of the GRB afterglow that electron acceleration could be provided by a manifestly non-Fermi process that is based on acceleration in strong local magnetic-field aligned electric potential differences \citep{medspit09} . Indeed,  such local electric potentials being built up, might be the ultimate mechanism of electron acceleration \citep[cf.,][]{tt2001}. This process is know from magnetic reconnection \citep[cf., e.g.,][]{tjp09,tnb10} and from the strong magnetic field-aligned electric fields in the aurora which most probably are caused by shear flows in local turbulence. In the case of the Weibel instability such fields might arise from merging of current filaments, while in the shock front they will evolve naturally as a consequence of the cross-shock potential and the resulting strong electron cross-field drift current in the shock front, which should ultimately disrupt into electron phase space holes  by the action of the Buneman instability. The accelerated electrons experience a jitter motion and cause radiation.  We referred to radiation spectra of this kind which were calculated from following the jitter motion of electrons and  yielded satisfactory results on the shape of the emitted photon spectra. Future refinements of such calculations may indeed reproduce the observed spectra, though one should keep in mind the caveat that the simulated particles are just macro-particles which do not exhibit the individual motion of the accelerated electrons in the real natural objects. The calculated radiation spectra are kind of mock spectra the similarity of which to real spectra may be just incidental. Moreover, as long as the electron acceleration mechanism remains obscure one may not expect any fast progress. 

In the light of these remarks  \citep[cf. also][]{medvedev09a} we acknowledge without any reservation that the most recent numerical full particle simulations have substantially advanced our knowledge about relativistic collisionless shocks. Undoubtedly, they have contributed to a much deeper understanding of shock formation, shock structure, and the related acceleration and radiation processes. The technical improvements  expected in the near future concern a realistic mass ratio $\mu=1840$ in 3D simulations of electron-proton plasmas, substantial extensions of simulation boxes and simulation times, and increases of cell numbers, particle numbers and resolution.

\begin{acknowledgements}
We thank the Editors, M.C.E. Huber and T. Courvoisier for invitation to this review and, in addition, M.C.E. Huber for very careful reading of the manuscript and many valuable suggestions. We also thank the anonymous referees for very valuable comments which helped rectifying the paper, led to several crucial corrections, accounting for omissions and the addition of a number of previously missed references. The comment on the Boris pusher in Section 4 is the sole credit of the referees. RT acknowledges discussions over the years with F. Aharonian (Dublin, Ireland), A. Balogh (Imperial College London and International Space Science Institute Bern), J. Bleeker (Utrecht U), M. Gedalin (BGU Israel), M. Hoshino and C. H. Jaroschek (Tokyo U), S. Matsukiyo (Kyushu U) and P. Yoon (UMD, USA). Part of this work has been prepared during Visiting Scientist periods in 2006/2007at ISSI Bern. The hospitality of the ISSI staff and the support of the librarians, Irmela Schweizer and Andrea Fischer, is cordially acknowledged. 
\end{acknowledgements}

\bibliographystyle{spbasic}      
\bibliography{relshock-4-bib-SI}   
\vspace{-0.7cm}


\end{document}